\documentclass[pra,aps,reprint,superscriptaddress,twocolumn]{revtex4-2}
\pdfoutput=1
\usepackage{hyperref}

\usepackage{comment}
\usepackage{tikz}
\usepackage{lipsum}
\usepackage{graphicx,color}
\usepackage{amsmath}
\usepackage{bm}
\usepackage{amssymb}
\usepackage[all]{hypcap}
\usepackage{mathtools}
\usepackage{dsfont}
\usepackage{physics}
\usepackage{enumitem}
\usepackage{lipsum}
\usepackage{float}
\usepackage{tabularx}
\usepackage{framed,enumitem} 
\usepackage{adjustbox}
\usepackage{threeparttable}
\usepackage{float}
\makeatletter
\let\newfloat\newfloat@ltx
\makeatother
\usepackage{algorithm}
\usepackage{algpseudocode}
\begin{document}
	\title{Theory and Experimental Demonstration of Wigner Tomography of Unknown Unitary Quantum Gates}
	\author{Amit Devra}
	\email{amit.devra@tum.de}
	\affiliation{\footnotesize Technical University of Munich, TUM School of Natural Sciences, Department of Chemistry, Lichtenbergstra{\ss}e 4, 85748 Garching, Germany}
	\affiliation{\footnotesize Munich Centre for Quantum Science and Technology (MCQST), Schellingstra{\ss}e 4, 80799 M{\"u}nchen, Germany}
	\author{L\'eo Van Damme}
	\affiliation{\footnotesize Technical University of Munich, TUM School of Natural Sciences, Department of Chemistry, Lichtenbergstra{\ss}e 4, 85748 Garching, Germany}
	\affiliation{\footnotesize Munich Centre for Quantum Science and Technology (MCQST), Schellingstra{\ss}e 4, 80799 M{\"u}nchen, Germany}
	\author{Frederik vom Ende}
	\affiliation{\footnotesize Institut f{\"u}r Theoretische Physik, Freie Universit{\"a}t Berlin, Arnimallee 14, 14195 Berlin, Germany}
	\author{Emanuel Malvetti}
	\affiliation{\footnotesize Technical University of Munich, TUM School of Natural Sciences, Department of Chemistry, Lichtenbergstra{\ss}e 4, 85748 Garching, Germany}
	\affiliation{\footnotesize Munich Centre for Quantum Science and Technology (MCQST), Schellingstra{\ss}e 4, 80799 M{\"u}nchen, Germany}
	\author{Steffen J. Glaser}
	\email{glaser@tum.de}
	\affiliation{\footnotesize Technical University of Munich, TUM School of Natural Sciences, Department of Chemistry, Lichtenbergstra{\ss}e 4, 85748 Garching, Germany}
	\affiliation{\footnotesize Munich Centre for Quantum Science and Technology (MCQST), Schellingstra{\ss}e 4, 80799 M{\"u}nchen, Germany}
\begin{abstract}
We investigate the tomography of unknown unitary quantum processes within the framework of a finite-dimensional Wigner-type representation. This representation provides a rich visualization of quantum operators by depicting them as shapes assembled as a linear combination of spherical harmonics. These shapes can be experimentally tomographed using a scanning-based phase-space tomography approach. However, so far, this approach was limited to \textit{known} target processes and only provided information about the controlled version of the process rather than the process itself. To overcome this limitation, we introduce a general protocol to extend Wigner tomography to \textit{unknown} unitary processes. This new method enables experimental tomography by combining a set of experiments with classical post-processing algorithms introduced herein to reconstruct the unknown process. We also demonstrate the tomography approach experimentally on IBM quantum devices and present the specific calibration circuits required for quantifying undesired errors in the measurement outcomes of these demonstrations.
\end{abstract}
\maketitle
\section{Introduction}
\label{intro}
\begin{figure}[!]
	\centering
	\includegraphics[scale=1.1]{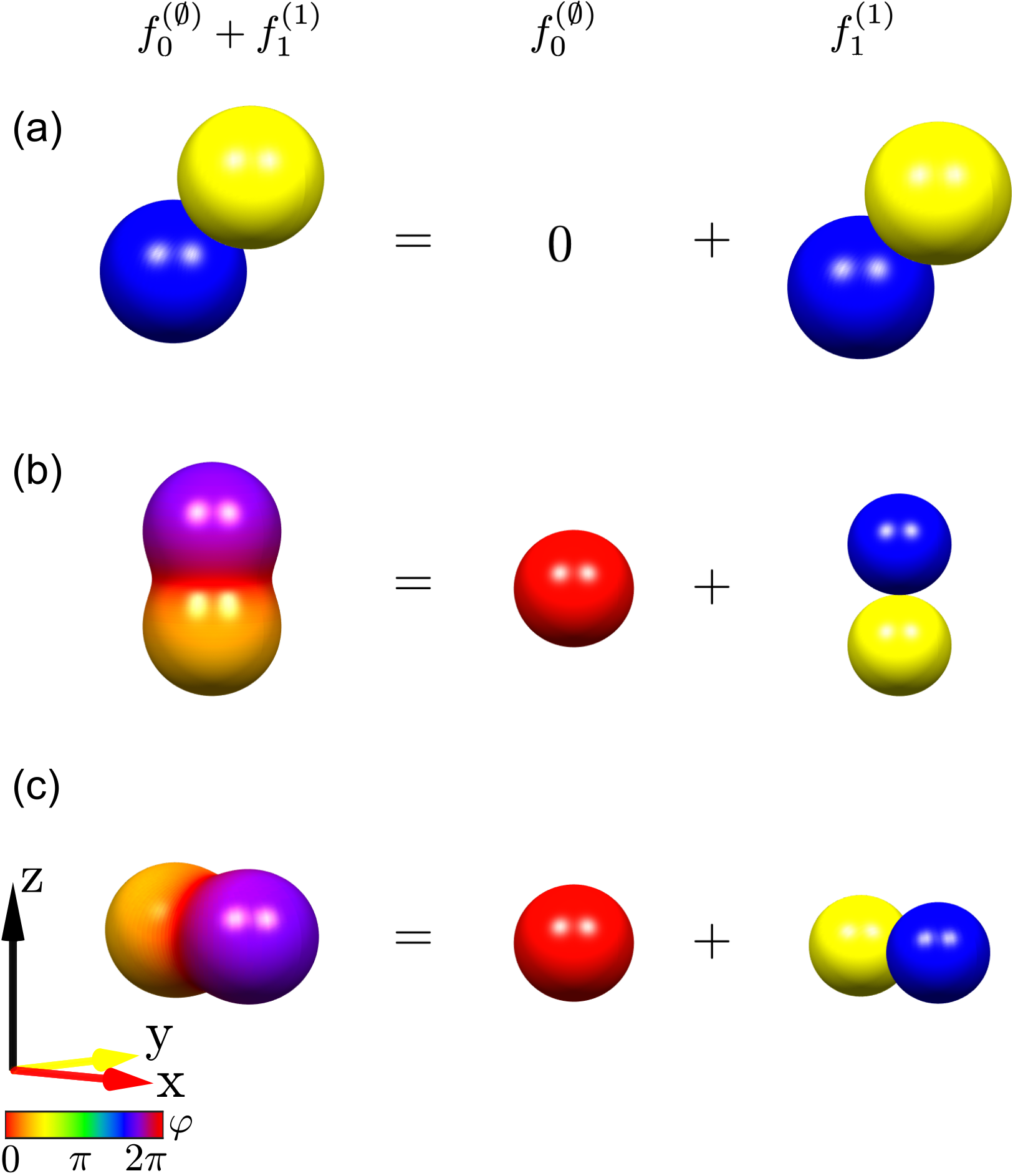}
	\caption{DROPS visualization of different single qubit quantum gates: (a) Hadamard (H), (b) $\frac{\pi}{2}$ phase shift (S), and (c) $\sqrt{\text{NOT}}$. This visualization is based on mapping: $A = \sum_{\ell\in L} A^{(\ell)} \longleftrightarrow  \bigcup_{\ell\in L} \{f^{(\ell)}\}$ of a quantum operator $A$ onto a set of spherical droplet functions $f^{(\ell)}$ with label $\ell$ (where $\ell=\emptyset$ and $\ell=1$ corresponds to identity and linear terms respectively). For single-qubit gates, the direction of the $\ell=1$ droplet directly reflects the rotation axis of the gate. In these three-dimensional polar plots of droplets $f^{(\ell)}(\theta,\phi)$, the distance from the center to a point on the surface and phase color $\varphi$ is described by $\abs{f^{(\ell)}(\theta,\phi)}$ and $\text{arg}[f^{(\ell)}(\theta,\phi)]$, respectively. See Appendix~\ref{Supp:visualization} for a detailed analysis of this example.}
	\label{fig:intro_example}
\end{figure}

The rapid growth in quantum computing and quantum information processing necessitates a detailed understanding of complex quantum systems. The DROPS (Discrete Representation of OPeratorS)~\cite{DROPS_main} representation is an approach that provides an intuitive visualization of the quantum system and aids in understanding quantum dynamics. DROPS is a generalization of continuous phase-space representation over a sphere, also known as Wigner representation for spins~\cite{Balint_continous,koczor2019phase,Leiner_2020,huber2024beads}. This visualization works by mapping quantum operators (such as density matrices, processes, Hamiltonians, etc.) onto a set of spherical functions, referred to as \textit{droplets}. As an example, Fig.~\ref{fig:intro_example} provides a glimpse into the DROPS representation by visualizing some single-qubit quantum gates. The theoretical description of this representation is summarized in Appendix~\ref{Supp:visualization}. This interactive DROPS visualization is available as a free software named SpinDrops~\cite{spinDROPS}. This approach has recently been expanded to the so-called BEADS representation~\cite{huber2024beads}, which makes it possible to directly obtain quantitative information about expectation values of interest. \\
\begin{figure*}[t]
	\centering
	\adjustbox{max width=\textwidth}{\includegraphics[scale=1]{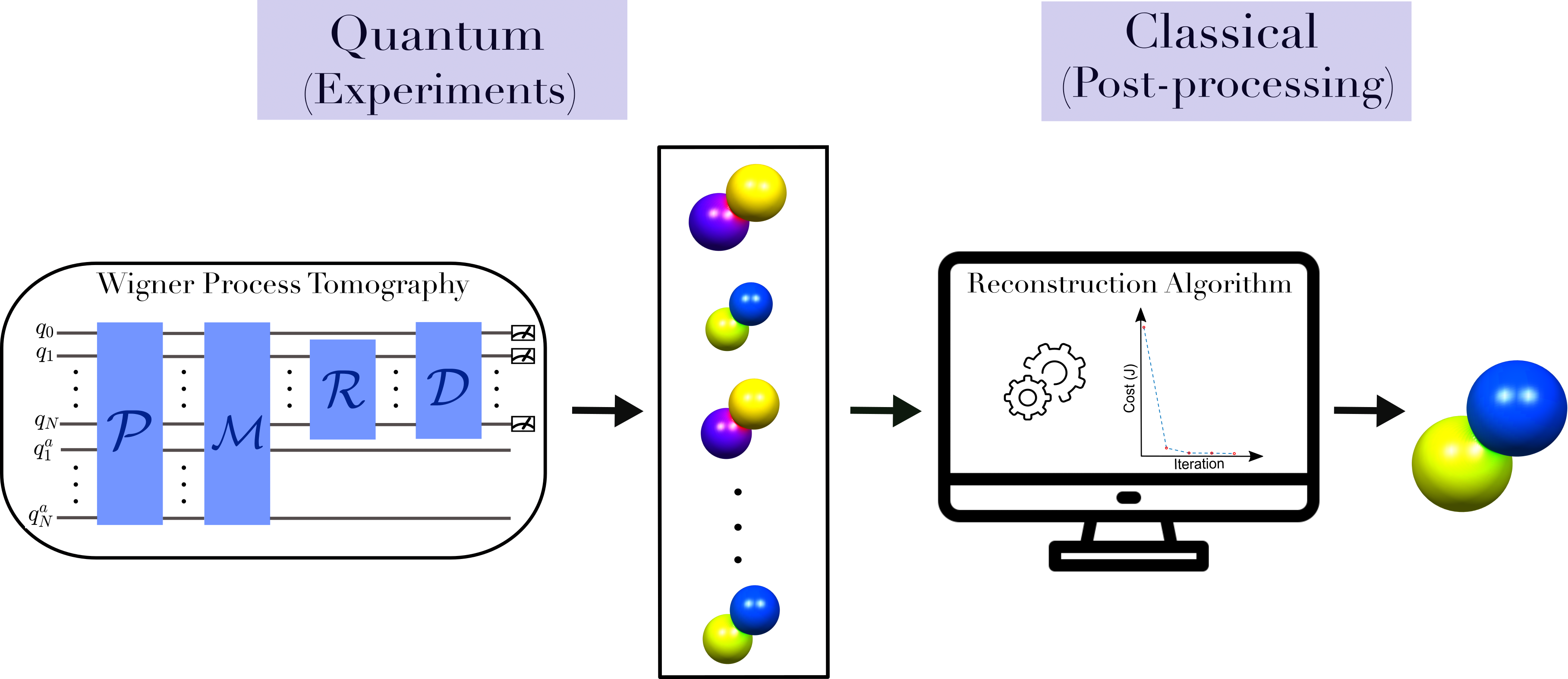}}
	\caption{Schematic overview of the Wigner tomography approach for the unknown processes. In the experimental part, the approach first experimentally scans the spherical droplets $f_{k}^{(\ell)}$ corresponding to the scaled versions of an unknown process $U_{k}^{[N]}$. These droplets are then combined in a post-processing step using a reconstruction algorithm to estimate the droplet corresponding to the unknown process $U$ with a high signal-to-noise ratio.}	
	\label{fig:approach}
\end{figure*}\\

The characteristic shapes of the droplets arising in the DROPS representation are related to experimentally measurable quantities. In our recent studies~\cite{leiner2017wigner,leiner2018wigner}, a scanning-based tomography approach was developed to experimentally tomograph spherical droplets corresponding to quantum states and unitary processes. These tomography techniques, known as Wigner state and process tomography, were initially demonstrated on an ensemble state-based NMR quantum processor. Later, they were adapted to pure-state quantum computing~\cite{Devra_WQST} and experimentally demonstrated on near-term quantum devices. However, so far, the Wigner process tomography approach was limited to \textit{known} target processes. This constraint arises from the requirement of mapping an $N$ qubit unitary process matrix $U^{[N]}$ onto an $N+1$ qubit Hermitian density matrix $\rho^{[N+1]}$ using a controlled process operation $cU^{[N+1]}$, which requires prior knowledge of the process $U^{[N]}$ for experimental execution~\cite{Arajo_2014,Barenco_elementary}. Additionally, experimentally implementing a controlled version of a unitary process matrix, i.e., $cU^{[N+1]}$, differs from implementing the unitary process $U^{[N]}$ itself. Therefore, the information obtained through process tomography pertains to the controlled version of $U^{[N]}$ rather than the process matrix $U^{[N]}$ itself~\cite{Devra_dissertation}. In contrast, the approach presented here does not require the experimental execution of the controlled version of the target process. \\

To address the existing limitation, we propose a quantum circuit that overcomes the constraints of the previous approach. This new circuit utilizes an additional set of $N$ ancilla qubits and maps the scaled versions of an \textit{unknown} $N$ qubit process matrix onto $N+1$ qubit density matrices. This mapping allows for the experimental tomography of spherical droplets corresponding to scaled process matrices. Subsequently, these experimentally tomographed droplets can be combined with a classical post-processing step to extract the unknown process matrix using the `reconstruction algorithm' introduced in this paper. Fig.~\ref{fig:approach} provides an illustrative sketch of the presented approach.\\

The paper is structured as follows. First, we provide a brief background of Wigner state and process tomography in Sec.~\ref{Sec.:Background} and discuss the existing limitations. In Sec.~\ref{Sec.: Mapping}, we examine the limitations (\textit{blind spots}) of the current naive circuit discussed in Ref.~\cite{zhou2011adding,Gavorova} in achieving a complete mapping and introduce a modified circuit to map scaled unknown process matrices onto density matrices. Sec.~\ref{Sec.:WQPT theory} offers a general method for an $N$ qubit system to experimentally tomograph the scaled unknown processes, followed by an algorithm in Sec.~\ref{Sec.:Reconstruction theory} to reconstruct the unknown process from experimentally tomographed droplets for a single-qubit system. Moreover, in Sec.~\ref{Sec.:Reconstruction theory}, we demonstrate how to integrate the reconstruction algorithm with an optimization routine and discuss the resulting benefits. In Sec.~\ref{Sec.:WQPT_Experimental}, we present the experimental implementation of the proposed tomography approach on IBM quantum devices, specifically for the case of a single-qubit system. We also discuss the calibration experiments designed to quantify the undesired errors in the measurement outcomes and present the experimental results. Sec.~\ref{Sec:Adaptive} discusses an adaptive method for the tomography approach, thereby reducing the total number of experiments. Sec.~\ref{Sec.:Reconstuction_extentsion} introduces a generalized reconstruction algorithm with a cost function valid for multi-qubit systems. We conclude by summarizing and discussing the tomography approach in Sec.~\ref{Sec.Discussion}. Further details are deferred to the appendices. 
\section{Background}
\label{Sec.:Background}
In this section, we briefly summarize Wigner tomography of quantum states and known processes~\cite{leiner2017wigner, leiner2018wigner, Devra_WQST, Devra_dissertation} using a scanning-based approach to establish a foundation for Wigner tomography of unknown processes.  
\subsection{Scanning approach and Wigner state tomography}
\label{Sec.:WQST}
An $N$ qubit quantum operator $A^{[N]}$ described by a complex $2^{N}\times2^{N}$ matrix can be represented by rank $j$ and label $\ell$ spherical functions $f^{(\ell)} = \sum_{j\in J(\ell)}f_{j}^{(\ell)}$ as described in Sec.~\ref{Supp:visualization} and illustrated with examples in Fig.~\ref{fig:intro_example}. We are interested in the experimental tomography of the spherical functions $f_{j}^{(\ell)}$ representing a quantum operator $A$ corresponding to a quantum state or process. This can be achieved experimentally using a scanning-based tomography approach~\cite{leiner2017wigner}. The scanning-based approach estimates the scalar product of a rotated axial tensor operator $T_{j,\alpha\beta}^{(\ell)[N]}$ with the operator $A^{[N]}$ for different azimuthal ($\alpha$) and polar ($\beta$) angles, as
\begin{equation}
	\label{Eq.1}
	f_{j}^{(\ell)}(\beta,\alpha) =  s_{j}\langle{T_{j,\alpha\beta}^{(\ell)[N]}}|{A^{[N]}}\rangle,
\end{equation}
which can be equivalently rewritten in shorthand notation as
\begin{equation}
	\label{Eq.1.1}
	f_{j}^{(\ell)}(\beta,\alpha) =  s_{j}\langle{T_{j,\alpha\beta}^{(\ell)[N]}}\rangle_{A^{[N]}},
\end{equation}
where the real coefficient $s_{j} = \sqrt{(2j+1)/(4\pi)}$ and the scalar product is expressed as:
\begin{equation}
	\label{Eq.2}
	\langle{T_{j,\alpha\beta}^{(\ell)[N]}}\rangle_{A^{[N]}} = \operatorname{tr}\big\{(T_{j,\alpha\beta}^{(\ell)[N]})A^{[N]}\big\}.
\end{equation}  
Note that the tensor operators $T_{j}^{(\ell)[N]}$ are Hermitian, i.e., $({T_{j,\alpha\beta}^{(\ell)[N]}})^\dagger = ({T_{j,\alpha\beta}^{(\ell)[N]}})$. The operator $T_{j,\alpha\beta}^{(\ell)[N]}$ is the rotated version of an axial tensor operator $T_{j0}^{(\ell)[N]}$ with rank $j$ and order $m=0$:
\begin{equation}
	\label{Eq.3}
	T_{j,\alpha\beta}^{(\ell)[N]} = R_{\alpha\beta}^{[N]}(T_{j0}^{(\ell)})^{[N]}(R_{\alpha\beta}^{[N]})^\dagger.
\end{equation}
The axial tensor operators $T_{j0}^{(\ell)[N]}$ in terms of Pauli operators are provided in Appendix~\ref{Supp:Axial tensor}. Moreover, the rotation operator is defined as follows:
\begin{equation}
	\label{Eq.4}
	R_{\alpha\beta}^{[N]} = \text{exp}(-i\alpha F_{z}^{[N]})\text{exp}(-i\beta F_{y}^{[N]}),
\end{equation}
where $F_{z} = \frac{1}{2}\sum_{k=1}^{N}\sigma_{kz}^{[N]}$ and $F_{y} = \frac{1}{2}\sum_{k=1}^{N}\sigma_{ky}^{[N]}$. Here, we use a shorthand notation $\sigma_{ka} = \mathds{1}\otimes\dots\otimes\mathds{1}\otimes\sigma_{a}\otimes\mathds{1}\otimes\dots\otimes\mathds{1}$, where $\sigma_{a}$ is located in the $k^{\textit{th}}$ position and $a\in\{x,y,z\}$. \\

In the case of state tomography, the operator of interest is an $N$ qubit density matrix $\rho^{[N]}$. The corresponding spherical droplet function can be experimentally measured by using Eq.~\eqref{Eq.1.1} for $A = \rho^{[N]}$:
\begin{equation}
	\label{Eq.5}
	f_{j}^{(\ell)}(\beta,\alpha) =  s_{j}\langle{T_{j,\alpha\beta}^{(\ell)[N]}}\rangle_{\rho^{[N]}}.
\end{equation}
Ref.~\cite{Devra_WQST} provides in-depth details and experimental procedures for Wigner state tomography on pure-state quantum devices. 
\subsection{Wigner tomography of known processes}
\label{Sec.:WQPT}
\begin{figure}[h!]
	\centering
	\includegraphics[scale=0.8]{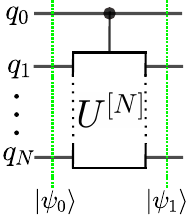}
	\caption{Schematic representation of the controlled process operation $cU^{[N+1]}$ to map a process matrix $U^{[N]}$ onto a density matrix $\rho^{[N+1]}$. Here, $q_0$ is an ancilla qubit and $q_{1},\dots,q_{N}$ are system qubits. See Table~\ref{Tab:Case_cU} for state $|\psi_{0}\rangle$ and $|\psi_{1}\rangle$.}
	\label{fig:cU_known}
\end{figure}

In general, process tomography aims to characterize quantum operations experimentally~\cite{nielsen2002quantum}, and in Wigner process tomography, one directly tomographs the spherical droplets corresponding to a quantum process or a gate experimentally. The scanning-based approach is used here by mapping an $N$ qubit \textit{unitary} process matrix onto an $N+1$ qubit (\textit{Hermitian}) density matrix using a controlled process operation $cU^{[N+1]}$, illustrated in Fig.~\ref{fig:cU_known}~\cite{leiner2018wigner, Devra_WQST}. Under $cU^{[N+1]}$, the unitary process $U^{[N]}$ only acts on the target qubits $q_{1},\dots,q_{N}$ if the ancilla qubit $q_{0}$ is in state $|1\rangle$. 
\begin{table}
	\caption{The state before and after the controlled-U ($cU^{[N+1]}$) circuit presented in Fig.~\ref{fig:cU_known} for different initial states $|\psi_{q_{0}}\rangle$ of the control qubit $q_0$. Initially, the system qubits $q_1\dots q_{N}$ are in state $|\psi_{s}\rangle$.}
	\centering
	\begin{tabular}{p{1cm} p{2cm} p{2cm} p{3cm}}
		\toprule
		\textrm{$|\psi\rangle$}&
		\textrm{$|\psi_{q_{0}}\rangle$} = $|0\rangle$&
		\textrm{$|\psi_{q_{0}}\rangle$} = $|1\rangle$&
		\textrm{$|\psi_{q_{0}}\rangle$} = $\frac{1}{\sqrt{2}}(|0\rangle+|1\rangle)$ 
		\\
		\hline
		$|\psi_0\rangle$ & $|0\psi_{s}\rangle$ & $|1\psi_{s}\rangle$ & $\frac{1}{\sqrt{2}}\big(|0\psi_{s}\rangle+|1\psi_{s}\rangle\big)$\\[1ex]
		$|\psi_1\rangle$ & $|0\psi_{s}\rangle$ & $|1(U\psi_{s})\rangle$ & $\frac{1}{\sqrt{2}}\big(|0\psi_{s}\rangle+|1(U\psi_{s})\rangle\big)$\\[0.6ex]
		\hline
		\hline
	\end{tabular}
	\label{Tab:Case_cU}
\end{table}
The controlled process $cU^{[N+1]}$ can be written in matrix form as:
\begin{equation}
	\label{Eq.6}
	{cU}^{[N+1]} = 
	\begin{pmatrix} 
		\mathds{1}^{[N]} 
		& 
		0^{[N]}  \\[1ex]
		0^{[N]}  & U^{[N]}
	\end{pmatrix}
\end{equation}
where the top diagonal block $\mathds{1}^{[N]}$ corresponds to the $2^{N}\times2^{N}$ identity matrix $\mathds{1}^{[N]}$ and the bottom diagonal block corresponds to a unitary process matrix $U^{[N]}$. The off-diagonal blocks $0^{[N]}$ are $2^{N}\times2^{N}$ dimensional zero matrices. The states $|\psi_{0}\rangle$ and $|\psi_{1}\rangle$ in Fig.~\ref{fig:cU_known} are provided in Table~\ref{Tab:Case_cU} considering different initial states $|\psi_{q_{0}}\rangle$ of the control qubit $q_{0}$. The mapping requires preparing the control qubit $q_{0}$ in an equal superposition state $\frac{1}{\sqrt{2}}(|0\rangle+|1\rangle)$ and system qubits $q_{1},\dots,q_{N}$ in a maximally mixed state~\cite{leiner2018wigner, Devra_WQST}. Therefore, the prepared initial density operator is
\begin{equation}
	\label{Eq.7}
	\rho_{0}^{[N+1]} = \frac{1}{2} \big(|{0}\rangle+|{1}\rangle\big)\big(\langle{0}|+\langle{1}|\big)  \otimes \frac{1}{2^{N}} (\mathds{1}^{[N]}).
\end{equation} 
After applying $cU^{[N+1]}$, the density matrix can be written in block form as
\begin{equation}
	\label{Eq.8}
	\rho_{U}^{[N+1]} = \frac{1}{2^{N+1}}
	\begin{pmatrix} 
		\mathds{1}^{[N]}  & 
		(U^{[N]})^{\dagger}  \\[1ex]
		U^{[N]}  & \mathds{1}^{[N]} 
	\end{pmatrix}.
\end{equation}
Using this technique, an $N$ qubit unitary operator $U^{[N]}$ and its conjugate transpose $(U^{[N]})^{\dagger}$ are imprinted onto the off-diagonal blocks of the density matrix $\rho_{U}^{[N+1]}$. Now, the spherical droplets corresponding to a unitary process $U^{[N]}$ can be experimentally measured using the modified version of Eq.~\eqref{Eq.1.1} as
\begin{equation}
	\label{Eq.9}
	f_{j}^{(\ell)}(\beta,\alpha) = s_{j}\langle{T_{j,\alpha\beta}^{(\ell)[N]}}\rangle_{U^{[N]}},
\end{equation}  
which can be equivalently re-written~\cite{Devra_WQST} as
\begin{equation}
	\label{Eq.10}
	f_{j}^{(\ell)}(\beta,\alpha) = s_{j}\langle{\sigma^{+}\otimes T_{j,\alpha\beta}^{(\ell)[N]}}\rangle_{\rho_{U}^{[N+1]}},
\end{equation} 
where $\sigma^{+} = \frac{1}{2}(\sigma_{x}+i\sigma_{y})$. We refer the reader to Ref.~\cite{Devra_WQST} for details and an experimental implementation of the presented scheme on a pure-state quantum computer. \\

As mentioned, mapping an \textit{unknown} process matrix onto a density matrix using the circuit shown in Fig.~\ref{fig:cU_known} is not possible, as the design of the controlled operation $cU^{[N+1]}$ requires prior knowledge of the process $U^{[N]}$~\cite{Arajo_2014, Gavorova}. In the next section, we discuss a new approach to map scaled versions of an unknown process matrix onto density matrices. From here onward, the operator $U^{[N]}$ refers to an \textit{unknown} quantum process unless specified otherwise.
\section{Mapping an unknown process matrix onto a density matrix}
\label{Sec.: Mapping}  
\begin{figure}[h]
	\centering
	\includegraphics[scale=0.8]{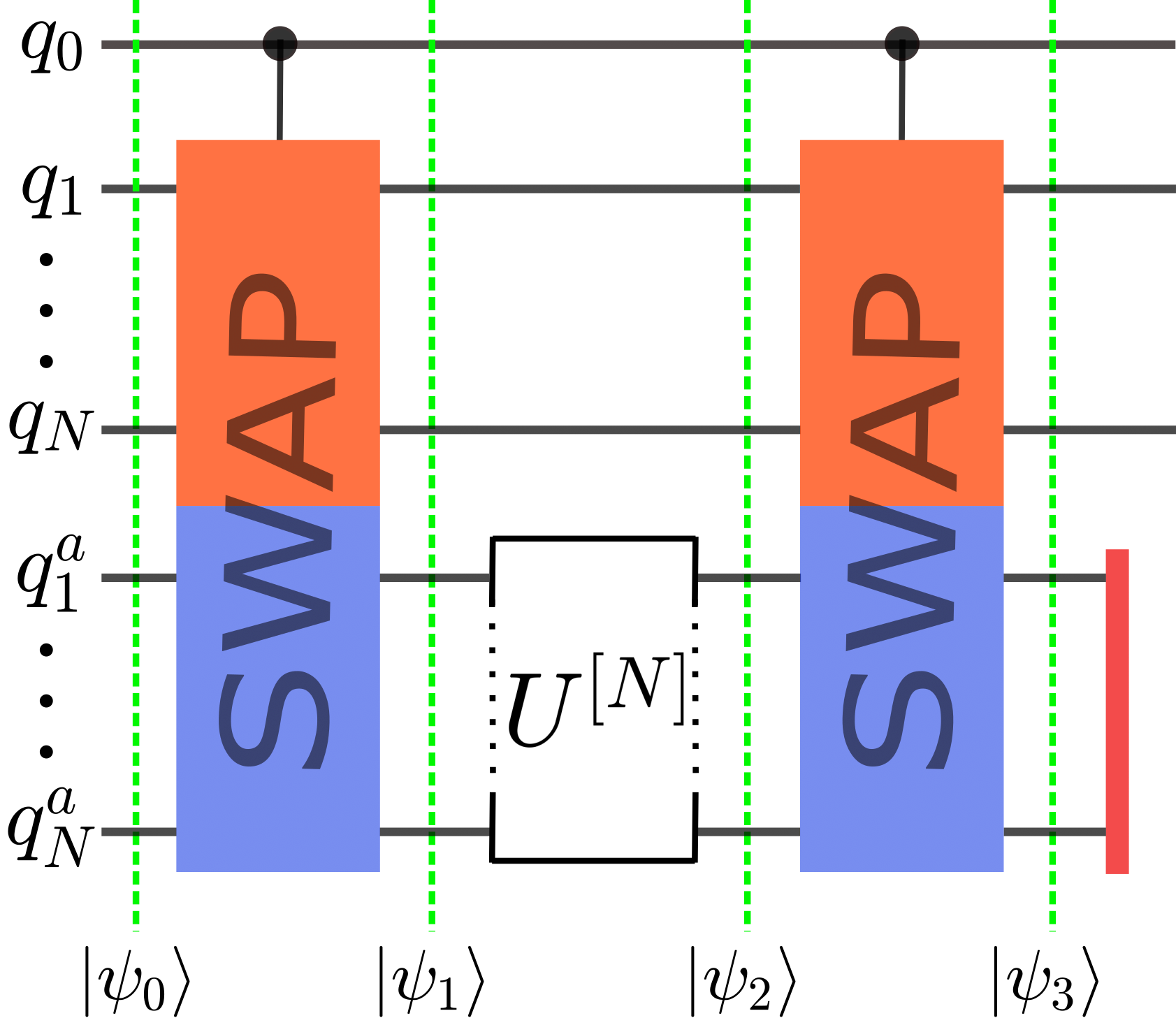}
	\caption{Schematic representation of a circuit for mapping (with inherent blind spots) an unknown unitary $U^{[N]}$ onto a density matrix. Here, $q_1,\dots,q_{N}$ are system qubits, while $q_{0}$ and $q_{1}^{a},\dots,q_{N}^{a}$ are ancilla qubits. The first and third gate of the circuit is a controlled-swap that swaps the upper block (orange or dark gray) with the lower block (blue or light gray) only if the control qubit $q_0$ is in the state $|1\rangle$. The red barrier after the second controlled-swap gate indicates partially tracing out the ancilla qubits $q_{1}^{a},\dots,q_{N}^{a}$. See Table~\ref{Tab:Case} for the state description after each gate for different initial states of the control qubit $q_{0}$.}
	\label{fig:Classic_control_circuit}
\end{figure}

We are interested in mapping an unknown quantum process matrix onto a density matrix. To achieve this, we have adopted a circuit discussed in Ref.~\cite{zhou2011adding,Gavorova}, as illustrated in Fig.~\ref{fig:Classic_control_circuit}. In this case, for an $N$ qubit system, mapping an unknown process matrix requires an additional $N+1$ qubits. Out of these $N+1$ ancillary qubits, $q_{0}$ is a control qubit. The initial state of system qubits ($q_{1},\dots ,q_{N}$) and the ancilla qubits ($q_{1}^{a},\dots ,q_{N}^{a}$) are represented by $|\psi_{s}\rangle$ and $|\psi_{a}\rangle$, respectively. Table~\ref{Tab:Case} illustrates how the state evolves after each circuit block, considering three different initial states ($|\psi_{q_{0}}\rangle$) of the control qubit $q_0$.
\begin{table*}
	\centering
	\caption{Evolution of the state after each gate of the quantum circuit presented in Fig.~\ref{fig:Classic_control_circuit} for different initial states $|\psi_{q_{0}}\rangle$ of the control qubit ($q_0$). Initially the system qubits $q_1\dots q_{N}$ are in state $|\psi_{s}\rangle$, whereas the ancilla qubits $q_{1}^{a}\dots q_{N}^{a}$ are in state $|\psi_{a}\rangle$.}
	\begin{threeparttable}
	\begin{tabular}{p{1.5cm} p{2.5cm} p{2.5cm} p{4.7cm}}
		\toprule
		\textrm{$|\psi\rangle$}&
		\textrm{$|\psi_{q_{0}}\rangle$} = $|0\rangle$&
		\textrm{$|\psi_{q_{0}}\rangle$} = $|1\rangle$&
		\textrm{$|\psi_{q_{0}}\rangle$} = $\frac{1}{\sqrt{2}}(|0\rangle+|1\rangle)$ 
		\\
		\hline
		$|\psi_0\rangle$ & $|0\psi_{s}\psi_{a}\rangle$ & $|1\psi_{s}\psi_{a}\rangle$ & $\frac{1}{\sqrt{2}}\big(|0\psi_{s}\psi_{a}\rangle+|1\psi_{s}\psi_{a}\rangle\big)$\\[1ex]
		
		$|\psi_1\rangle$ & $|0\psi_{s}\psi_{a}\rangle$ & $|1\psi_{a}\psi_{s}\rangle$ & $\frac{1}{\sqrt{2}}\big(|0\psi_{s}\psi_{a}\rangle+|1\psi_{a}\psi_{s}\rangle\big)$\\[1ex]
		
		$|\psi_2\rangle$ & $|0\psi_{s}(U\psi_{a})\rangle$ & $|1\psi_{a}(U\psi_{s})\rangle$ & $\frac{1}{\sqrt{2}}\big(|0\psi_{s}(U\psi_{a})\rangle+|1\psi_{a}(U\psi_{s})\rangle\big)$ \\[1ex]
		
		$|\psi_3\rangle$ & $|0\psi_{s}(U\psi_{a})\rangle$ & $|1(U\psi_{s})\psi_{a}\rangle$ & $\frac{1}{\sqrt{2}}\big(|0\psi_{s}(U\psi_{a})\rangle+|1(U\psi_{s})\psi_{a}\rangle\big)$\\[1ex]
		
		$|\psi_4\rangle^{\ddagger}$ & $|0\psi_{s}\rangle$ & $|1(U\psi_{s})\rangle$ & non-pure state$^{*}$\\[0.6ex]
		\hline
		\hline
	\end{tabular}
	     \begin{tablenotes}
	     	\item[$\ddagger$]\small{The state vector description can only be used when $|\psi_{q_{0}}\rangle$ is either $|0\rangle$ or $|1\rangle$.} 
     		\item[*] \small{See Eq.~\eqref{Eq.11} for the density matrix description.}
		\end{tablenotes}
    	     \end{threeparttable}%
	\label{Tab:Case}
\end{table*}

Specifically, it demonstrates that when the control qubit $q_0$ is in either $|0\rangle$ or $|1\rangle$, i.e., in a computation basis state (\textit{classical} state), the resulting state $|\psi_{4}\rangle$--- after partially tracing out the ancilla qubits $q_{1}^{a},\dots,q_{N}^{a}$--- is identical to what would be expected in the case of a controlled process circuit (Fig.~\ref{fig:cU_known}), as outlined in Table~\ref{Tab:Case_cU}.\\

Although it is tempting to assume that the block CSWAP--$U^{[N]}(q_{1}^{a},\dots,q_{N}^{a})$--CSWAP shown in Fig.~\ref{fig:Classic_control_circuit} implements $cU^{[N+1]}$ for arbitrary $U^{[N]}$, this is not true if the qubit $q_{0}$ is not in a computational basis state~\cite{zhou2011adding,Gavorova}. When $q_{0}$ is in an equal superposition state, i.e., $|\psi_{q_{0}}\rangle=\frac{1}{\sqrt{2}}(|0\rangle+|1\rangle)$, the state following the second controlled swap gate ($|\psi_{3}\rangle$) is entangled (see Table~\ref{Tab:Case}). In this particular scenario, partially tracing out the ancilla qubits $q_{1}^{a},\dots,q_{N}^{a}$ from the remaining qubits $q_{0},q_{1},\dots,q_{N}$ will result in a loss of information.\\

To facilitate the mapping of a process matrix onto a density matrix, the control qubit $q_{0}$ is required to be initially in an equal superposition state, while the qubits $q_{1},\dots,q_{N}$ and $q_{1}^{a},\dots,q_{N}^{a}$ are required to be in a maximally mixed state, i.e., $\rho_{s}^{[N]} = |\psi_s\rangle\langle\psi_s| = \frac{1}{2^N}\mathds{1}^{[N]}$ and $\rho_{a}^{[N]} = |\psi_a\rangle\langle\psi_a| = \frac{1}{2^N}\mathds{1}^{[N]}$.  The final density matrix after tracing out the ancilla qubits $q_{1}^{a},\dots,q_{N}^{a}$ (indicated by the red barrier in the circuit depicted in Fig.~\ref{fig:Classic_control_circuit}) is
\begin{equation}
	\label{Eq.11}
	\tilde{\rho}_{4}^{[N+1]} = 
	\frac{1}{2^{N+1}}
	\begin{pmatrix} 
		\mathds{1}^{[N]} 
		& 
		c^{*}\cdot (U^{[N]})^{\dagger} \\[0.5ex]
		c\cdot U^{[N]}  & \mathds{1}^{[N]} 
	\end{pmatrix},
\end{equation}
with scaling factor
\begin{equation}
	\label{Eq.12}
	c = \frac{1}{2^{N}}\operatorname{tr}((U^{[N]})^{\dagger}).
\end{equation}
In Eq.~\eqref{Eq.11}, $c^{*}$ is the complex conjugate of $c$. Here, the unknown process $U^{[N]}$ is mapped onto the density matrix $\tilde{\rho}_{4}^{[N+1]}$ up to a scaling factor $c$ (c.f. Eq.~\eqref{Eq.8}). For conciseness, here we only present the state after partially tracing out the ancilla qubits $q_{1}^{a},\dots,q_{N}^{a}$ and provide detailed calculations in Appendix~\ref{Supp:density_scaling}. The scaling factor is a result of the partial trace, leading to a loss of information, as mentioned earlier. For example, in the case of a general single-qubit rotation~\cite{nielsen2002quantum} with rotation angle $\gamma$ and rotation axis $\hat{n} = (n_x, n_y, n_z)$, given by 
\begin{equation}
	\label{Eq.13}
	U^{[1]} = \cos\Big(\frac{\gamma}{2}\Big)\mathds{1} - i \cdot \sin\Big(\frac{\gamma}{2}\Big)(n_{x}\sigma_{x}+n_{y}\sigma_{y}+n_{z}\sigma_{z}),
\end{equation}
the scaling factor is simply
\begin{equation}
	\label{Eq.14}
	c = \cos\Big(\frac{\gamma}{2}\Big),
\end{equation}
because the Pauli matrices $\sigma_{x}$, $\sigma_{y}$, and $\sigma_{z}$ are traceless. \\

The mapping of a unitary operator $U$ onto the off-diagonal blocks of $\tilde{\rho}_{4}^{[2]}$ in Eq.~\eqref{Eq.11} is lossless (corresponding to $\abs{c}=1$) if and only if the rotation angle is a multiple of $2 \pi$ (i.e., $\gamma = 2n\pi$ for $n\in\mathbb{Z}$). However, in the general case, the off-diagonal blocks of the density matrix $\tilde{\rho}_{4}^{[2]}$ are scaled down ($\abs{c}<1$), which would result in a loss of signal-to-noise ratio in the experimental tomography results. For example, for the $\sqrt{\text{NOT}}$ gate (which has a rotation angle of $\gamma = \frac{\pi}{2}$), the scaling factor is only $c = \sqrt{\frac{1}{2}}$. In the case of processes for which  $\abs{c}$ approaches 0, the protocol given in Fig.~\ref{fig:Classic_control_circuit} has actual blind spots, where no information about the process of interest is obtained. This occurs whenever the rotation angle $\gamma$ is close to an odd multiple of $\pi$ (i.e., if $\gamma = (2n+1)\pi$ for $n\in\mathbb{Z}$), which is the case for many standard quantum gates like NOT (X), Y, Z, and Hadamard (H) gates. To overcome these limitations and remove the blind spots inherent to this approach, we extend the circuit presented in Fig.~\ref{fig:Classic_control_circuit} and discuss it in the next section.

\subsection{Modified circuit for mapping}    
\label{Sec.:modified circuit} 
\begin{figure}[h]
	\centering
	\includegraphics[scale=0.8]{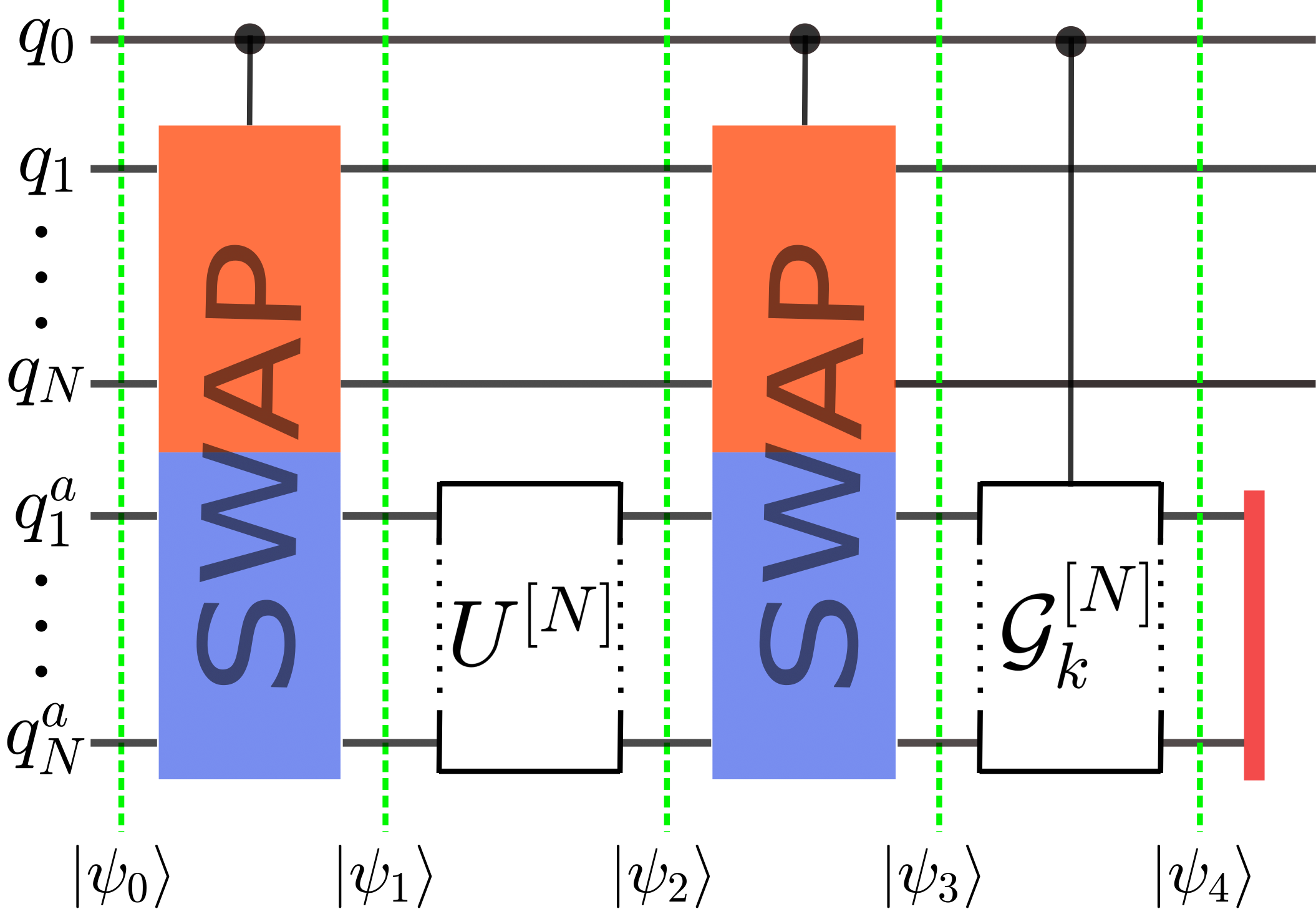}
	\caption{Circuit for mapping unknown scaled process matrices $\epsilon_{k}U^{[N]}$ onto density matrices without introducing blind spots by repeating the experiments for different gates $\mathcal{G}_{k}$. Here, $q_1,\dots ,q_{N}$ are system qubits and qubits $q_{0}$ and $q_{1}^{a},\dots ,q_{N}^{a}$ are ancilla qubits. See Table~\ref{Tab:Cases_quantum} for the state evolution after each gate.}
	\label{fig:cU_unknownWR}
\end{figure}

To remove the \textit{blind spots} caused by the circuit depicted in Fig.~\ref{fig:Classic_control_circuit}, we propose the extended quantum circuit shown in Fig.~\ref{fig:cU_unknownWR}. In this new circuit design, we introduce a controlled rotation denoted as $c\mathcal{G}^{[N]}_k$, which exclusively operates on the ancilla qubits $q_{1}^{a},\dots,q_{N}^{a}$ when the control qubit $q_{0}$ assumes the state $|1\rangle$. For a system comprising $N$ qubits  $q_{1},\dots,q_{N}$, the circuit shown in Fig.~\ref{fig:cU_unknownWR} is reiterated with varying rotations $\mathcal{G}^{[N]}_k$, extending up to $4^{N}$ repetitions in the standard approach. Here, ‘$4^N$' refers to the number of elements in the Pauli operator basis for a system consisting of $N$ qubits. For instance, for a single system qubit ($N=1$), $k$ ranges from 1 to 4, since $\mathcal{G}^{[1]}_{k}\in\{\sigma_{x},\sigma_{y},\sigma_{z},\mathds{1}\}$, i.e., $\mathcal{G}^{[1]}_{1} = \sigma_{x}$, $\mathcal{G}^{[1]}_{2} = \sigma_{y}$, $\mathcal{G}^{[1]}_{3} = \sigma_{z}$, and $\mathcal{G}^{[1]}_{4} = \mathds{1}$. Similarly, for a system with two qubits ($N=2$), $k = 1,...,16$, as $\mathcal{G}^{[2]}_{k}\in\{\sigma_{x}\otimes\mathds{1},\dots,\mathds{1}\otimes\sigma_{x},\dots,\sigma_{z}\otimes\sigma_{z},\mathds{1}\otimes\mathds{1}\}$. However, in general $\mathcal{G}_{k}$ can be any multi-qubit rotation (see Sec.~\ref{Sec:Adaptive}).
\begin{table}
	\caption{Evolution of a quantum state after each gate of the quantum circuit presented in Fig.~\ref{fig:cU_unknownWR}. Considering initially the control qubit $q_{0}$ to be in state $|\psi_{q_{0}}\rangle = \frac{1}{\sqrt{2}}(|0\rangle+|1\rangle)$, the system qubits $q_1\dots q_{N}$ in the state $|\psi_{s}\rangle$, and the ancilla qubits $q_{1}^{a}\dots q_{N}^{a}$ in the state $|\psi_{a}\rangle$. See supplementary Sec.~\ref{Supp:density_scaling} for density matrix calculations.}
	\centering
	\begin{tabular}{p{1.5cm} p{5.5cm}}
		\toprule
		\textrm{$|\psi\rangle$}&
		\textrm{$|\psi_{q_{0}}\rangle$} = $\frac{1}{\sqrt{2}}(|0\rangle+|1\rangle)$ 
		\\
		\hline
		$|\psi_0\rangle$ &  $\frac{1}{\sqrt{2}}\big(|0\psi_{s}\psi_{a}\rangle+|1\psi_{s}\psi_{a}\rangle\big)$\\[1ex]
		
		$|\psi_1\rangle$ &  $\frac{1}{\sqrt{2}}\big(|0\psi_{s}\psi_{a}\rangle+|1\psi_{a}\psi_{s}\rangle\big)$\\[1ex]
		
		$|\psi_2\rangle$ &  $\frac{1}{\sqrt{2}}\big(|0\psi_{s}(U\psi_{a})\rangle+|1\psi_{a}(U\psi_{s})\rangle\big)$ \\[1ex]
		
		$|\psi_3\rangle$ & 
		$\frac{1}{\sqrt{2}}\big(|0\psi_{s}(U\psi_{a})\rangle+|1(U\psi_{s})\psi_{a}\rangle\big)$\\[1ex]
		
		$|\psi_4\rangle$ &
		$\frac{1}{\sqrt{2}}\big(|0\psi_{s}(U\psi_{a})\rangle+|1(U\psi_{s})(\mathcal{G}_{k}\psi_{a})\rangle\big)$\\[0.6ex]
		\hline
		\hline
	\end{tabular}
	\label{Tab:Cases_quantum}
\end{table}

Table~\ref{Tab:Cases_quantum} describes the state evolution after each circuit block displayed in Fig.~\ref{fig:cU_unknownWR}. This analysis is conducted under the initial conditions in which the control qubit $q_{0}$ is in the state $|\psi_{q_{0}}\rangle = \frac{1}{\sqrt{2}}(|0\rangle+|1\rangle)$, the system qubits $q_1\dots q_{N}$ are in the state $|\psi_{s}\rangle$, and the ancilla qubits $q_{1}^{a}\dots q_{N}^{a}$ are in the state $|\psi_{a}\rangle$. In this case, when the qubits $q_{1},\dots,q_{N}^{a}$ are initially in a maximally mixed state, i.e, $\rho_{s}^{[N]} = \rho_{a}^{[N]} = \frac{1}{2^N}\mathds{1}^{[N]}$, the density matrix, after tracing out the ancilla qubits $q_{1}^{a}\dots q_{N}^{a}$ (indicated by the red barrier in Fig.~\ref{fig:cU_unknownWR}) is
\begin{equation}
	\label{Eq.15}
	\tilde{\rho}_{5}^{[N+1]} = \frac{1}{2^{N+1}}
	\begin{pmatrix} 
		\mathds{1}^{[N]} 
		& 
		\epsilon^{*}_{k}(U^{[N]})^{\dagger}  \\[1ex]
		\epsilon_{k}U^{[N]}  & \mathds{1}^{[N]} 
	\end{pmatrix},
\end{equation}
which can be rewritten as:
\begin{equation}
	\label{Eq.16}
	\rho_{U_{k}}^{[N+1]} = 
	\frac{1}{2^{N+1}}
	\begin{pmatrix} 
		\mathds{1}^{[N]}  &  (U_{k}^{[N]})^{\dagger}  \\[1ex]
		U_{k}^{[N]}  & \mathds{1}^{[N]} 
	\end{pmatrix}, 
\end{equation} 
where $U_{k}^{[N]} = \epsilon_{k}U^{[N]}$ and the scaling factor
\begin{equation}
	\label{Eq.17}
	\epsilon_{k} = \frac{1}{2^{N}} \langle U^{[N]}|\mathcal{G}_{k}^{[N]}\rangle.
\end{equation} 
In Eq.~\eqref{Eq.15}, $\epsilon^{*}_{k}$ represents the complex conjugate of $\epsilon_{k}$. Here, depending on the rotation $\mathcal{G}_{k}^{[N]}$, different scaled process matrices $U^{[N]}_{k}$ are mapped onto the off-diagonal blocks of the density matrix $\rho_{U_{k}}^{[N+1]}$ in Eq.~\eqref{Eq.16}. In the case where $\mathcal{G}^{[N]} = \mathds{1}^{[N]}$, the density matrix $\tilde{\rho}_{5}^{[N+1]}$ in Eq.~\eqref{Eq.15} is identical to $\tilde{\rho}_{4}^{[N+1]}$ in Eq.~\eqref{Eq.11} as the scaling factor $c$ in Eq.~\eqref{Eq.12} is also identical to $\epsilon$ in Eq.~\eqref{Eq.17}. We provide the detailed calculation in Appendix~\ref{Supp:density_scaling}.\\

If the unitary operator $U^{[N]}$ happens to be equal to $\mathcal{G}_{k}^{[N]}$ for a specific $k$, then $\epsilon_{k} = 1$, resulting in an exact (lossless) mapping of process $U^{[N]}$ onto a density matrix $\rho_{U_{k}}^{[N+1]}$. This is also evident in the state vector formalism provided in Table~\ref{Tab:Cases_quantum}. In this context, when $\mathcal{G}_{k} = U$, the state $|\psi_{4}\rangle$ modifies to:
\begin{equation}
	\label{Eq.18}
	\begin{split}
	|\psi_{4}\rangle' & = \frac{1}{\sqrt{2}}(|0\psi_{s}(U\psi_{a})\rangle+|1(U\psi_{s})(U\psi_{a})\rangle)\\ 
	& = \frac{1}{\sqrt{2}}(|0\psi_{s}\rangle+|1(U\psi_{s})\rangle)\otimes (U|\psi_{a}\rangle),
	\end{split}
\end{equation}
indicating that the ancilla qubits $q_{1}^{a}\dots q_{N}^{a}$ are fully separable from the qubits $q_{0},q_{1}\dots q_{N}$. However, in general, where $U^{[N]}$ differs from $\mathcal{G}_{k}^{[N]}$, the state $|\psi_{4}\rangle$ remains entangled, leading to the mapping of scaled process matrices  $U_{k}^{[N+1]}$ onto density matrices $\rho_{U_{k}}^{[N+1]}$, as shown in Eq.~\eqref{Eq.16}.

\subsection{Scaling factors for single-qubit system}
\label{Sec.: scaling_factor_single}
\begin{figure}[h]
	\centering
	\adjustbox{max width=\textwidth}{\includegraphics[scale=0.9]{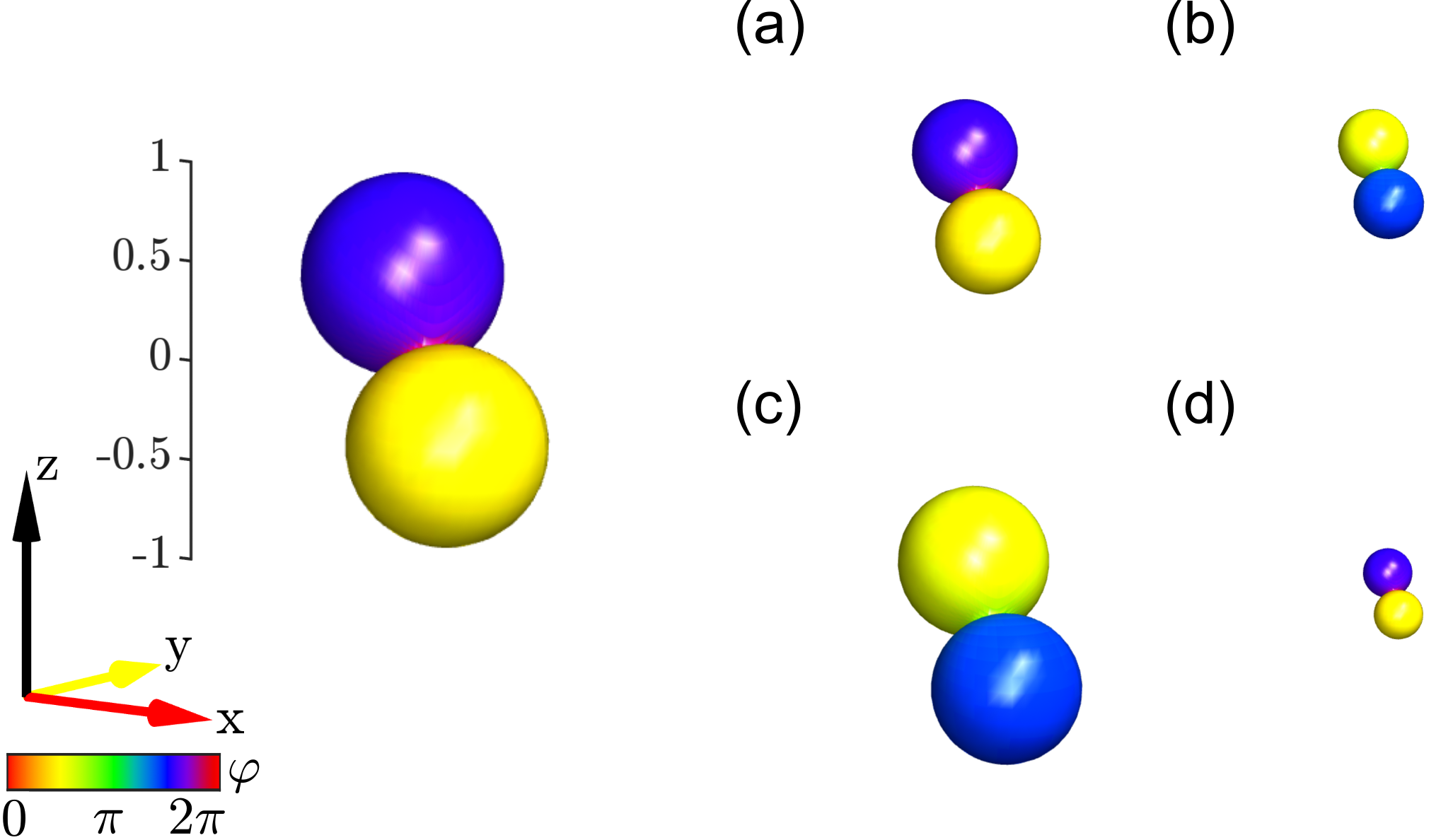}}
	\caption{Spherical droplet representing a process matrix $U^{[1]}$ (on the left) with quaternion components $A = 0.5198$, $B = -0.3462$, $C = -0.7424$, and $D = 0.2425$. The resulting droplets corresponding to scaled process matrices (a) $\hat{U}_{1}^{[1]} = AU^{[1]}$, (b) $\hat{U}_{2}^{[1]} = BU^{[1]}$, (c) $\hat{U}_{3}^{[1]} = CU^{[1]}$, and (d) $\hat{U}_{4}^{[1]} = DU^{[1]}$ associated with different controlled rotations $c\mathcal{G}$, with $\mathcal{G}$ =  $\sigma_{x}$, $\sigma_{y}$, $\sigma_{z}$, and $\mathds{1}$ respectively, are displayed on the right. Refer to Table~\ref{Tab:scaling_factor} for scaling factors $\epsilon_{k}$ corresponding to different rotations $\mathcal{G}^{[1]}_{k}$ and the resulting process matrices $\hat{U}_{k}^{[1]}$.}
	\label{fig:scaling_example}
\end{figure}

For the single-qubit (i.e., $N=1$) case, the scaling factors $\epsilon_{k}$ corresponding to different rotations $\mathcal{G}_{k}$ are provided in Table~\ref{Tab:scaling_factor} for a single-qubit unitary process given by:
\begin{equation}
	\label{Eq.19}
	U^{[1]} = 
	\begin{pmatrix} 
		u_{11} &
		u_{12} \\[1ex]
		u_{21}  & 
		u_{22} 
	\end{pmatrix} = 
	\begin{pmatrix} 
		D+iC & B+iA  \\[1ex]
		-B+iA  & D-iC 
	\end{pmatrix}.
\end{equation}
Here, $U^{[1]}$ is written using the usual representation of quaternions as complex $2\times 2$ matrices~\cite{BLUMICH1985356}, and for this matrix to be special unitary, the real components $A, B, C, $ and $D$ must satisfy the condition $A^{2}+B^{2}+C^{2}+D^{2}=1$. The second column of Table~\ref{Tab:scaling_factor} presents the scaling factors in terms of matrix elements $u_{11}$, $u_{12}$, $u_{21}$, and $u_{22}$, while the third column expresses them in terms of quaternion components $A$, $B$, $C$, and $D$. The resulting scaled matrices $U^{[1]}_{k}$ are displayed in the fourth column. For simplicity and to keep a consistent droplet color, the first three scaled process matrices $U^{[1]}_{k=1:3}$ are divided by $-i$, resulting in
\begin{equation}
	\label{Eq.20}
	\hat{U}^{[1]}_{k = 1:3} = \frac{1}{(-i)}(U^{[1]}_{k=1:3}) = i\cdot U^{[1]}_{k=1:3},
\end{equation}
and $\hat{U}^{[1]}_{4} = U^{[1]}_{4}$. The new scaled process matrices $\hat{U}^{[1]}_{k}$ are presented in the last column of Table~\ref{Tab:scaling_factor}. Fig.~\ref{fig:scaling_example} shows an example using droplets for the scaling factors of a single-qubit system. 

\begin{table}[h]
\centering
\caption{Scaling factors ($\epsilon_{k}$) for a single-qubit system ($N=1$) corresponding to different controlled rotations with $\mathcal{G}^{[1]}_{k}\in\{\sigma_{x},\sigma_{y},\sigma_{z},\mathds{1}\}$ and the resulting scaled process matrices $U^{[1]}_{k}$. The second column provides the scaling factor in terms of matrix elements $u_{11}$, $u_{12}$, $u_{21}$, and $u_{22}$, while the third column expresses them in terms of quaternion components $A$, $B$, $C$, and $D$. The last column presents the scaled matrices $\hat{U}^{[1]}_{k}$ defined in Eq.~\eqref{Eq.20}. See Fig.~\ref{fig:scaling_example} for an example.} 
	\begin{tabular}{p{0.8cm} p{2.2cm} p{1.2cm} p{2.2cm} p{0.9cm}} 
		\toprule
		$\mathcal{G}^{[1]}_{k}$ & $\epsilon_{k}$ & $\epsilon_{k}$ & $U^{[1]}_{k} = \epsilon_{k}U^{[1]}$ & $\hat{U}^{[1]}_{k}$\\[0.5ex]
		\hline
		$\sigma_{x}$ & $\frac{1}{2} (u_{12}^{*}+u_{21}^{*})$ & $-(i)A$ & $-(i)AU^{[1]}$ & $AU^{[1]}$ \\[0.7ex]
		
		$\sigma_{y}$ & $\frac{i}{2} (u_{21}^{*}-u_{12}^{*})$ & $-(i)B$ & $-(i)BU^{[1]}$ & $BU^{[1]}$\\[0.7ex]
		
		$\sigma_{z}$ & $\frac{1}{2} (u_{11}^{*}-u_{22}^{*})$ & $-(i)C$ & $-(i)CU^{[1]}$ & $CU^{[1]}$\\[0.7ex]
		
		$\mathds{1}$ & $\frac{1}{2} (u_{11}^{*}+u_{22}^{*})$ & $D$ & $DU^{[1]}$ & $DU^{[1]}$\\[0.5ex]
		\hline
		\hline
	\end{tabular}
	\label{Tab:scaling_factor}
\end{table}

Now, using the quantum circuit shown in Fig.~\ref{fig:cU_unknownWR}, the scaled process matrices $U_{k}^{[N]} = \epsilon_{k}U^{[N]}$ of an $N$ qubit unknown process matrix $U^{[N]}$ are mapped onto the $N+1$ qubit density matrices $\rho_{U_{k}}^{[N+1]}$ (see Eq.~\eqref{Eq.16}). The following section formalizes the Wigner process tomography, aiming to experimentally tomograph the droplets corresponding to the scaled process matrices $U_{k}^{[N]}$.

 \section{Theory of Wigner tomography of unknown process}
\label{Sec.:WQPT theory}
\begin{figure}[h]
	\includegraphics[scale=0.75]{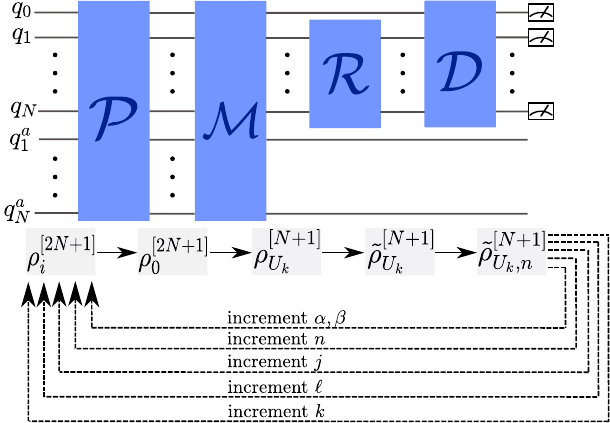}
	\caption{Schematic of Wigner tomography for unknown processes. The algorithm consists of four key blocks: Preparation ($\mathcal{P}$), Mapping ($\mathcal{M}$), Rotation ($\mathcal{R}$), and Detection-associated rotations ($\mathcal{D}$) followed by measurements. The $\mathcal{M}$ block is replaced with the circuit presented in Fig.~\ref{fig:cU_unknownWR}. The lower part of the figure shows the evolution of the density matrix after each block. The algorithm is repeated for different parameters.}
	\label{fig:wqpt_algo}
\end{figure}

We first extend the Wigner tomography algorithm presented in Ref.~\cite{Devra_WQST} to unknown processes and then explain each step individually. A schematic diagram of the algorithm is shown in Fig.~\ref{fig:wqpt_algo}. A spherical droplet function $f_{j,k}^{(\ell)}$ representing a scaled quantum process $U_{k}^{[N]}$ can be experimentally tomographed using the following steps:
\begin{enumerate}
	\item \textbf{Preparation ($\mathcal{P}$)}: Prepare ancilla qubit $q_{0}$ in the superposition state $\frac{1}{\sqrt{2}}(|0\rangle+|1\rangle)$ and effectively create the remaining qubits $q_{1},\dots,q_{N},q_{1}^{a},\dots,q_{N}^{a}$ in a maximally mixed state. 
	\item \textbf{Mapping ($\mathcal{M}$)}: Implement the circuit presented in Fig.~\ref{fig:cU_unknownWR} to map scaled unknown processes $U_{k}^{[N]}$ onto density matrices $\rho_{U_{k}}^{[N+1]}$ for different $k$.
	\item \textbf{Rotation ($\mathcal{R}$)}: Rotate the system qubits $q_{1},\dots,q_{N}$ inversely for scanning.
	\item \textbf{Detection-associated rotations ($\mathcal{D}$)}: Apply local unitary operations to measure the required expectation values of axial tensor operators $T_{j0}^{(\ell)[N]}$ (see Appendix~\ref{Supp:Axial tensor}) that are not directly measurable.
\end{enumerate}
These four steps are repeated for a set of polar $\beta\in[0,\pi]$ and azimuthal $\alpha\in[0,2\pi]$ angles and for different $n$, rank $j$, label $\ell$, and controlled rotations ($c\mathcal{G}_{k}$). Now we explain each step of the algorithm.    

\textit{Step 1:} Considering that the algorithm starts with the state $\rho_{i} = |00\dots0\rangle\langle00\dots0|$, qubit $q_{0}$ is prepared in the equal superposition state $\frac{1}{\sqrt{2}}(|0\rangle+|1\rangle)$ by applying a Hadamard gate. A maximally mixed state of qubits $q_{1},\dots,q_{N},q_{1}^{a},\dots,q_{N}^{a}$ can be effectively created by temporally averaging the results of experiments conducted on all possible computational basis states~\cite{preskill1998lecture,Devra_WQST,Devra_dissertation,knill_temporal}. This can be achieved by appropriately applying local NOT gates. We discuss this for the case of a single system qubit in Sec.~\ref{Sec.:single-qubit_experiments}.\\

\textit{Step 2:} The circuit presented in Fig.~\ref{fig:cU_unknownWR} is implemented to map scaled process matrices $U_{k}^{[N]}$ onto the density matrices $\rho_{U_{k}}^{[N+1]}$ as shown in Eq.~\eqref{Eq.13}.\\

\textit{Step 3:} Since here we are interested in 
measuring a droplet function $f_{j,k}^{(\ell)}$ representing a scaled process matrix $U_{k}^{[N]}$, Eq.~\eqref{Eq.10} takes the form:
\begin{equation}
	\label{Eq.16a}
	f_{j,k}^{(\ell)}(\beta,\alpha) = s_{j}\langle{\sigma^{+}\otimes T_{j,\alpha\beta}^{(\ell)[N]}}\rangle_{\rho_{U_{k}}^{[N+1]}}.
\end{equation}
Now, instead of rotating the axial tensor operators $T_{j0}^{(\ell)[N]}$, we make this experimentally convenient by equivalently rotating the density matrix in the inverse direction, such that:
\begin{equation}
	\label{Eq.17a}
	f_{j,k}^{(\ell)}(\beta,\alpha) = s_{j}\langle{\sigma^{+}\otimes T_{j0}^{(\ell)[N]}}\rangle_{\tilde{\rho}_{U_{k}}^{[N+1]}}\,,
\end{equation}
where
\begin{equation}
	\label{Eq.18a}
	\tilde{\rho}_{U_{k}}^{[N+1]} = (R_{\alpha\beta}^{[N+1]})^{-1}\rho_{U_{k}}^{[N+1]}R_{\alpha\beta}^{[N+1]},
\end{equation}
and $R_{\alpha\beta}^{[N+1]} = \mathds{1}^{[1]}\otimes R_{\alpha\beta}^{[N]}$, i.e., the rotation operator $R_{\alpha\beta}^{[N]}$ only acts on the system qubits $q_{1}\dots q_{N}$. The rotation operator $R_{\alpha\beta}^{[N]}$ is the same as shown in Eq.~\eqref{Eq.4}. Using the relation $\sigma^{+} = \frac{1}{2}(\sigma_{x}+i\sigma_{y})$, Eq.~\eqref{Eq.10} can be rewritten as:
\begin{equation}
	\label{Eq.21a}
        \small
		f_{j,k}^{(\ell)}(\beta,\alpha)  = \frac{s_{j}}{2}\big(\langle{\sigma_{x} \otimes T_{j0}^{(\ell)[N]}}\rangle_{{\tilde{\rho}}_{U_{k}}^{[N+1]}} + i\langle{\sigma_{y} \otimes T_{j0}^{(\ell)[N]}}\rangle_{{\tilde{\rho}}_{U_{k}}^{[N+1]}}\big).
\end{equation}

\textit{Step 4:} This step is used for measuring expectation values of Pauli operators that are not directly observable. This can be achieved by using local unitary rotations $u_{n}$ (detection-associated rotations) ~\cite{Devra_WQST}. The explicit values for $T_{j0}$ transformed in Pauli operators for one and two qubits are provided in Appendix~\ref{Supp:Axial tensor}.\\

By employing Eq.~\eqref{Eq.21a}, we can experimentally tomograph the droplet functions $f_{j,k}^{(\ell)}$ corresponding to the scaled matrices $U_{k}^{[N]}$. This method provides information about the scaled unknown matrices $U^{[N]}_{k}$, and we have developed a methodology to combine these experimentally tomographed, scaled droplets to reconstruct an unknown process $U^{[N]}$ with high signal-to-noise ratio. Consequently, the entire Wigner tomography approach for an unknown process comprises two integral components: a quantum aspect (corresponds to experiments) and a classical aspect (concerning the post-processing of experimental data), as also illustrated in the schematic in Fig.~\ref{fig:approach}.

\section{Reconstruction of an unknown process from scaled process droplets}
\label{Sec.:Reconstruction theory}
In this section, we present a methodology to reconstruct an unknown process $U$ from the experimentally tomographed droplet functions $f_{j,k}^{(\ell)}$ corresponding to the scaled processes $U_{k}^{[N]}$ for a single system qubit ($N=1$). \\

Let us first rewrite Eq.~\eqref{Eq.21a} for a single-qubit system ($N=1$). The possible values of rank $j$ are: $j=0$ for  label $\ell=\emptyset$, and $j=1$ for label $\ell=1$ (refer to Appendix~\ref{Supp:Axial tensor}). Hence, single qubit scaled processes $U_{k}^{[1]}$ represented by spherical functions $f_{0,k}^{(\emptyset)}$ and $f_{1,k}^{(1)}$ can be experimentally tomographed using:
\begin{equation}
	\label{Eq.22a}
        \small
	\begin{aligned}
		f_{0,k}^{(\emptyset)}(\beta,\alpha)  =  \frac{1}{2}\sqrt{\frac{1}{4\pi}}\big(\langle{\sigma_{x} \otimes T_{00}^{(\ell)[1]}}\rangle_{{\tilde{\rho}}_{U_{k}}^{[2]}} + i\langle{\sigma_{y} \otimes T_{00}^{(\ell)[1]}}\rangle_{{\tilde{\rho}}_{U_{k}}^{[2]}}\big),\\
		f_{1,k}^{(1)}(\beta,\alpha) =  \frac{1}{2}\sqrt{\frac{3}{4\pi}}\big(\langle{\sigma_{x} \otimes T_{10}^{(1)[1]}}\rangle_{{\tilde{\rho}}_{U_{k}}^{[2]}}+ i \langle{\sigma_{y} \otimes T_{10}^{(1)[1]}}\rangle_{{\tilde{\rho}}_{U_{k}}^{[2]}}\big).
	\end{aligned}
\end{equation}
Substituting the explicit values of the spherical tensor operators in terms of Pauli operators from Appendix~\ref{Supp:Axial tensor} simplifies the above equation as:
\begin{equation}
	\label{Eq.23}
	\begin{aligned}
		f_{0,k}^{(\emptyset)}(\beta,\alpha)  = {} & \frac{1}{4}\sqrt{\frac{1}{2\pi}}\big(\langle{\sigma_{0x}}\rangle_{{\tilde{\rho}}_{U_{k}}^{[2]}} + i\langle{\sigma_{0y}}\rangle_{{\tilde{\rho}}_{U_{k}}^{[2]}}\big),\\
		f_{1,k}^{(1)}(\beta,\alpha) = {} & \frac{1}{4}\sqrt{\frac{3}{2\pi}}\big(\langle{\sigma_{0x}\sigma_{1z}}\rangle_{{\tilde{\rho}}_{U_{k}}^{[2]}} + i \langle{\sigma_{0y}\sigma_{1z}}\rangle_{{\tilde{\rho}}_{U_{k}}^{[2]}}\big),
	\end{aligned}
\end{equation}
where $\sigma_{0x} = \sigma_{x}\otimes\mathds{1}$ and $\sigma_{0x}\sigma_{1z} = \sigma_{x}\otimes\sigma_{z}$, for example.
\subsection{Reconstruction algorithm}
\label{Sec.:Reconstruction algo}
As discussed in Sec.~\ref{Sec.: scaling_factor_single}, in the case of a single-qubit system ($N=1$), the index $k$ ranges from 1 to 4, representing the different rotations: $\mathcal{G}^{[1]}_{1} = \sigma_{x}$, $\mathcal{G}^{[1]}_{2} = \sigma_{y}$, $\mathcal{G}^{[1]}_{3} = \sigma_{z}$, and $\mathcal{G}^{[1]}_{4} = \mathds{1}$. The resulting rank $j = 0$ and $j = 1$ droplet functions corresponding to the scaled process $U_{k}^{[1]}$ are experimentally measured using Eq.~\eqref{Eq.23}. For simplicity, we combine these two droplet functions, i.e.: 
\begin{equation}
	\label{Eq.28}
	f_{k} = f_{0,k}^{(\emptyset)}+f_{1,k}^{(1)}.
\end{equation}
To keep a consistent droplet color, we use $\hat{U}^{[1]}_{k}$ defined in Eq.~\eqref{Eq.20} (also see Table~\ref{Tab:scaling_factor}), which modifies the experimental droplet functions as follows:
\begin{equation}
	\label{Eq.29}
	\begin{aligned}
	\hat{f}_{1} = {} & i(f_{0,1}^{(\emptyset)}+f_{1,1}^{(1)}) = if_{1} \longleftrightarrow \hat{U}_{1}^{[1]} \approx AU_{a}^{[1]}\\
	 \hat{f}_{2} = {} & i(f_{0,2}^{(\emptyset)}+f_{1,2}^{(1)}) = if_{2} \longleftrightarrow \hat{U}_{2}^{[1]}  \approx BU_{a}^{[1]}\\	
	\hat{f}_{3} = {} & i(f_{0,3}^{(\emptyset)}+f_{1,3}^{(1)}) = if_{3} \longleftrightarrow \hat{U}_{3}^{[1]} \approx CU_{a}^{[1]}\\	
	\hat{f}_{4} = {} & f_{0,4}^{(\emptyset)}+f_{1,4}^{(1)} = f_{4} \longleftrightarrow \hat{U}_{4}^{[1]} \approx DU_{a}^{[1]}.
	\end{aligned}
\end{equation}
Here $U_{a}^{[1]}$ is the actual (experimental) unknown process composed of quaternion components $A$, $B$, $C$, and $D$ as described in Eq.~\eqref{Eq.19}.\\

The goal of the algorithm is to reconstruct a spherical droplet with a high signal-to-noise ratio corresponding to an unknown process from the scaled droplets $\hat{f}_{k}$ described in Eq.~\eqref{Eq.29}. A naive approach to achieve this is to simply compute an average of the scaled process droplets $\hat{f}_{k}$, such that the combined droplet is $\hat{f}_{\text{comb}} = \frac{1}{4}(\hat{f}_{1}+\hat{f}_{2}+\hat{f}_{3}+\hat{f}_{4})$. However, this naive approach does not necessarily lead to a droplet with a high signal-to-noise ratio~\cite{Devra_dissertation}. To address this, we adopt the principle of matched filtering~\cite{ernst1987principles,spencer2010equivalence}. This technique maximizes the signal-to-noise ratio of the reconstructed droplet by combining scaled droplets with different weights. Specifically, it assigns a higher weight to the droplet with a larger signal (size) and a lower weight to the droplet with a smaller signal (size). This allows a combination of droplets with different weights, resulting in a reconstructed droplet with a significantly improved signal-to-noise ratio compared to the naive equal-weighted droplet combination approach described above. Here, we present an approach to estimate quaternion components \(A\), \(B\), \(C\), and \(D\) from experimental scaled droplets $\hat{f}_{k}$ to reconstruct the unknown process with a high signal-to-noise ratio. In Algorithm~\ref{alg:quaternion_estimation}, we first outline the steps of the algorithm for a single system qubit ($N=1$) and subsequently elaborate via an example.

\begin{algorithm*}
\caption{Estimation of quaternion components from experimental scaled droplets $\hat{f}_{k}$.}
\label{alg:quaternion_estimation}
\begin{flushleft}
\textbf{Aim}: To estimate quaternion components $A$, $B$, $C$, and $D$ corresponding to an actual unknown process $U_{a}^{[1]}$ from experimental scaled droplets $\hat{f}_{k}$.\\
\textbf{Input}: Experimentally tomographed droplet functions $\hat{f}_{k}$ representing scaled processes $\hat{U}_{k}^{[1]}$, with $k = 1:4$.
\end{flushleft}
\begin{algorithmic}[1]
\State Compute the correlation matrix $M$.
\State Estimate zero-order (i.e., iteration number $i = 0$) values of quaternion components $A_{i}$, $B_{i}$, $C_{i}$, and $D_{i}$ from the correlation matrix $M$.
\State Compute weighted droplets: $\hat{f}^{[i]}_{w,1} = A_{i}\hat{f}_{1}$, $\hat{f}^{[i]}_{w,2} = B_{i}\hat{f}_{2}$, $\hat{f}^{[i]}_{w,3} = C_{i}\hat{f}_{3}$, and $\hat{f}^{[i]}_{w,4} = D_{i}\hat{f}_{4}$.
\State Combine the weighted droplet functions: $\hat{f}^{[i]}_{\text{comb}} = \hat{f}^{[i]}_{w,1}+\hat{f}^{[i]}_{w,2}+\hat{f}^{[i]}_{w,3}+\hat{f}^{[i]}_{w,4}$.
\State Estimate a unitary matrix $U_{\text{est},i}$ from the combined droplet function $\hat{f}^{[i]}_{\text{comb}}$.
\State Estimate new quaternion components $A_{i+1}$, $B_{i+1}$, $C_{i+1}$, and $D_{i+1}$ from $U_{\text{est},i}$.
\State Input new values of quaternion components $A_{i+1}$, $B_{i+1}$, $C_{i+1}$, and $D_{i+1}$ into Step 3.
\end{algorithmic}
\begin{flushleft}
\textbf{Termination}: Repeat Steps 3 to 7 until the change in quaternion values over iterations is not significant.\\
\textbf{Output}: Estimated droplet function $f_{\text{est}}$ representing a unitary process $U_{\text{est}}$. 
\end{flushleft}
\end{algorithm*}

The first step of the algorithm involves computing a correlation matrix that provides the correlation between different scaled droplet functions $\hat{f}_{k}$. In the second step, this is used to estimate the zero-order (i.e., iteration $i=0$) values of quaternion components; see Appendix~\ref{Supp:correlation_matrix}. The correlation matrix can be computed using the discretized scalar products between different spherical droplet functions, as discussed in Appendix~\ref{Supp:correlation_matrix}. \\
\begin{figure*}[t]
	\centering
	\adjustbox{max width=\textwidth}{\includegraphics[scale=1.1]{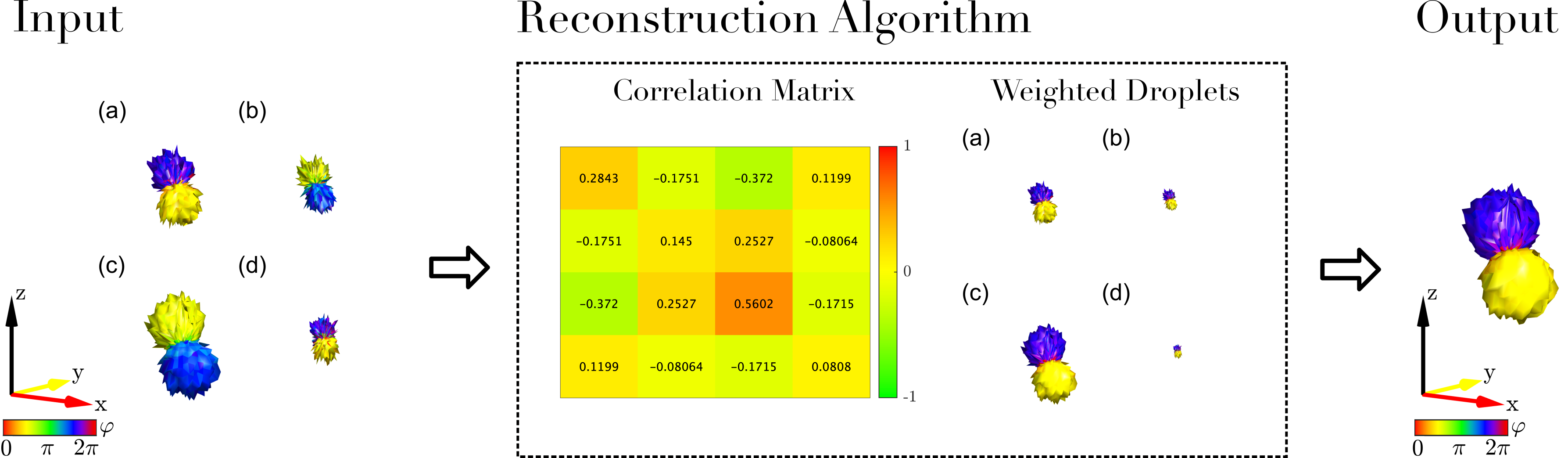}}
	\caption{Illustration of the reconstruction of an unknown process from the tomographed droplets with shot noise using the algorithm presented in Algorithm~\ref{alg:quaternion_estimation}. The input tomographed scaled process droplets: (a) $\hat{f}_{1}$, (b) $\hat{f}_{2}$, (c) $\hat{f}_{3}$, and (d) $\hat{f}_{4}$ are shown on the left. Inside the reconstruction algorithm (middle panel): first, a correlation matrix is computed, providing the zero-order estimates of the quaternion components $A_{0}$, $B_{0}$, $C_{0}$, and $D_{0}$. These zero-order estimates are then multiplied with droplet functions, resulting in the weighted droplet functions: $\hat{f}^{[0]}_{w,1} = A_{0}\hat{f}_{1}$, $\hat{f}^{[0]}_{w,2} = B_{0}\hat{f}_{2}$, $\hat{f}^{[0]}_{w,3} = C_{0}\hat{f}_{3}$, and $\hat{f}^{[0]}_{w,4} = D_{0}\hat{f}_{4}$. The resulting output droplet on the rightmost panel corresponds to the process with quaternion components $A_{3} = 0.5196$, $B_{3} = -0.3443$, $C_{3} = -0.7416$, and $D_{3} = 0.2479$, obtained after $i = 3$ iterations with a fidelity of 0.9999.}
	\label{fig:8}
\end{figure*}

Computing a weighted combination of experimental droplets $\hat{f}_{k}$ in step 3 is based on the principle of matched filtering~\cite{ernst1987principles, spencer2010equivalence}, as discussed earlier, and it is a key step of the reconstruction algorithm. For an iteration $i$, the weighted scaled droplet functions $\hat{f}_{k}$ are combined as described in Step 4 of Algorithm~\ref{alg:quaternion_estimation}, such that
\begin{equation}
	\label{Eq.30}
	\begin{split}
		\hat{f}^{[i]}_{\text{comb}} & = \hat{f}^{[i]}_{w,1}+\hat{f}^{[i]}_{w,2}+\hat{f}^{[i]}_{w,3}+\hat{f}^{[i]}_{w,4}\\
		& = A_{i}\hat{f}_{1}+B_{i}\hat{f}_{2}+C_{i}\hat{f}_{3}+D_{i}\hat{f}_{4}.
	\end{split}
\end{equation}
Where $A_{i}$, $B_{i}$, $C_{i}$, and $D_{i}$ are the estimated quaternion components in iteration $i$. The actual unknown process $U_{a}$ is composed of quaternion components $A$, $B$, $C$, and $D$ (see Eq.~\eqref{Eq.19}), and  $f_{a}$ is the spherical droplet corresponding to it, such that, $\hat{f}_{1} = Af_{a}$, $\hat{f}_{2} = Bf_{a}$,  $\hat{f}_{3} = Cf_{a}$, and $\hat{f}_{4} = Df_{a}$ (see Eq.~\eqref{Eq.29}). After substituting these transformations, Eq.~\eqref{Eq.30} takes the form:
\begin{equation}
	\label{Eq.31}
	\hat{f}^{[i]}_{\text{comb}} = A_{i}(Af_{a})+B_{i}(Bf_{a})+C_{i}(Cf_{a})+D_{i}(Df_{a}).
\end{equation}
Therefore, a combination of experimental droplets $\hat{f}_{k}$ with varying weights $A_{i}$, $B_{i}$, $C_{i}$, and $D_{i}$ provides a highly accurate estimate of the actual droplet function $f_{a}$. Additionally, if these weights are the same as the actual quaternion components, i.e., $A_{i} = A$, $B_{i} = B$, $C_{i} = C$, and $D_{i} = D$, the resulting combined droplet $\hat{f}^{[i]}_{\text{comb}}$ of Eq.~\ref{Eq.31} equals the actual droplet function $f_{a}$:
\begin{equation}
    \label{Eq.32}
    \hat{f}^{[i]}_{\text{comb}} = f_{a},
\end{equation} 
where $A^{2}+B^{2}+C^{2}+D^{2}=1$. \\

We demonstrate the algorithm using a numerical example in Fig.~\ref{fig:8}, simulated with the number of shots $N_{s} = 300$. Here, the actual quaternion values are $A = 0.5198$, $B = -0.3462$, $C = -0.7424$, and $D = 0.2425$; see Fig.~\ref{fig:scaling_example}. In this example, the algorithm was terminated after $i=3$ iterations due to a non-significant change in the values of quaternion components over iterations. Additional supporting information for this example is provided in the plots in Appendix~\ref{Supp:fig:8_plots}. Several additional examples demonstrating the reconstruction algorithm for a single qubit system are also presented in the doctoral thesis of Devra~\cite{Devra_dissertation}.      
\subsection{Reconstruction algorithm with optimization}
\label{Sec.:Reconstruction algo optimization}
The reconstruction algorithm presented above provides a reliable estimate of quaternion components. However, while this approach is powerful, it does not ensure that the solution is optimal. Here, we elaborate on combining the reconstruction algorithm with an optimization routine and discuss the benefits of this integrated approach.\\

We use the zero-order estimates (corresponding to $i=0$) of quaternion components $A_{0}$, $B_{0}$, $C_{0}$, and $D_{0}$ obtained from the correlation matrix as an initial guess for the optimization to minimize the following cost function:
\begin{equation}
\label{Eq.33}
    \begin{aligned}
        J = {} & {||\hat{U}_{1} - A_{i}U_{\text{est},i} ||}^{2}+       
            {||\hat{U}_{2} - B_{i}U_{\text{est},i} ||}^{2}+\\
          & {||\hat{U}_{3} - C_{i}U_{\text{est},i} ||}^{2} +            
            {||\hat{U}_{4} - D_{i}U_{\text{est},i} ||}^{2}.
    \end{aligned}
\end{equation}  
Here $\hat{U}_{1}$, $\hat{U}_{2}$, $\hat{U}_{3}$, and $\hat{U}_{4}$ are scaled matrices, i.e., the matrix representations of the experimentally obtained scaled droplet functions $\hat{f}_{1}$, $\hat{f}_{2}$, $\hat{f}_{3}$, and $\hat{f}_{4}$, respectively. The process matrix estimation from the droplet functions can be derived using the approach described in Ref.~\cite{Devra_WQST}. $A_{i}$, $B_{i}$, $C_{i}$, and $D_{i}$ are the quaternion components estimated in iteration $i$. The operator $U_{\text{est},i}$ represents the estimated unitary process matrix for iteration $i$, composed of quaternion components $A_{i}$, $B_{i}$, $C_{i}$, and $D_{i}$, as defined in Eq.~\eqref{Eq.19}. The cost function $J$ quantifies the closeness of the estimated process matrix to the experimentally obtained scaled process matrices. To minimize this cost function, we utilize the gradient-based optimization function, \texttt{fminunc} in Matlab. In Appendix~\ref{Supp:recons_opt_example}, we provide an example to showcase the reconstruction algorithm with optimization and compare it to the one without optimization.\\

\begin{figure}[!]
	\centering
        \adjustbox{max width=\textwidth}{\includegraphics[scale=0.9]{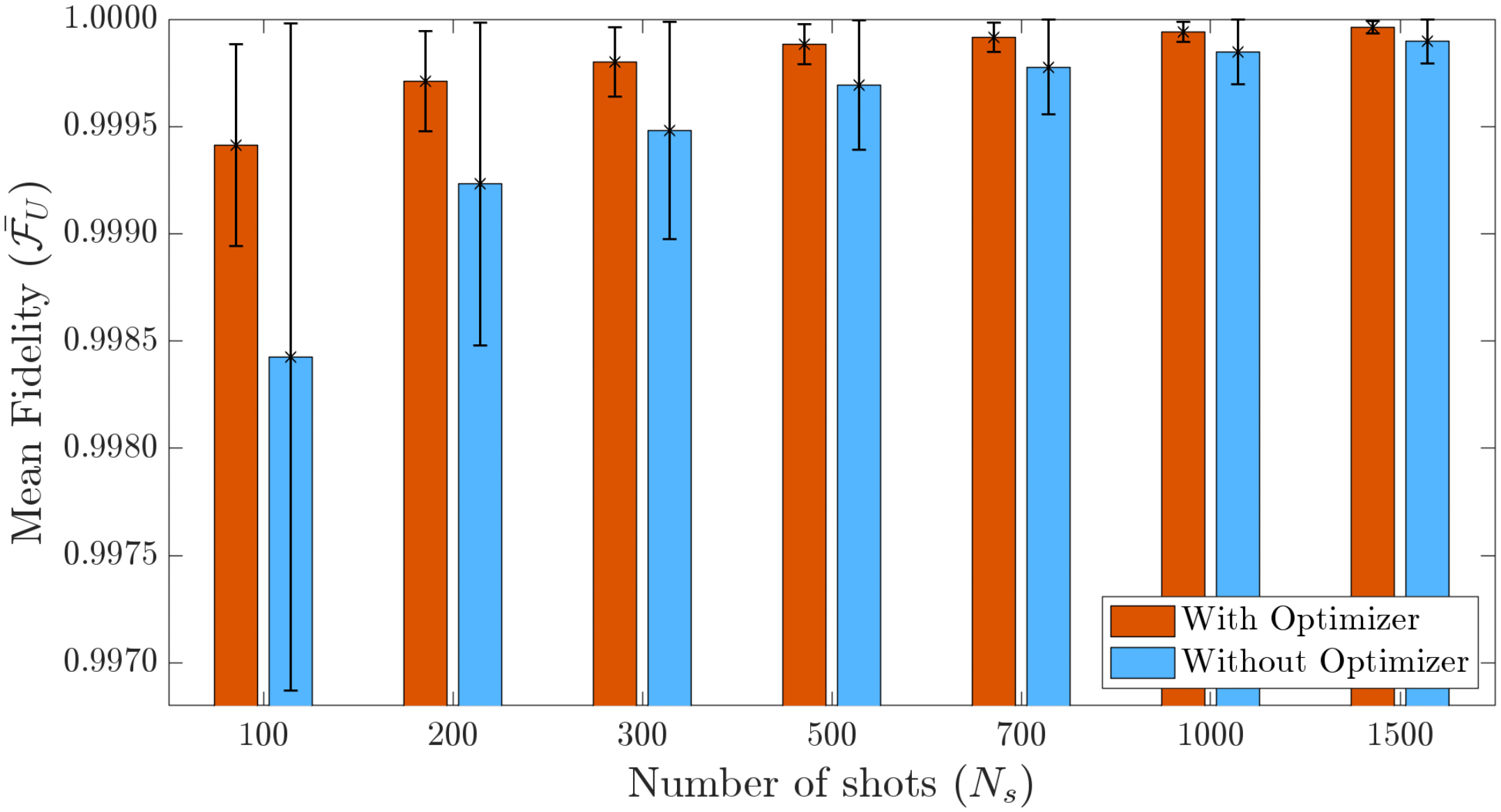}}
	\caption{Plot of mean fidelity $\bar{\mathcal{F}_{U}}$ with different number of shots $N_{s}$ for reconstruction algorithm approach with and without optimizer. The mean is computed over 50 noise instances for 100 random gates for each $N_{s}$. The standard deviation for each $N_{s}$ is represented by the vertical bars.}
	\label{fig:9}
\end{figure}

We also conducted a numerical study to evaluate the performance of the reconstruction algorithm in the presence of shot noise with and without optimization. In this study, we compared the process fidelity $\mathcal{F}_{U}$ (computed using the definition in Eq.~\eqref{Eq:D3}) of the reconstructed process matrix with the target process matrix for different numbers of shots, denoted as $N_{s}$. The Lebedev sampling scheme with 50 grid points~\cite{LEBEDEV197610} was employed for scanning purposes, and this same scheme was utilized in the experiments presented in the next section. For each value of the number of shots $N_{s}$, 100 random gates were generated, and for each of these gates, 50 different noise instances were created. The mean and standard deviation of the process fidelity were computed using both approaches. The plot depicting the mean fidelity ($\bar{\mathcal{F}_{U}}$) for both approaches with different numbers of shots ($N_{s}$) is shown in Fig.~\ref{fig:9}. The plot also includes the standard deviation for each approach, represented by vertical bars across varying numbers of shots.\\

The cost function described in Eq.~\eqref{Eq.33}, which relies on the quaternion components, is valid only for a single-qubit system and is not scalable to multiple qubits. We discuss the extension of this approach in Sec.~\ref{Sec.:Reconstuction_extentsion}.

\section{Experimental implementation}
\label{Sec.:WQPT_Experimental}
In this section, we describe the details for experimentally implementing Wigner process tomography for unknown processes. We present experimental results for a single system qubit, i.e., $N = 1$, performed on an IBM quantum device for a pure state on an individual, well-defined quantum system. The quantum circuits presented here are general and can be directly used on any circuit-based quantum computer. 

\subsection{Single qubit system}
\label{Sec.:single-qubit_experiments}
\begin{figure}[!]
	\centering
	\includegraphics[scale=0.8]{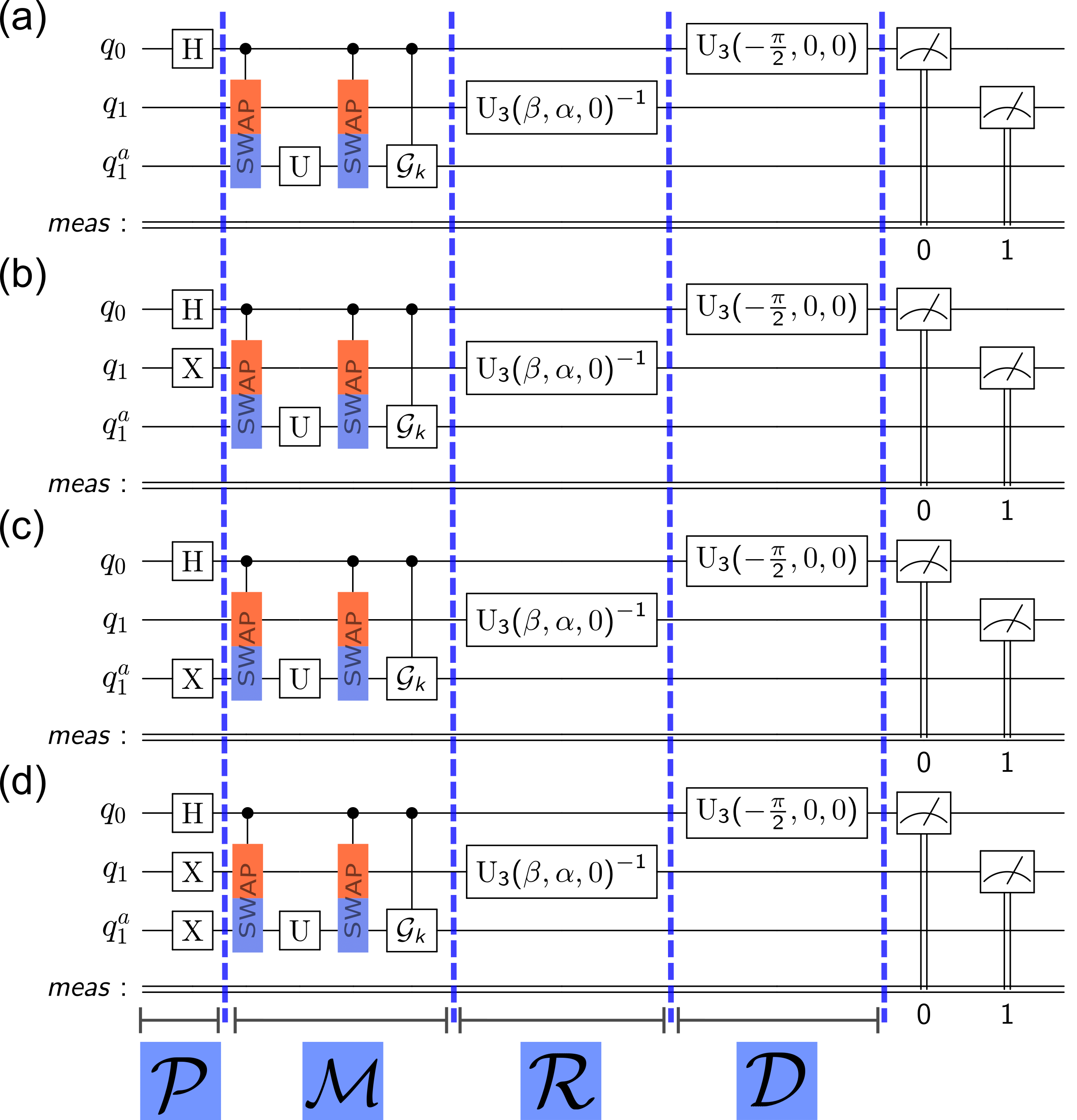}
	\caption{Quantum circuit set for performing tomography of a single qubit unknown process $U$. The initial state of the qubits is $|\psi_{i}\rangle = |000\rangle$. The four blocks of algorithm, namely the preparation ($\mathcal{P}$), the mapping ($\mathcal{M}$), the rotation ($\mathcal{R}$), and the detection-associated rotations ($\mathcal{D}$) are shown here explicitly. The circuits shown here only measure expectation values $\langle{\sigma_{0x}}\rangle$ and $\langle{\sigma_{0x}\sigma_{1z}}\rangle$, therefore the number of circuit extends by replacing $\mathrm{U}_{3}(-\frac{\pi}{2},0,0)$ with $\mathrm{U}_{3}(\frac{\pi}{2},0,\frac{\pi}{2})$ to measure $\langle{\sigma_{0y}}\rangle$ and $\langle{\sigma_{0y}\sigma_{1z}}\rangle$. Each circuit is then repeated for different rotations $\mathcal{G}_{k}\in\{\mathds{1},\sigma_{x},\sigma_{y},\sigma_{z}\}$ and scanning angles $\beta$ and $\alpha$. The $\mathrm{U}_{3}$ gate used in these circuits is discussed in Appendix~\ref{Supp:U3}.}
	\label{fig:10}
\end{figure}

For Wigner tomography of a single-qubit unknown process, three qubits are required ($q_{0}$, $q_{1}$, $q_{1}^{a}$). Here, $q_{1}$ is the system qubit, and $q_{0}$ and $q_{1}^{a}$ are ancilla qubits. The spherical droplet functions $f_{0,k}^{(\emptyset)}$ and $f_{1,k}^{(1)}$ corresponding to scaled process matrices $U_{k}^{[1]}$ for different controlled rotations $\mathcal{G}_{k}\in\{\mathds{1},\sigma_{x},\sigma_{y},\sigma_{z}\}$ can be tomographed by combining the experimentally measured expectation values for a set of scanning angles $\beta$ and $\alpha$, as expressed in Eq.~\eqref{Eq.23}. The general quantum circuits for performing Wigner tomography are shown in Fig.~\ref{fig:10}. This figure explicitly highlights the four steps of the algorithm discussed in Sec.~\ref{Sec.:WQPT theory}. Now, we explain these steps.
\\

In the preparation step ($\mathcal{P}$), we initiate the control qubit $q_{0}$ in an equal superposition state using the Hadamard gate ($\mathrm{H}$). To prepare qubits $q_{1}$ and $q_{1}^{a}$ in a maximally mixed state ($\rho_{mm}^{[2]} = \frac{1}{4}\mathds{1}^{[2]}$), we employ the temporal averaging approach~\cite{knill_temporal, preskill1998lecture, Devra_WQST}. This approach was initially introduced for a single qubit~\cite{Devra_WQST}, and here, we extend it to prepare qubits $q_{1}$ and $q_{1}^{a}$ in a maximally mixed state. This extension involves repeating the experiment for the four computational basis states ($|00\rangle$, $|01\rangle$, $|10\rangle$, and $|11\rangle$) and averaging the measurement outcomes:
\begin{equation}
    \label{Eq.34}
    \begin{aligned}
          \rho_{mm}^{[2]} = {} & \frac{1}{4}\mathds{1}^{[2]} \\
            = & \frac{1}{4}(|00\rangle\langle{00}|+|01\rangle\langle{01}|+|10\rangle\langle{10}|+|11\rangle\langle{11}|).
    \end{aligned}
\end{equation} 
The different computational basis states are created by applying NOT ($\mathrm{X}$) gates in the circuits shown in Fig.~\ref{fig:10}, assuming that the initial state of all qubits is $|\psi_{i}\rangle = |000\rangle$. In these circuits, the state of qubits $q_{1}$ and $q_{1}^{a}$ after the preparation step is (a) $|00\rangle$, (b) $|01\rangle$, (c) $|10\rangle$, and (d) $|11\rangle$. Therefore, the (temporally) averaged expectation values of the circuits presented in Fig.~\ref{fig:10} provide $\langle{\sigma_{0x}}\rangle$ and $\langle{\sigma_{0x}\sigma_{1z}}\rangle$ for a controlled rotation $\mathcal{G}_{k}$.\\

In the mapping step ($\mathcal{M}$), we substitute the circuit presented in Fig.~\ref{fig:cU_unknownWR} for $N=1$. For the scanning part of the rotation step ($\mathcal{R}$), the $\mathrm{U}_{3}$ gate is employed to rotate the system qubit $q_{1}$ by an angle $\beta$ around the $y$ axis followed by a rotation of $\alpha$ around the $z$ axis. The $\mathrm{U}_{3}$ gate used in these circuits is discussed in Appendix~\ref{Supp:U3}. \\

The circuits shown in Fig.~\ref{fig:10} provide expectation values $\langle{\sigma_{0x}}\rangle$ and $\langle{\sigma_{0x}\sigma_{1z}}\rangle$ for a controlled rotation $\mathcal{G}_{k}$. The other required expectation values (see Eq.~\eqref{Eq.23}), $\langle{\sigma_{0y}}\rangle$ and $\langle{\sigma_{0y}\sigma_{1z}}\rangle$, are computed similarly by replacing $\mathrm{U}_{3}(-\frac{\pi}{2},0,0)$ with $\mathrm{U}_{3}(\frac{\pi}{2},0,\frac{\pi}{2})$ in the detection-associated rotations step ($\mathcal{D}$). These additional circuits are not shown here. Hence, there are eight quantum circuits for each of the four rotations $\mathcal{G}_{k}\in\{\mathds{1},\sigma_{x},\sigma_{y},\sigma_{z}\}$, making it a total of 32 ($=8\times4$) quantum circuits for tomography, which are evaluated for a set of scanning angles $\beta$ and $\alpha$. Now, we discuss the implementation of these quantum circuits on an experimental device.

\subsection{Calibration experiments}
\label{Sec.:Calibration}
For implementing a quantum circuit on an experimental device, each quantum gate is decomposed into hardware native gates, which are then implemented using control pulses. Compared to the tomography experiments of known processes (c.f. Ref.~\cite{Devra_WQST}), we found two major experimental problems in the first implementations of the tomography experiments for unknown processes:
\begin{enumerate}[label=(\Alph*)]
\item The experimentally measured expectation values were scaled down uniformly by about 50\% compared to the simulated expectation values, resulting in a corresponding scaling of experimental droplet functions. (The experimental scaling mentioned here should not be confused with the scaling factor $\epsilon$ presented in Sec.~\ref{Sec.:modified circuit}.) 
\item Preliminary experiments showed a high fidelity of the (re-scaled) tomographed droplet functions except for a relatively large rotation of the rank $j = 1$ droplet (which is not rotational symmetric) by about $20^{\circ}$ around the $z$ axis, independent of the process of interest that was tomographed.
\end{enumerate} 
Problem (A) is consistent with decoherence during the pair of relatively long CSWAP gates and experimental imperfections of a large number of single-qubit and two-qubit gates needed to implement the CSWAP gates. For instance, on IBM devices, a single CSWAP gate has an overall duration of several microseconds. It consists of about ten controlled two-qubit gates and about thirty single-qubit gates (the actual numbers depend on the actual device properties). The resulting uniform scaling of the experimentally measured droplet function is automatically taken into account by the normalization steps of the reconstruction algorithm (see Sec.~\ref{Sec.:Reconstruction theory}). However, this additional scaling of the droplet reduces the signal-to-noise ratio of the tomography experiments, at least for noisy near-term quantum devices.\\

\begin{figure}[!]
	\centering
	\includegraphics[scale=0.7]{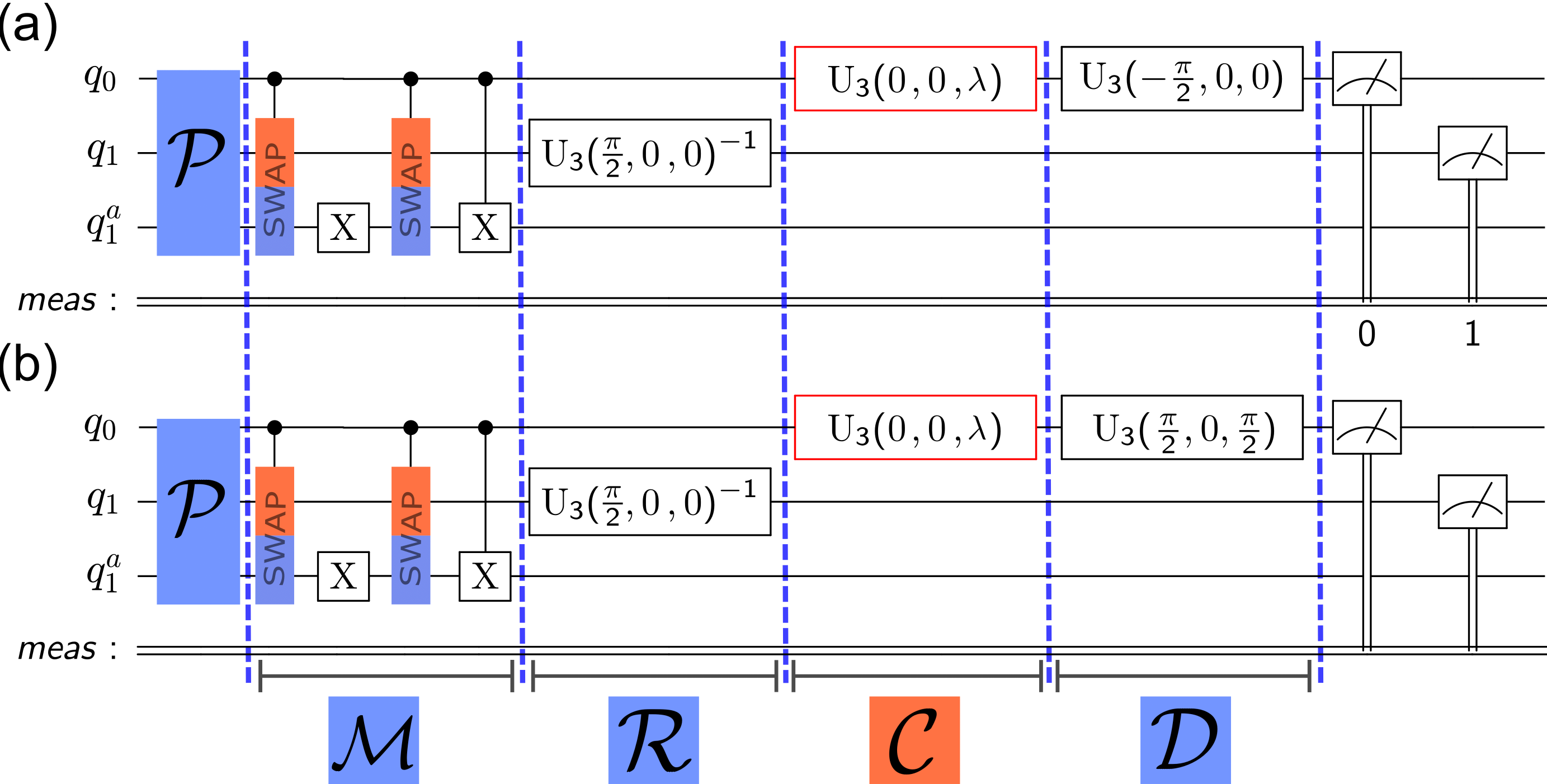}
	\caption{Circuits for performing calibration experiments to quantify the phase shift of qubit $q_{0}$. These circuits are adapted from the tomography circuits presented in Fig.~\ref{fig:10}. For both circuits, the $\mathcal{P}$ block is identical to the preparation block shown in Fig.~\ref{fig:10}. In comparison to the circuits shown in Fig.~\ref{fig:10}, here the mapping block $\mathcal{M}$ contains $U = \mathrm{X}$ (NOT) gate, $c\mathcal{G} = c\mathrm{X}$, and rotation block $\mathcal{R}$ has values $\beta = \frac{\pi}{2}$ and $\alpha = 0$. An additional calibration block $\mathcal{C}$ is introduced here, which consists of $\mathrm{U}_{3}(0,0,\mathrm{\lambda})$ (corresponds to $RZ(\lambda)$) on qubit $q_{0}$, followed by a detection-associated rotation block $\mathcal{D}$ to estimate $\langle{\sigma_{0x}\sigma_{1z}}\rangle$ and $\langle{\sigma_{0y}\sigma_{1z}}\rangle$ from circuits (a) and (b), respectively. See Fig.~\ref{fig:12} for measurement results.}
\label{fig:11}
\end{figure}

\begin{figure}[h]
	\centering
	\includegraphics[scale=0.65]{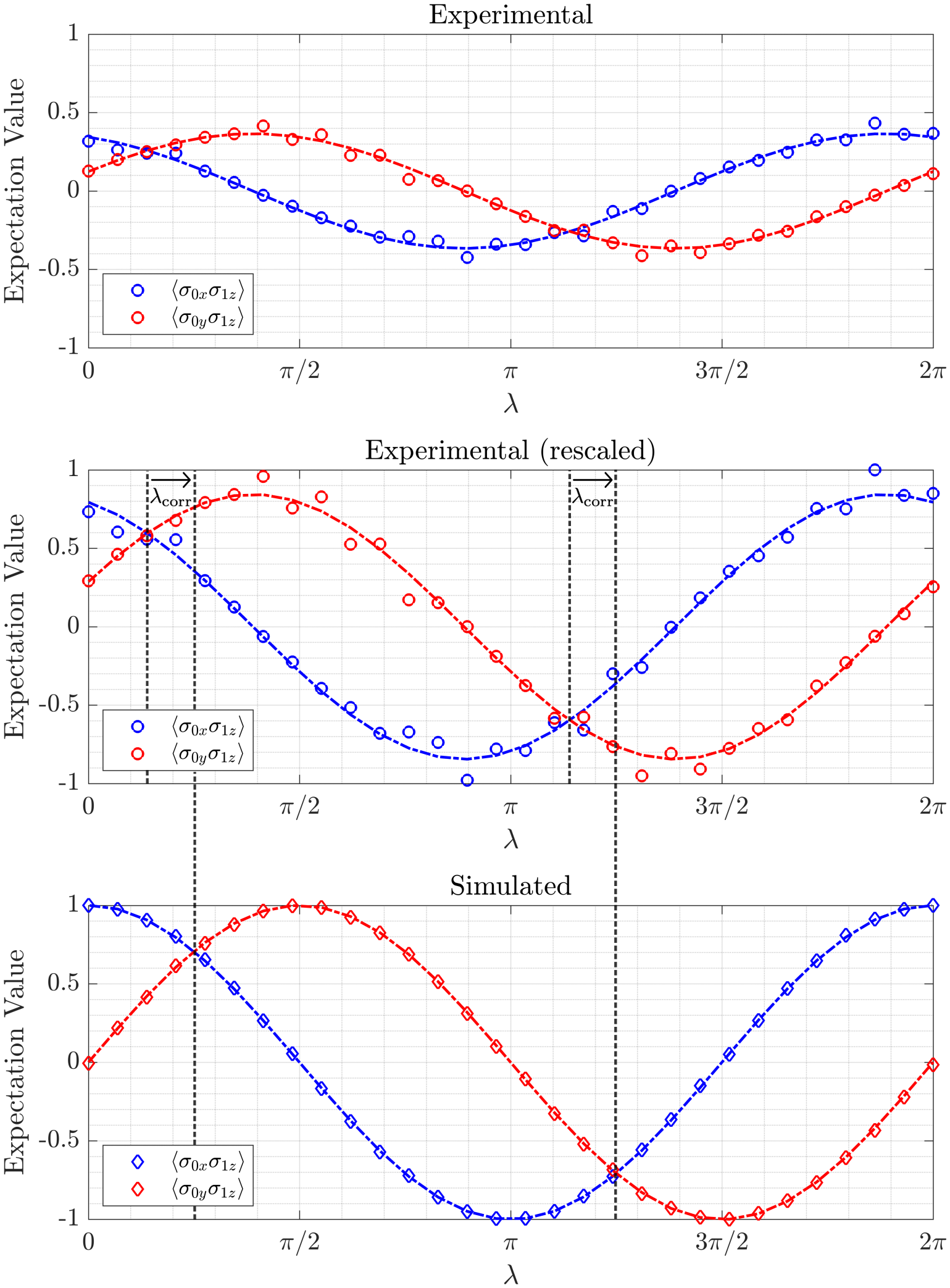}
	\caption{Experimental (top panel), re-scaled experimental (middle panel), and simulated (bottom panel) plot of the expectation values $\langle{\sigma_{0x}\sigma_{1z}}\rangle$ and $\langle{\sigma_{0y}\sigma_{1z}}\rangle$ as a function of rotation around the $z$ axis with a rotation angle $\lambda$. The plot is an outcome of the calibration circuits presented in Fig.~\ref{fig:11}. The expectation values in the top panel are obtained experimentally using the \texttt{ibmq$\_$mumbai} device. The re-scaled normalized experimental values in the middle panel are plotted for a direct visual comparison with the simulated values. The dashed lines highlight the phase shift $\lambda_{\text{corr}}$ of the experimental results compared to the simulation.}
	\label{fig:12}
\end{figure}

Problem (B) is also a result of the long duration of the CSWAP gates, which amplifies the effect of small detunings of the ancilla qubit $q_{0}$. In general, phase errors arise from the frequency detuning of the drive during the physical implementation of single and two-qubit gates, as discussed in Ref.~\cite{werninghaus2022experimental}. This effect can be mitigated through calibration experiments. Preceding experiments of unknown processes with a simplified tomography experiment of an X (NOT) gate can be performed, which enables precise determination of this phase error by fitting phase-shifted sine and cosine functions to the data; see Figs.~\ref{fig:11} and \ref{fig:12}.\\

For quantifying the unwanted additional phase shift of qubit $q_{0}$, we used the calibration circuits shown in Fig.~\ref{fig:11}. These circuits are derived from the tomography circuits presented in Fig.~\ref{fig:10}, with $U = \mathrm{X}$ (NOT) gate, $c\mathcal{G} = c\mathrm{X}$ in the mapping block $\mathcal{M}$, scanning angles $\beta = \frac{\pi}{2}$ and $\alpha = 0$ in rotation block $\mathcal{R}$. An additional calibration block $\mathcal{C}$ is introduced here which consist of $\mathrm{U}_{3}(0,0,\mathrm{\lambda})$ (corresponds to $RZ(\lambda)$) on qubit $q_{0}$. The $\mathcal{C}$ block is followed by a detection-associated rotation block $\mathcal{D}$ to experimentally estimate $\langle{\sigma_{0x}\sigma_{1z}}\rangle$ and $\langle{\sigma_{0y}\sigma_{1z}}\rangle$ from circuits (a) and (b), respectively. In this case where $U = \mathrm{X}$, the rank $j = 0$ droplet is zero; therefore, the expectation values $\langle{\sigma_{0x}}\rangle$ and $\langle{\sigma_{0y}}\rangle$ are zero for all scanning angles $\beta$ and $\alpha$. The preparation block $\mathcal{P}$ of both circuits in Fig.~\ref{fig:11} corresponds to the preparation block shown in Fig.~\ref{fig:10}, i.e., applying the Hadamard gate to prepare qubit $q_{0}$ in an equal superposition state and applying suitable NOT gates to prepare qubits $q_{1}$ and $q_{1}^{a}$ in a maximally mixed state using temporal averaging.\\

The calibration circuits shown in Fig.~\ref{fig:11} are repeated for $RZ(\lambda)$ with $\lambda\in [0,2\pi]$. The plot of the expectation values $\langle{\sigma_{0x}\sigma_{1z}}\rangle$ and $\langle{\sigma_{0y}\sigma_{1z}}\rangle$ as a function of angle $\lambda$ is shown in Fig.~\ref{fig:12}. The simulated plot in the bottom panel shows the ideal values of expectation values, whereas the expectation values in the top panel are obtained experimentally using \texttt{ibmq$\_$mumbai} device. A plot of normalized experimental values is shown in the middle panel to provide a direct visual comparison with the simulated values in the bottom panel. The experimental expectation values have a phase shift of $\lambda_{\text{corr}} = 0.3473$ radians $(\approx 20^{\circ})$ compared to the simulated values. This phase shift corresponds to the unwanted extra rotation of qubit $q_{0}$. We compensate for this extra rotation of qubit $q_{0}$ in the tomography experiments by applying $-RZ(\lambda_{\text{corr}})$ on qubit $q_{0}$ right before the $\mathcal{D}$ (detection-associated rotation) block of circuits in Fig.~\ref{fig:10}.

\subsection{Experimental results}
\label{Sec: Experimental results}
We experimentally implemented the tomography circuits depicted in Fig.~\ref{fig:10} with an additional $-RZ(\lambda_{\text{corr}})$ rotation as discussed above on the \texttt{ibmq$\_$mumbai} device. For these experiments, we employed a Lebedev $n=50$ grid~\cite{LEBEDEV197610} for scanning angles $\beta$ and $\alpha$, and each experiment was repeated for each grid point for $N_{s} = 4096$ shots.\\ 

\begin{figure}[!]
	\centering
	\includegraphics[scale=1]{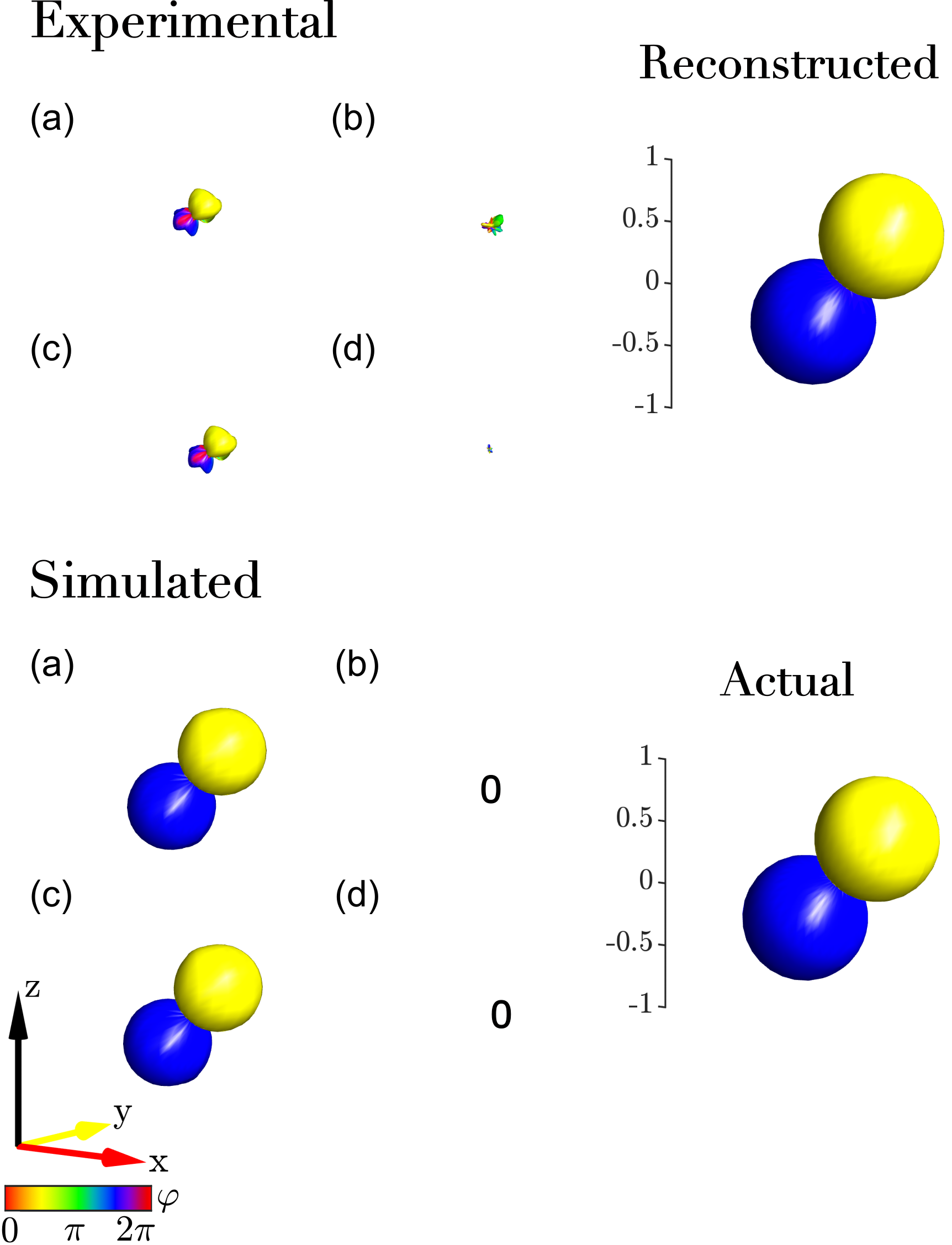}
	\caption{The experimental and simulated results for the tomography of the $U = \mathrm{H}$ (Hadamard) gate. quaternion components corresponding to a Hadamard gate are $A = C = -0.7071$ and $B = D = 0$. The left panel shows the simulated and experimental scaled process droplets (a) $\hat{f}_{1}$, (b) $\hat{f}_{2}$, (c) $\hat{f}_{3}$, and (d) $\hat{f}_{4}$. The right panel shows the simulated actual droplets corresponding to $\mathrm{H}$ gate and the reconstructed droplet from the experimental droplets. The reconstructed droplet is obtained with a fidelity of 0.9974. See Appendix~\ref{Supp:Hadamard_gate} for more details.}
	\label{fig:13}
\end{figure}

The experimental results presented in Fig.~\ref{fig:13} focus on the ``unknown'' process $U = \mathrm{H}$ (Hadamard) gate. The quaternion values associated with a Hadamard gate are $A = C = -0.7071$ and $B = D = 0$. In this case, there are two non-zero scaled process matrices $\hat{U}_{k}$ for $k=1$ and $k=3$ corresponding to $\mathcal{G} = \sigma_{x}$ and $\mathcal{G} = \sigma_{z}$, respectively (see Table~\ref{Tab:scaling_factor}). The simulated and the experimental scaled droplets corresponding to different controlled rotations are shown in the left panel of Fig.~\ref{fig:13}. The right panel shows the simulated actual droplet corresponding to the Hadamard gate and the reconstructed droplet from the experimental droplets. The expectation values, along with other details (such as correlation matrix and plots of cost and non-fidelity with iterations) of the reconstruction algorithm, are presented in Appendix~\ref{Supp:Hadamard_gate}.
\\

\begin{figure}[!]
	\centering
	\includegraphics[scale=1]{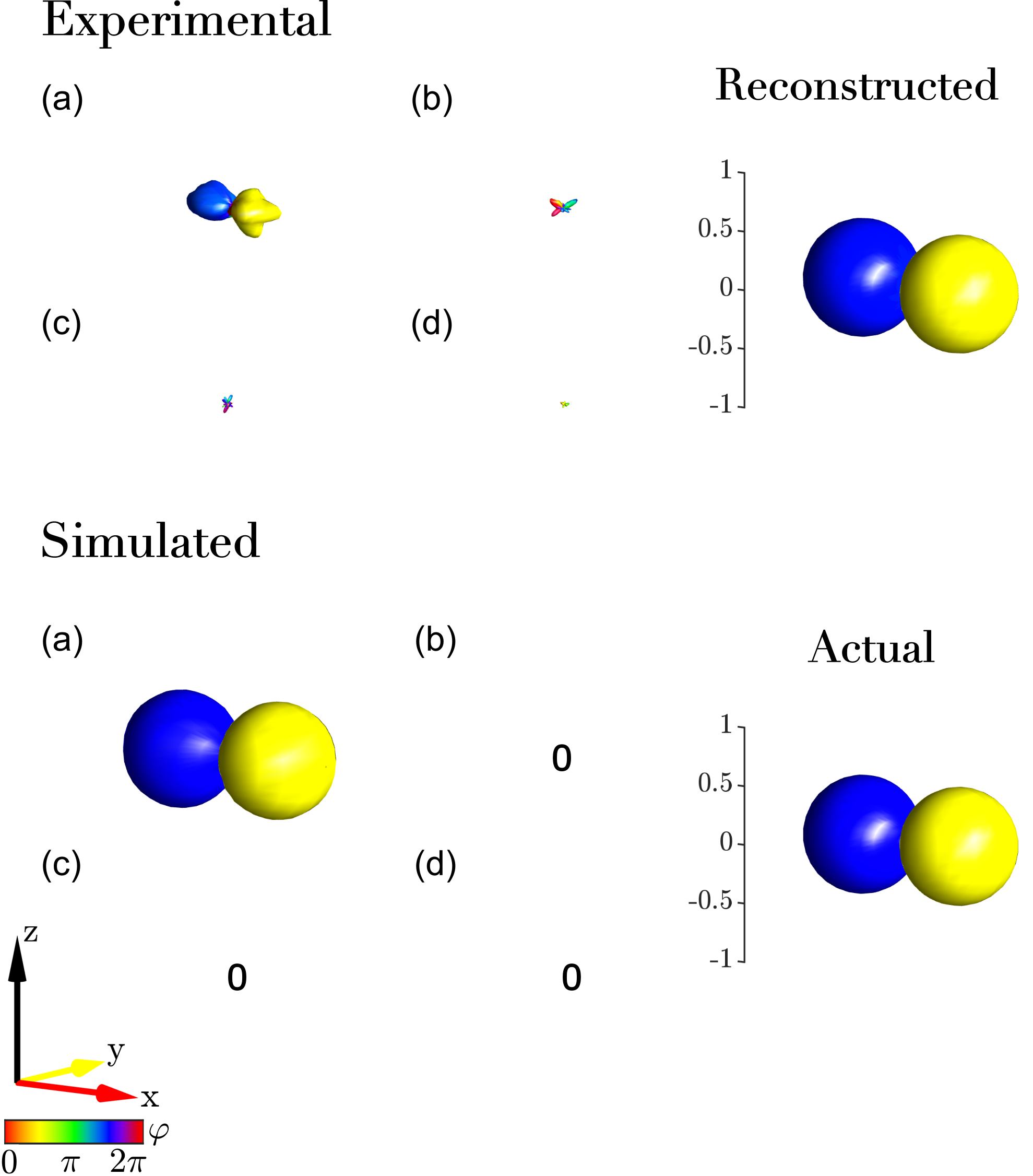}
	\caption{The experimental and simulated results for the tomography of the $U = \mathrm{X}$ (NOT) gate. quaternion components corresponding to a $\mathrm{X}$ gate are $A=1$ and $B = C = D = 0$. The left panel shows the simulated and experimental scaled process droplets (a) $\hat{f}_{1}$, (b) $\hat{f}_{2}$, (c) $\hat{f}_{3}$, and (d) $\hat{f}_{4}$. The right panel shows the simulated actual droplets corresponding to the $\mathrm{X}$ gate and the reconstructed droplet from the experimental droplets. The reconstructed droplet is obtained with a fidelity of 0.9991. See Appendix~\ref{Supp:not_gate} for more details.}
	\label{fig:14}
\end{figure}

Similarly, the experimental result for a $U = \mathrm{X}$ (NOT) gate is presented in Fig.~\ref{fig:14}. The quaternion values associated with a NOT gate are $A=1$ and $B = C = D = 0$. According to Table~\ref{Tab:scaling_factor}, ideally there is only one non-zero scaled process matrix $\hat{U}_{k} = \epsilon_{k}U$ for $k=1$, corresponding to the $\mathcal{G} = \sigma_{x}$ rotation. The simulated and experimental scaled process droplets $\hat{f}_{k}$ are shown in Fig.~\ref{fig:14} along with the reconstructed and simulated actual droplets. The experimental and simulated expectation values, along with the supporting plots, are presented in Appendix~~\ref{Supp:not_gate}. 
\\

\begin{figure}[!]
	\centering
	\includegraphics[scale=1]{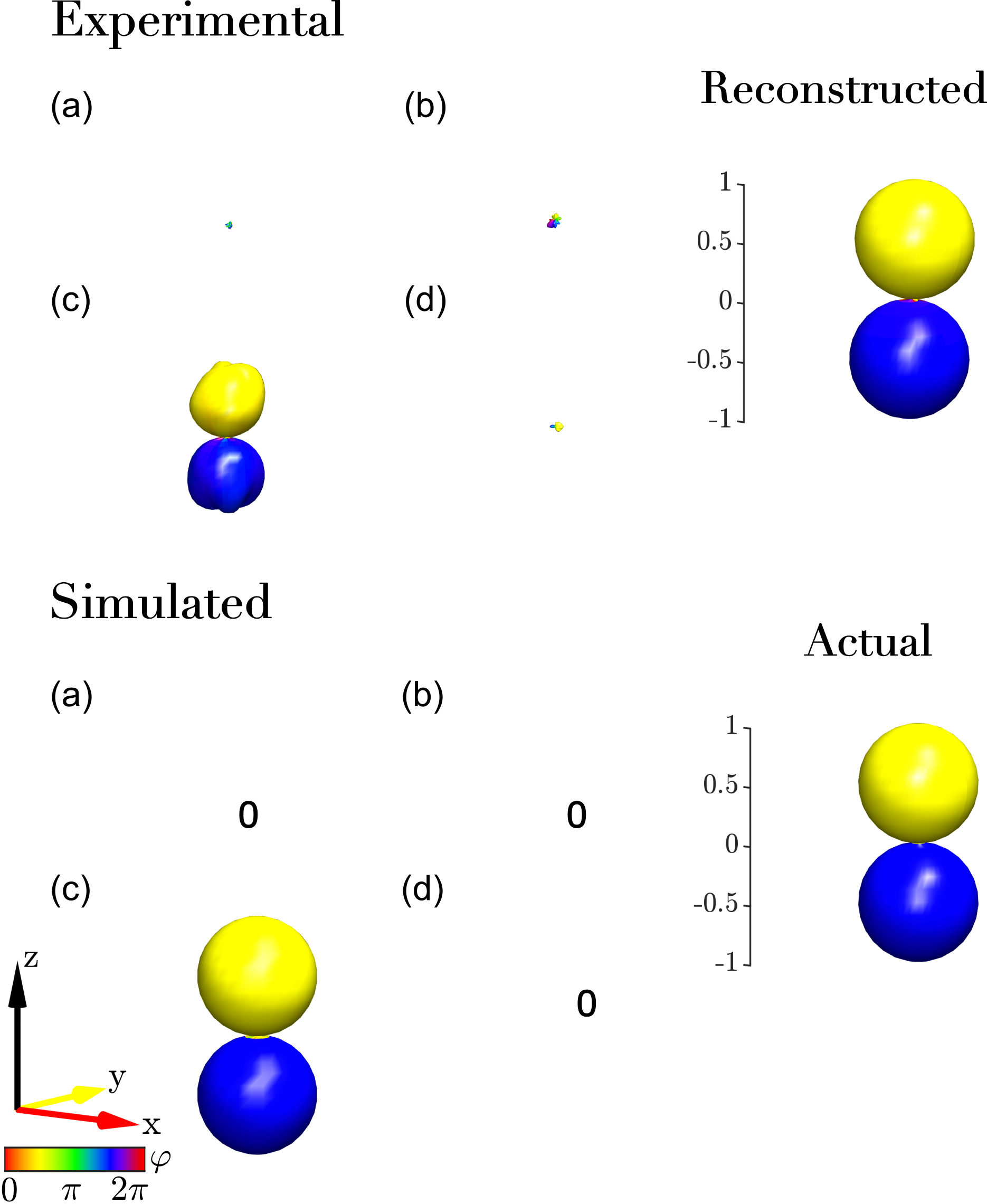}
	\caption{The experimental and simulated results for the tomography of the $U = \mathrm{Z}$ gate. quaternion components corresponding to a $\mathrm{Z}$ gate are $C=1$ and $A = B = D = 0$. The left panel shows the simulated and experimental scaled process droplets (a) $\hat{f}_{1}$, (b) $\hat{f}_{2}$, (c) $\hat{f}_{3}$, and (d) $\hat{f}_{4}$. The right panel shows the simulated actual droplets corresponding to the $\mathrm{X}$ gate and the reconstructed droplet from the experimental droplets. The reconstructed droplet is obtained with a fidelity of 0.9966. See Appendix~\ref{Supp:Z_gate} for more details.}
	\label{fig:15}
\end{figure}

The resulting experimental and simulated droplets for a $U = \mathrm{Z}$ gate are shown in Fig.~\ref{fig:15}. The quaternion values associated with a Z gate are $C = 1$ and $A=B=D=0$. Similar to the NOT gate, there is only one non-zero scaled process matrix $\hat{U}_{k}$ for $k=3$, corresponding to $\mathcal{G} = \sigma_{z}$ (see Table~\ref{Tab:scaling_factor}). The experimental expectation values and supporting plots for the reconstruction algorithm for this case are provided in Appendix~\ref{Supp:Z_gate}. 
\section{Adaptive approach for Wigner process tomography}
\label{Sec:Adaptive}
An essential step for the Wigner process tomography approach presented herein is the circuit illustrated in Fig.~\ref{fig:cU_unknownWR}, which is designed to map $N$ qubit scaled unknown unitary process matrices, $U_{k}^{[N]} = \epsilon_{k}U^{[N]}$ onto $N+1$ qubit Hermitian density matrices $\rho_{U_{k}}^{[N+1]}$ (see Eq.~\eqref{Eq.16}). This is achieved by iteratively applying the circuit $k$ times for different rotations $\mathcal{G}^{[N]}_{k}$. An explicit form of the scaling factors $\epsilon_{k}$ is given in Eq.~\eqref{Eq.17}. In general, if the unitary process $U^{[N]}$ is completely unknown, i.e., no prior partial information is available, the circuit may be repeated up to $4^{N}$ times, with $\mathcal{G}^{[N]}_{k}$ encompassing all elements of the Pauli operator basis, as discussed in Sec.~\ref{Sec.:modified circuit}. However, if partial information about the process $U$ is given, it can be used to design an appropriate rotation $\mathcal{G}$, thereby reducing the number of circuit repetitions, $k$. \\

\begin{figure}[t]
	\centering
	\adjustbox{max width=\textwidth}{\includegraphics[scale=0.88]{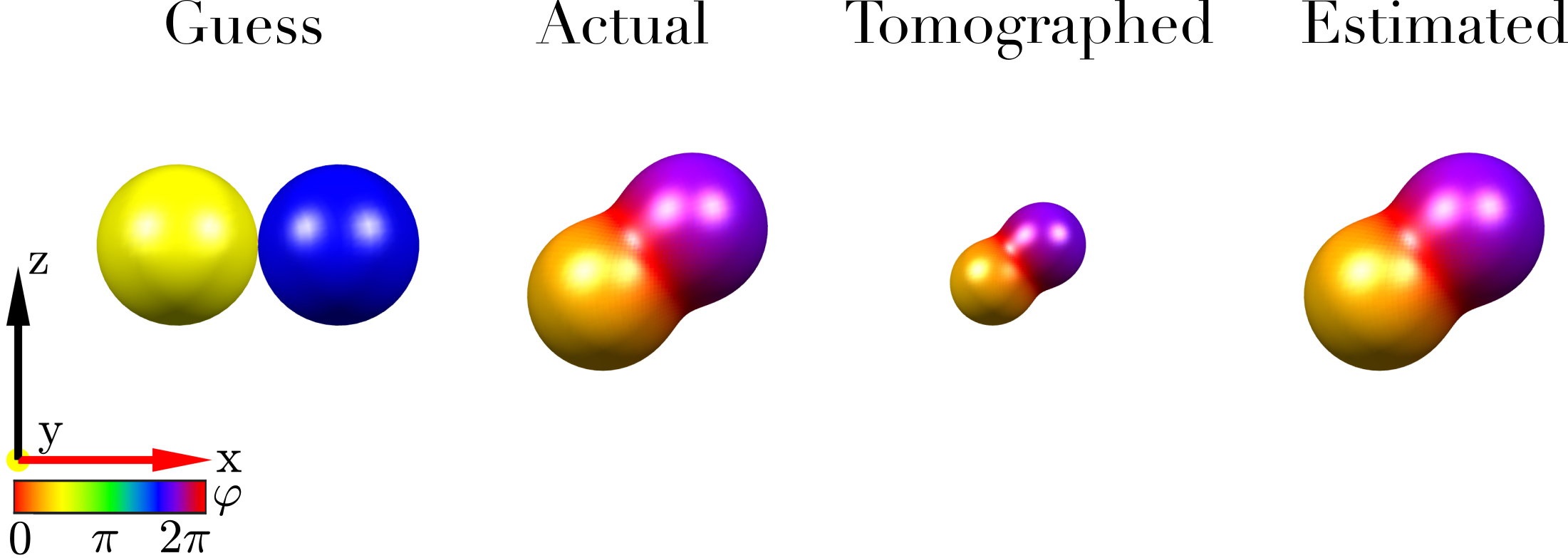}}
	\caption{Example of reconstructing unknown actual processes using the adapted approach. The first column shows the droplets corresponding to the presumed (guessed) process $U_{\text{gue}} = \text{NOT}$. The second column illustrates the simulated droplet corresponding to the actual experimental process $U_{a}$, showcasing variations in rotation axis and angle (see main text) compared to the $U_{\text{gue}}$. The third column presents the tomographed droplet, and the fourth column displays the reconstructed actual droplet estimated from the tomographed droplet.}
	\label{fig:16}
\end{figure}

\begin{figure}[h]
	\centering
	\adjustbox{max width=\textwidth}{\includegraphics[scale=0.8]{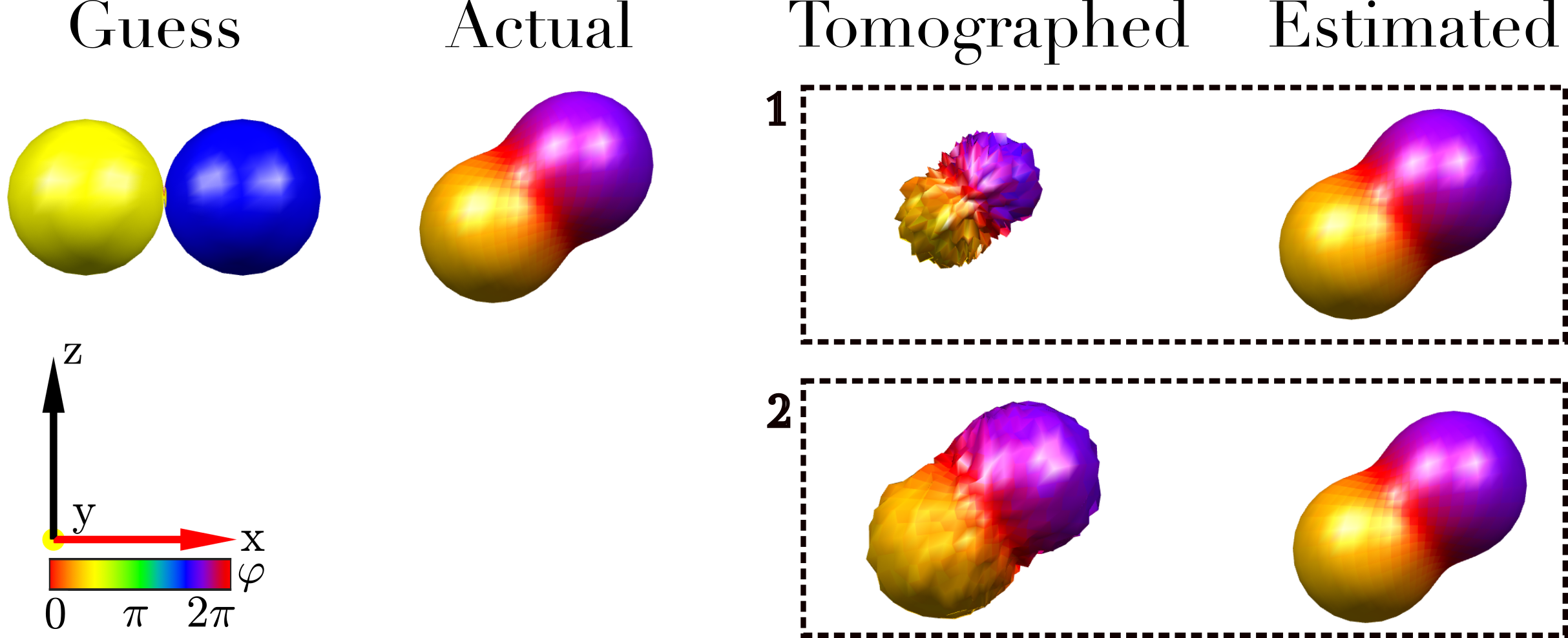}}
	\caption{Demonstration of the reconstruction of the actual unknown process from noisy tomographed droplets using the iterative adapted approach. The first column displays the droplet corresponding to the guessed process, $U_{\text{gue}} = \text{NOT}$ gate. The second column illustrates the droplet corresponding to the simulated actual experimental process $U_{a}$. The third column presents the tomographed droplets for iterations \textbf{1} and \textbf{2}, along with the estimated droplets from the tomographed data.}
	\label{fig:17}
\end{figure}

For example, for a single-qubit ($N=1$) system, presented in Sec.~\ref{Sec.:WQPT_Experimental}, the number of circuit repetitions is $k = 4$, with rotations $\mathcal{G}^{[1]}_{1} = \sigma_{x}$, $\mathcal{G}^{[1]}_{2} = \sigma_{y}$, $\mathcal{G}^{[1]}_{3} = \sigma_{z}$, and $\mathcal{G}^{[1]}_{4} = \mathds{1}$, which can be reduced to a few repetitions in this adapted approach if a reasonable guess of the process is available. If a desired target gate $U_{\text{tar}}$ is to be implemented, a reasonable guess for the actually realized gate $U_{a}$ by a circuit will be the $U_{\text{tar}}$. We illustrate the adaptive approach in Fig.~\ref{fig:16} using noise-less examples and an iterative version of the adaptive approach in Fig.~\ref{fig:17} using noisy data.\\

In Fig.~\ref{fig:16}, the first column shows the guessed process droplet $f_{\text{gue}}$ corresponding to the guessed process, which in this case is a $U_{\text{gue}} = \text{X} \;\text{(NOT)}$ gate (with rotation angle $\gamma = \pi$ and rotation axis $\vec{n} = (1,0,0)$; see Eq.~\eqref{Eq.13}). The second column displays the actual process droplet $f_{a}$ corresponding to the actual process $U_{a}$ realized experimentally. The third column exhibits the tomographed scaled process droplet $\hat{f}$ corresponding to the scaled process matrix $\hat{U} = \epsilon U_{a}$, with 
\begin{equation}
	\label{Eq.35}
	\epsilon = \frac{1}{2} \operatorname{tr}((U_{a})^{\dagger}U_{\text{gue}}).
\end{equation}
In this adaptive approach, the circuits are performed only once with $\mathcal{G} = U_{\text{gue}}$. Finally, the last column shows the estimated droplet ($f_{\text{est}}$) after reconstruction, which involves estimating the process matrix of a tomographed droplet $\hat{f}$ and normalizing it.\\ 
\\
In the specific example presented in Fig.~\ref{fig:16}, the actual process $U_{a}$ has a different rotation axis ($\vec{n} = (0.8,0,0.6)$) as well as a different rotation angle ($\gamma = 0.5 \pi$) compared to the guessed process. This discrepancy is evident by the orientation and the shape of the actual droplet $f_{a}$ shown in the second column compared to the guess droplet $f_{\text{gue}}$ shown in the first column. Here, the tomographed droplet shown in the third column is scaled by a factor of $\epsilon = 0.5657$ (computed using Eq.~\eqref{Eq.35}) compared to the actual droplet $f_{a}$ in the second column. The fourth column shows the estimated droplet $f_{\text{est}}$, which reconstructs the actual unknown process droplet $f_{a}$ precisely (c.f. droplets of the second and the fourth column).\\

For example, Fig.~\ref{fig:16} showcases the efficacy of the adapted approach to estimate the unknown process $U_{a}$ from the tomographed droplet, utilizing only a single repetition of the tomography circuits with $\mathcal{G} = U_{\text{gue}}$. In many cases, even in the presence of noise, one iteration may be enough to obtain a reasonable estimate of the actual process droplet $U_{a}$ from the tomographed droplet $\hat{f}$, provided the signal-to-noise ratio is sufficiently high. For example, Fig.~\ref{fig:17} shows the case of Fig.~\ref{fig:16} with shot noise, and in this case, the estimated process droplet $\hat{f}_{\text{est}}^{[1]}$ is obtained with a fidelity of $0.9864$ (see top panel of Fig.~\ref{fig:17}).\\
\\

If this reconstructed fidelity is not good enough for a specific application, the estimate can be further refined using an iterative approach. Here, we employ a second iteration \textbf{2} (see bottom panel of Fig.~\ref{fig:17}), where the tomographed droplet $\hat{f}^{[2]}$ corresponding to the scaled process $\hat{U}_{2} = \epsilon_{2}U_{a}$ is obtained by choosing the estimated process droplet $\hat{f}_{\text{est}}^{[1]}$ from iteration \textbf{1} as the $\mathcal{G}$ gate, where the resulting scaling factor is $\epsilon_{2} = 0.9864$ (see Eq.~\eqref{Eq.35}). This results in a tomographed droplet with an improved signal-to-noise ratio from which the process droplet $\hat{f}_{\text{est}}^{[2]}$  is estimated with a fidelity of 0.9981.\\


Such an adaptive approach, in general, is experimentally less resource intensive in comparison to the general approach presented in Fig.~\ref{fig:cU_unknownWR}, particularly in terms of circuit repetitions $k$. Despite this efficiency, it still provides sufficient information about the actual unknown process, as demonstrated by single-qubit examples in Fig.~\ref{fig:16} and Fig.~\ref{fig:17}. The adaptive approach can also be extended to multi-qubit unknown gates, where prior partial information, i.e., guess processes, can be leveraged to design the $\mathcal{G}$ gate to efficiently estimate the actual process $U_{a}$ while requiring fewer repetitions $k$. It is important to note that the initial guess $U_{\text{gue}}$ in no way biases the outcome of the estimated process $U_{\text{est}}$, as illustrated in Fig.~\ref{fig:16} and Fig.~\ref{fig:17}. The accuracy of this initial guess only affects the scaling factors (and hence the signal to noise) of the experimentally tomographed droplet of the first iteration.\\

\section{Generalized reconstruction algorithm}
\label{Sec.:Reconstuction_extentsion}
The reconstruction algorithm is designed to rebuild an unknown process matrix from experimentally tomographed droplet functions $f_{k}$. The algorithm outlined in Sec.~\ref{Sec.:Reconstruction algo} relies on quaternions, where the initial estimates of the quaternion components are derived from the correlation matrix described in the same section and are further refined using the optimization routine discussed in Sec.~\ref{Sec.:Reconstruction algo optimization}. However, this quaternion-based approach is limited to a single qubit system. Here, we introduce a method that is scalable to multiple qubits.\\

A general $N$-qubit unitary process matrix can be expressed in terms of Pauli operators as:
\begin{equation}
    \label{Eq.36}
    U^{[N]} = \sum_{k=1}^{4^{N}} c_{k}\sigma^{[N]}_{k},
\end{equation}
where $c_{k}$ are complex coefficients and $\sigma_{k}$ are the Pauli operators. For instance, for $N=1$, $k$ ranges from 1 to 4, as $\sigma^{[1]}_{k} \in \{\sigma_{x},\sigma_{x},\sigma_{x},\mathds{1}\}$, i.e., $\sigma^{[1]}_{1} = \sigma_{x}$, $\sigma^{[1]}_{2} = \sigma_{y}$, $\sigma^{[1]}_{3} = \sigma_{z}$, and $\sigma^{[1]}_{4} = \mathds{1}$, leading to
\begin{equation}
    \label{Eq.37}
    U^{[1]} = c_{1}\sigma_{x}+c_{2}\sigma_{y}+c_{3}\sigma_{z}+c_{4}\mathds{1}.
\end{equation}
Similarly, for a system with two qubits ($N=2$), $k$ spans from 1 to 16, as $\sigma^{[2]}_{k}\in\{\sigma_{x}\otimes\mathds{1},\dots,\mathds{1}\otimes\sigma_{x},\dots,\sigma_{z}\otimes\sigma_{z},\mathds{1}\otimes\mathds{1}\}$. The scaling factors $\epsilon_{k}$ (see Eq.~\eqref{Eq.17}) resulting from the mapping presented in Sec.~\ref{Sec.:modified circuit} in terms of complex coefficients $c_{1}$, $c_{2}$, $c_{3}$, and $c_{4}$ are provided in Table~\ref{Tab:scaling_generalized}. 
\begin{table}[h]
\centering
\caption{The scaling factors ($\epsilon_{k}$) for a single-qubit system ($N=1$) corresponding to different controlled rotations with $\mathcal{G}^{[1]}_{k}\in\{\sigma_{x},\sigma_{y},\sigma_{z},\mathds{1}\}$ and the resulting scaled process matrices $U^{[1]}_{k}$. The second column presents the scaling factor in terms of matrix elements $u_{11}$, $u_{12}$, $u_{21}$, and $u_{22}$, while the third column expresses them in terms of complex coefficients $c_{1}$, $c_{2}$, $c_{3}$, and $c_{4}$.} 
	\begin{tabular}{p{0.8cm} p{2.2cm} p{1.2cm} p{2cm}} 
		\toprule
		$\mathcal{G}^{[1]}_{k}$ & $\epsilon_{k}$ & $\epsilon_{k}$ & $U^{[1]}_{k} = \epsilon_{k}U^{[1]}$\\[0.5ex]
		\hline
		$\sigma_{x}$ & $\frac{1}{2} (u_{12}^{*}+u_{21}^{*})$ & $c_{1}^{*}$ & $c_{1}^{*}U^{[1]}$\\[0.7ex]
		
		$\sigma_{y}$ & $\frac{i}{2} (u_{21}^{*}-u_{12}^{*})$ & $c_{2}^{*}$ & $c_{2}^{*}U^{[1]}$\\[0.7ex]
		
		$\sigma_{z}$ & $\frac{1}{2} (u_{11}^{*}-u_{22}^{*})$ & $c_{3}^{*}$ & $c_{3}^{*}U^{[1]}$\\[0.7ex]
		
		$\mathds{1}$ & $\frac{1}{2} (u_{11}^{*}+u_{22}^{*})$ & $c_{4}^{*}$ & $c_{4}^{*}U^{[1]}$\\[0.5ex]
		\hline
		\hline
	\end{tabular}
	\label{Tab:scaling_generalized}
\end{table}\\

Now, for a single-qubit system ($N=1$), the following cost function can be used to estimate the unknown process matrix:
\begin{equation}
    \label{Eq.38}
    \begin{aligned}
        J_{1} = {} & {||U^{[1]}_{1} - c_{1,i}^{*}U^{[1]}_{\text{est},i} ||}^{2}+       
            {||{U}^{[1]}_{2} - c_{2,i}^{*}U^{[1]}_{\text{est},i}||}^{2}+\\
          & {||{U}^{[1]}_{3} - c_{3,i}^{*}U^{[1]}_{\text{est},i}||}^{2} +            
            {||{U}^{[1]}_{4} - c_{4,i}^{*}U^{[1]}_{\text{est},i} ||}^{2},
    \end{aligned}
\end{equation}
where ${U}_{1}$, ${U}_{2}$, ${U}_{3}$, and ${U}_{4}$ are scaled matrices, representing the matrix forms of the experimentally obtained scaled droplet functions ${f}_{1}$, ${f}_{2}$, ${f}_{3}$, and ${f}_{4}$, respectively. The estimated process matrix $U_{\text{est},i}$ for iteration $i$ is composed of coefficients $c_{1,i}$, $c_{2,i}$, $c_{3,i}$, and $c_{4,i}$, as defined in Eq.~\eqref{Eq.37}. The cost defined in Eq.~\eqref{Eq.33} is equivalent to that defined in Eq.~\eqref{Eq.38}, and the correlation matrix can still be used to derive the zero-order initial estimates for coefficients $c_{1}$, $c_{2}$, $c_{3}$, and $c_{4}$ for the optimization. The cost function defined in Eq.~\eqref{Eq.38} can be generalized to the system of $N$ qubits, as follows:
\begin{equation}
    \label{Eq.39}
        J_{\text{gen}} = \sum_{k=1}^{4^N} {||U^{[N]}_{k} - c_{k,i}^{*}U^{[N]}_{\text{est},i} ||}^{2}. 
\end{equation}
Here, $U_{k}$ are the scaled process matrices corresponding to the experimentally obtained droplet functions $f_{k}$ for different controlled rotations $c\mathcal{G}_{k}$, as discussed in Sec.~\ref{Sec.:WQPT theory}. Designing a correlation matrix to derive the initial values of coefficients $c_{k}$ for an $N$ qubit system might be too cumbersome. Therefore, a reasonable guess can also be used to minimize the cost function defined above.\\

The generalized cost function defined in Eq.~\eqref{Eq.39} can be directly used to reconstruct an $N$ qubit unitary process matrix from the scaled process droplets, providing a generalization to the reconstruction algorithm presented in Sec.~\ref{Sec.:Reconstruction theory}.  
\section{Conclusion and Discussion}
\label{Sec.Discussion}
In this work, we presented a general approach for the tomography of unknown processes in the context of Wigner representations. This work extends the previously developed Wigner tomography methods~\cite{leiner2017wigner,leiner2018wigner,Devra_WQST} to unknown quantum processes by addressing the challenge of mapping an unknown process matrix onto a larger density matrix. We introduce a novel method to map scaled versions of an unknown unitary process matrix onto density matrices and also highlight the limitations (blind spots) of the naive circuit in providing a complete mapping using state vector and density matrix formalism (see Sec.~\ref{Sec.: Mapping}).\\

The tomography approach presented here consists of two parts: experimental acquisition and classical post-processing of data. The experimental part provides partial information (i.e., droplets corresponding to scaled processes) about the experimental process. In the post-processing stage, the information contained in these experimental scaled process droplets is combined to reconstruct the unknown process. We introduced a reconstruction algorithm for a single-qubit system and demonstrated its potential to rebuild the experimental process matrix with a high signal-to-noise ratio from noisy droplets (see Sec.~\ref{Sec.:Reconstruction theory}). Additionally, we showed how the reconstruction algorithm can be integrated with an optimization routine, highlighting the benefits gained from this combined approach. The reconstruction algorithm can also be extended to an $N$ qubit system, as discussed in Sec.~\ref{Sec.:Reconstuction_extentsion}.\\

We experimentally demonstrated the Wigner tomography approach for unknown quantum processes using the superconducting qubit-based IBM quantum devices. For a single-qubit system, our tomography approach requires two three-qubit CSWAP gates (see Sec.~\ref{fig:10}), resulting in a relatively deep circuit. During the experimental implementation of these circuits on IBM devices, we detected a significant phase shift in qubit $q_0$. We quantified the extra unwanted rotation of the qubit $q_{0}$ using the calibration circuits discussed in Sec.~\ref{Sec.:Calibration} and compensated for it in the main tomography experiments (see results in Sec.~\ref{Sec: Experimental results}). The compilation of the CSWAP gates in the hardware-native gate set leads to a long gate duration, during which the effect of small errors per gate accumulates, resulting in amplified errors. To reduce these errors, gates can be designed using noise-robust pulse-level control techniques~\cite{Niklas_cphase,KHANEJA2005296,Devra2018,Sugny_composite,Damme_robust}. Additionally, existing error-suppression techniques, such as dynamical decoupling~\cite{Lidar_DD,Suter_DD}, can be employed to suppress the errors accumulated during the idle time of the qubits. However, these methods generally require a good pulse-level control on the quantum devices~\cite{Lidar_DD}. The calibration circuit approach presented herein also functions as an effective error-suppression method for current noisy devices, offering a viable alternative to resource-intensive error mitigation methods.\\

The Wigner tomography approach for unknown processes fills a gap in the general theory work of Wigner tomography but is resource-intensive in comparison to the conventional tomography approaches~\cite{nielsen2002quantum} both in terms of the number of qubits and the number of experiments. In general, for Wigner tomography of an $N$ qubit unknown process, $2N+1$ qubits are required. The total number of experiments in terms of the number of controlled-$\mathcal{G}$ operations can be significantly cut down by employing an adaptive approach discussed in Sec.~\ref{Sec:Adaptive}. In the near term, for devices with a limited number of qubits, where a scalable error correction is not viable, the long experimental implementation time of the CSWAP gates used in this approach also results in a loss of signal-to-noise ratio of the tomographed droplets due to decoherence effects and experimental imperfections. \\

The unknown process tomography approach presented herein serves as a tool for analyzing a ``black box'' containing one or more unknown gates whose inner workings are hidden. Here, we consider three different scenarios:
\begin{itemize}
	\item In a \textbf{first} scenario, a quantum computer is considered that is capable of executing all required gates for the tomography with high fidelity. In this case, the black box can be plugged directly into the tomography circuits to perform the complete process tomography to extract full information about the black box.
	\item In a \textbf{second} scenario, consider a case where some partial information about the black box may be available. Here, the adaptive version of the tomography approach (see Sec.~\ref{Sec:Adaptive}) can be employed to extract full information about the black box, where such partial information can be leveraged to design the $\text{c}\mathcal{G}$ rotation to efficiently estimate the actual process with fewer experiments.
	\item In a \textbf{third} scenario, consider a case where the target process is fully known and where the tomography approach is used for a ``quality check'', i.e., to assess the experimental performance of a given target gate $U_\text{tar}$ on a particular quantum device assuming that the error is small. In this case, only {\it one} tomography experiment (with $\text{c}\mathcal{G}= \text{c}U_\text{tar}$, see Fig.~\ref{fig:10}) is sufficient to estimate the performance of the actual gate, i.e., experimental implementation of the target gate. 
\end{itemize}

\begin{figure}[h]
	\centering
	\adjustbox{max width=\textwidth}{\includegraphics[scale=0.72]{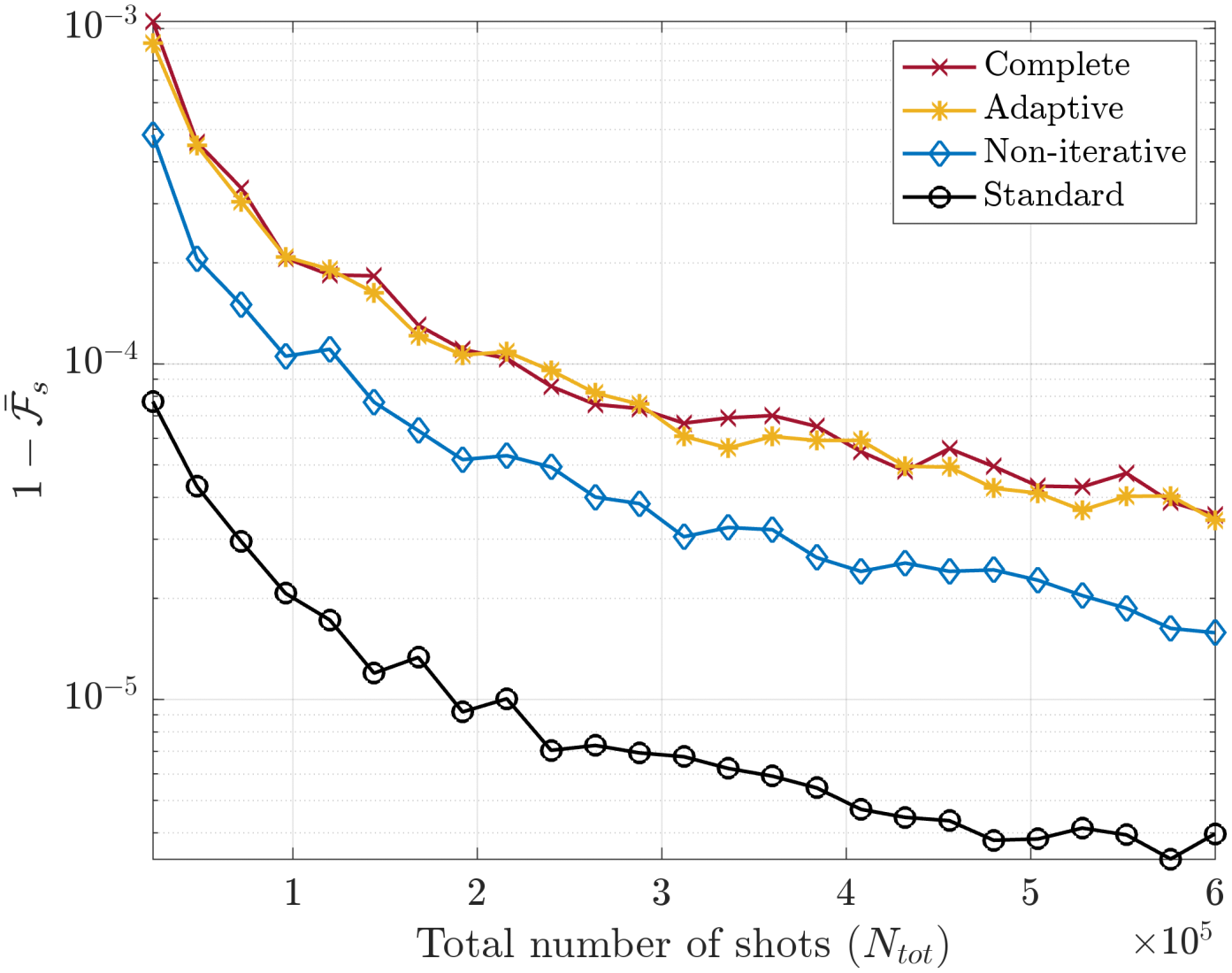}}
	\caption{Comparison of different versions of the Wigner tomography method (see text) and the standard process tomography approach. The corresponding plot with standard deviation is shown in Fig.~\ref{fig: G.1} of Appendix~\ref{Supp:comparison}.}
	\label{fig:18}
\end{figure}

To evaluate these approaches, we conducted a numerical study comparing different versions of the Wigner tomography method and the standard process tomography approach~\cite{nielsen2002quantum} for a single-qubit system. The results are presented in Fig.~\ref{fig:18}. The simulation considers noise arising only from a finite number of shots $N_s$ and shows the error in mean fidelity $\mathcal{\bar{F}}_{s}$ for reconstructing a random target unitary process matrix $U_\text{tar}$ using different tomography techniques under the same total number of shots $N_{tot}$.  Each data point represents the mean error across 100 distinct noise instances. \\

The total number of shots is defined as $N_{tot} = N_{p} \cdot N_{s}$, where $N_{p}$ is the number of experiments required for a given tomography method, and $N_{s}$ is the number of shots per experiment. For example, the standard tomography approach~\cite{nielsen2002quantum} requires $N_{p} = 12$, whereas complete Wigner tomography (first scenario) requires $N_{p} = 208$. The latter value is based on a Lebedev grid with $n = 26$, requiring $2$ circuits for process tomography (considering four circuits to prepare maximally mixed states with shots evenly distributed, see Fig.\ref{fig:10}), repeated $4$ times for different rotations $\mathcal{G}$. Differences in $N_p$ across methods are compensated by adjusting $N_s$ to maintain the same $N_{tot}$.  

For the adaptive approach (second scenario), Fig.~\ref{fig:18} shows the results after two iterations. In the first iteration, the rotation axis of $U_{\text{gue}}$ is assumed to be tilted by $30^{\circ}$ away from the rotation axis of the target operator $U_\text{tar}$ (whereas the rotation angles of $U_{\text{gue}}$ and $U_\text{tar}$ are assumed to be identical). The operator $\mathcal{G}$ is chosen to be identical to $U_{\text{gue}}$. The total number of shots, $N_{tot}$, was evenly split between two iterations, with $N_{tot}/2$ allocated to each. This ensures that the data points presented in Fig.~\ref{fig:18} effectively correspond to the same total number of shots across all approaches. \\

For the non-iterative approach (third scenario), only one experiment with rotation $\mathcal{G} = U_\text{tar}$ was performed. In this simulation, the rotation axis of $U_{a}$ is tilted away from the rotation axis of $U_\text{tar}$ only by a small angle of $2^{\circ}$ (while the rotation angle of the actual operator $U_{a}$ and of the target operator $U_\text{tar}$ is assumed to be identical). In the case of the standard process tomography approach, Qiskit and Forest-Benchmarking functions were used to reconstruct the channel matrix and to compute the unitary closest to a quantum process specified by a channel matrix~\cite{oi2003interference}. For all the approaches, the process fidelity was computed using the definition shown in Eq.~\eqref{Eq:D3}.\\

The results show that the non-iterative Wigner tomography method, where only a small error between the target operator $U_\text{tar}$ and the actual operator $U_\text{a}$ is assumed (third scenario), performs better than other Wigner tomography approaches while requiring one-fourth the number of experiments $N_{p}$ compared to the complete Wigner tomography approach (first scenario) for the same $N_{tot}$. However, the standard tomography approach achieves higher precision than Wigner tomography methods for the same $N_{tot}$. The plot in Fig.~\ref{fig:18} is replotted in Fig.~\ref{fig: G.2} of Appendix~\ref{Supp:comparison} with a reduction of the number of shots $N_s$ for the standard approach by a factor of 4 (purple dashed line) and by a factor of 8 (green dashed line).  This illustrates that the standard approach only requires between four and eight times less total number of shots to reach a comparable precision as the non-iterative Wigner tomography method (blue solid line). This difference possibely arises because, for a single-qubit process tomography, Wigner tomography requires measuring two-qubit expectation values (see Eq.~\eqref{Eq.23}), while the standard approach only needs one-qubit expectation values~\cite{nielsen2002quantum}. 
           
\section{Code availability}
\label{Sec.Code}
The Matlab and Python codes used for the tomography approach are available in the GitHub repository~\cite{WQPT_unknwon}. This repository also contains the code for plotting the spherical droplets from the Lebedev grid and for performing standard process tomography. The Python code uses Qiskit~\cite{gadi_aleksandrowicz_2019_2562111}, an open-source development for simulations and experiments that is directly usable on IBM quantum computers and can be modified for other quantum computing platforms.
   
\acknowledgments
This project has received funding from the European Union’s Horizon 2020 research and innovation program under the Marie-Sklodowska-Curie grant agreement No 765267 (QuSCo). We would like to thank Niklas J.~Glaser and Santana Lujan for the discussions regarding the Qiskit codes and Max Werninghaus for his useful comments on the experimental results. This research was a part of the Bavarian excellence network \textsc{enb}
via the International PhD Programme of Excellence
\textit{Exploring Quantum Matter} (\textsc{exqm}) as well as the \textit{Munich Quantum Valley} of the Bavarian
State Government with funds from Hightech Agenda \textit{Bayern Plus}. We acknowledge Sabine Tornow (CODE, UniBW) for providing access to the IBMQ hardware. We acknowledge the use of IBM Quantum services for this work. The views expressed are those of the authors and do not reflect the official policy or position of IBM or the IBM Quantum team.

\onecolumngrid

\clearpage
\appendix
\section{Visualization of operators}
\label{Supp:visualization}
Visualizing a quantum system is an important approach to understanding the underlying physics and dynamics. A general approach for visualization of quantum operators such as density operators, time-evolution operators (i.e., quantum processes), Hamiltonians, etc., was introduced by Garon et al.~\cite{DROPS_main} for coupled spin systems. The approach was later extended for up to six coupled spins in a paper by Leiner et al.~\cite{Leiner_2020}. This is a general visualization approach and is not just limited to the system consisting of two-level systems (i.e., qubits)~\cite{Leiner_2020}. Here, we briefly summarize the approach presented in Ref.~\cite{DROPS_main}.\\

The visualization approach is based on a bijective mapping of a quantum operator $A$ onto a set of spherical operators $f$:
\begin{equation}
	\label{Eq.S_1}
	A = \sum_{\ell\in L} A^{(\ell)} \longleftrightarrow  \bigcup_{\ell\in L} \{f^{(\ell)}\}.
\end{equation}
Where, set $L$ contains labels $\ell$ describing \textit{linearity}. A quantum operator $A^{(\ell)}$ can be decomposed into a linear combination of spherical tensor operators $T_{jm}^{(\ell)}$ with rank $j$, and order $m$~\cite{Racah,wigner1931gruppentheorie,biedenharn1981angular,silver2013irreducible}. Similarly, a spherical function $f^{(\ell)}$ can be decomposed into a linear combination of spherical harmonics $Y_{jm}$:
\begin{equation}
	\label{Eq.S_2}
	A^{(\ell)} = \sum_{j \in J(\ell)}\sum_{m=-j}^{j} c_{jm}^{(\ell)} T^{(\ell)}_{jm} \longleftrightarrow f^{(\ell)} = \sum_{j \in J(\ell)}\sum_{m=-j}^{j} c_{jm}^{(\ell)} Y_{jm}.
\end{equation}
This correspondence between irreducible spherical tensor operators ($T_{jm}^{(\ell)}$) and spherical harmonics $Y_{jm}$ is the foundation of the DROPS representation. A basis transformation for axial (rank $j$ and order $m=0$) spherical tensor operators $T_{j0}^{(\ell)}$ into Pauli operators is provided in Appendix~\ref{Supp:Axial tensor}. A general basis transformation from spherical tensor operators $T_{jm}$ to Pauli operators is given in Refs.~\cite{DROPS_main,Devra_WQST}.\\

Using this mapping, we move from matrices to shapes. An operator $A^{(\ell)}$ is decomposed into spherical tensor operators $T_{jm}^{(\ell)}$, mapped to complex spherical harmonics $Y_{jm}(\theta,\phi) = r(\theta,\phi)\text{exp}(i\varphi(\theta,\phi))$, and plotted as spherical \textit{droplets} $f^{(\ell)}(\theta,\phi)$. For example, the decomposition of each quantum gate plotted in Fig.~\ref{fig:intro_example} into Pauli ($\sigma$) and spherical tensor ($T_{jm}^{(\ell)}$) basis reads as follows:
\begin{equation}
	\label{Eq.S_3}
	\text{H} = \sqrt{\frac{1}{2}}
	\begin{pmatrix} 
		i & i  \\[1ex]
		i  & -i 
	\end{pmatrix} = i\sqrt{\frac{1}{2}}(\sigma_{x}+\sigma_{z}) =  i\sqrt{\frac{1}{2}}(T_{1-1}^{(1)}-T_{11}^{(1)}+\sqrt{2}T_{10}^{(1)}),
\end{equation}
\begin{equation}
	\label{Eq.S_4}
	\text{S} = \sqrt{\frac{1}{2}}
	\begin{pmatrix} 
		1-i & 0  \\[1ex]
		0  & 1+i 
	\end{pmatrix} = \sqrt{\frac{1}{2}}(\mathds{1}-i\sigma_{z}) = (T_{00}^{(\emptyset)}-iT_{10}^{(1)}),
\end{equation}
\begin{equation}
	\label{Eq.S_5}
	\sqrt{\text{NOT}} = \sqrt{\frac{1}{2}}
	\begin{pmatrix} 
		1 & -i  \\[1ex]
		-i  & 1 
	\end{pmatrix} = \sqrt{\frac{1}{2}}(\mathds{1}-i\sigma_{x}) = \sqrt{\frac{1}{2}}(\sqrt{2}T_{00}^{(\emptyset)}-iT_{1-1}^{(1)}+iT_{11}^{(1)}).
\end{equation}
Therefore, for a single qubit, the droplet components are $f_{0}^{(\emptyset)} \longleftrightarrow T_{00}^{(\emptyset)}$ and $f_{1}^{(1)} \longleftrightarrow \sum_{m=-j}^{j} T_{1m}^{(1)}$. For instance, for the Hadamard (H) gate, the component $T_{00}^{(\emptyset)}$ is zero, therefore $f_{0}^{(\emptyset)}$ is zero (see Fig.~\ref{fig:intro_example}).   

\section{Axial tensor operators}
\label{Supp:Axial tensor}
Axial tensor operators ($T_{j0}$) with rank $j$, and order $m=0$ are expressed through Pauli operators for single ($N=1$) and two ($N=2$) qubit systems in Table~\ref{Tab:AxialTensors}.
\begin{table}[h]
	\centering
	\begin{tabular}{p{0.3cm} p{0.4cm} p{0.4cm} p{6cm}}
		\hline
		\hline
		$N$&\textrm{$\ell$}&\textrm{$j$}&\textrm{$T_{j0}^{(\ell)}$}\\[0.8ex]
		\hline
		1 & $\emptyset$ & 0 & $T_{00}^{(\emptyset)} = \frac{1}{\sqrt{2}}(\mathds{1})$\\ [0.7ex]
		
		& 1 & 1 & $T_{10}^{(1)} = \frac{1}{\sqrt{2}}(\sigma_{z})$\\[0.7ex]
		
		2 & $\emptyset$ & 0 & $T_{00}^{(\emptyset)} = \frac{1}{2}\mathds{1}$\\[0.7ex]
		
		& 1 & 1 & $T_{10}^{(1)} = \frac{1}{2}(\sigma_{1z})$\\[0.7ex]
		
		& 2 & 1 & $T_{10}^{(2)} = \frac{1}{2}(\sigma_{2z})$\\[0.7ex]
		
		& 12 & 0 & $T_{00}^{(12)} = \frac{1}{2\sqrt{3}}(\sigma_{1x}\sigma_{2x}+\sigma_{1y}\sigma_{2y}+\sigma_{1z}\sigma_{2z})$\\[0.7ex]
		
		& 12 & 1 & $T_{10}^{(12)}  = \frac{1}{2\sqrt{2}}(\sigma_{1x}\sigma_{2y}-\sigma_{1y}\sigma_{2x})$\\[0.7ex]
		
		& 12 & 2 & $T_{20}^{(12)} =\frac{-1}{2\sqrt{6}}(\sigma_{1x}\sigma_{2x}+\sigma_{1y}\sigma_{2y}-2\sigma_{1z}\sigma_{2z})$\\
		\hline
		\hline
	\end{tabular}
	\caption{Axial tensor operator $T_{j0}^{(\ell)}$ for one ($N=1$) and two ($N=2$) qubit systems, see Refs.~\cite{DROPS_main,Devra_WQST}.}
	\label{Tab:AxialTensors}
\end{table}
\newpage
\section{Density matrix analysis}
\label{Supp:density_scaling}
This section presents the density matrix analysis for the circuit presented in Fig.~\ref{fig:Classic_control_circuit} and Fig.~\ref{fig:cU_unknownWR}. The circuit presented in Fig.~\ref{fig:Classic_control_circuit} yields the same result as the circuit in Fig.~\ref{fig:cU_unknownWR} when $\mathcal{G}_{k} = \mathds{1}$. The density matrix $\rho_{0} = |\psi_{0}\rangle\langle\psi_{0}|$ corresponding to the state $|\psi_{0}\rangle$ in Table~\ref{Tab:Cases_quantum} is:
\begin{equation}
	\label{Eq.S1}
	\rho_{0}^{[2N+1]} = \frac{(|0\rangle+|1\rangle)(\langle0|+\langle1|)}{2}\otimes\rho_{s}^{[N]}\otimes\rho_{a}^{[N]} = 
	\frac{1}{2}
	\begin{pmatrix} 
		\rho_{s}^{[N]}\otimes\rho_{a}^{[N]} 
		& 
		\rho_{s}^{[N]}\otimes\rho_{a}^{[N]}  \\[1ex]
		\rho_{s}^{[N]}\otimes\rho_{a}^{[N]}  & \rho_{s}^{[N]}\otimes\rho_{a}^{[N]} 
	\end{pmatrix},	
\end{equation}
where $\rho_{s} = |\psi_{s}\rangle\langle\psi_{s}|$ and $\rho_{a} = |\psi_{a}\rangle\langle\psi_{a}|$) represents the initial density matrix of the system $q_{1},\dots,q_{N}$ and the ancilla $q_{1}^{a},\dots,q_{N}^{a}$ qubits, respectively. The density matrix $\rho_{1} = |\psi_{1}\rangle\langle\psi_{1}|$, i.e., after the first controlled-swap gate of the circuit presented in Fig.~\ref{fig:cU_unknownWR} is:
\begin{equation}
	\label{Eq.S2}
	\begin{split}
			\rho_{1}^{[2N+1]} & = \begin{pmatrix} 
							\mathds{1}^{[2N]} & 0^{[2N]}  \\[1ex]
							0^{[2N]}  & \text{swap}^{[2N]}
						\end{pmatrix} \rho_{0}^{[2N+1]} 
				\begin{pmatrix} 
					\mathds{1}^{[2N]} & 0^{[2N]}  \\[1ex]
					0^{[2N]}  & \text{swap}^{[2N]}
				\end{pmatrix}^{\dagger}\\
		 & = \frac{1}{2}
		 \begin{pmatrix} 
			\rho_{s}^{[N]}\otimes\rho_{a}^{[N]} 
			& 
			(\rho_{s}^{[N]}\otimes\rho_{a}^{[N]})\text{swap}^{[2N]}  \\[1ex]
			\text{swap}^{[2N]}(\rho_{s}^{[N]}\otimes\rho_{a}^{[N]})  & \rho_{a}^{[N]}\otimes\rho_{s}^{[N]}
		\end{pmatrix}.
	\end{split} 
\end{equation}
Here, the controlled-swap gate is written in a block-diagonal matrix form. The matrix form of the swap gate for $N=1$ is:
\begin{equation}
	\label{Eq.S3}
	\text{swap}^{[2]} = 
        \begin{pmatrix} 
		1 & 0 & 0 & 0  \\[1ex]
            0 & 0 & 1 & 0  \\[1ex] 
            0 & 1 & 0 & 0  \\[1ex] 
		0 & 0 & 0 & 1 	 
	\end{pmatrix}.
\end{equation} 
After the operation $\mathds{1}^{[N+1]}\otimes U^{[N]}$ (second gate of the circuit in Fig.~\ref{fig:cU_unknownWR}) one obtains
\begin{equation}
	\label{Eq.S4}
       \begin{split}
		\rho_{2}^{[2N+1]} & = 
		\begin{pmatrix} 
			\mathds{1}^{[N]}\otimes U^{[N]} & 0^{[2N]}  \\[1ex]
			0^{[2N]}  & \mathds{1}^{[N]}\otimes U^{[N]}
		\end{pmatrix} \rho_{1}^{[2N+1]}
		\begin{pmatrix} 
			\mathds{1}^{[N]}\otimes U^{[N]} & 0^{[2N]}  \\[1ex]
			0^{[2N]}  & \mathds{1}^{[N]}\otimes U^{[N]}
		\end{pmatrix}^{\dagger}\\
		& =  \frac{1}{2}
		{\begin{pmatrix} 
			\rho_{s}^{[N]}\otimes U^{[N]}\rho_{a}^{[N]}(U^{[N]})^{\dagger} 
			& 
			(\rho_{s}^{[N]}\otimes U^{[N]}\rho_{a}^{[N]})\text{swap}^{[2N]}(\mathds{1}^{[N]}\otimes (U^{[N]})^{\dagger})  \\[1ex]
			(\mathds{1}^{[N]}\otimes U^{[N]})\text{swap}^{[2N]}(\rho_{s}^{[N]}\otimes\rho_{a}^{[N]})(U^{[N]})^{\dagger})  & \rho_{a}^{[N]}\otimes U^{[N]}\rho_{s}^{[N]}(U^{[N]})^{\dagger}
		\end{pmatrix}}.
	\end{split}
\end{equation}
The density matrix after applying the second controlled-swap gate (overall third gate of the circuit in Fig.~\ref{fig:cU_unknownWR}) is
\begin{equation}
	\label{Eq.S5}
	\rho_{3}^{[2N+1]} = 
	\frac{1}{2}
	\begin{pmatrix} 
		\rho_{s}^{[N]}\otimes U^{[N]}\rho_{a}^{[N]}(U^{[N]})^{\dagger} 
		& 
		\rho_{s}^{[N]}(U^{[N]})^{\dagger}\otimes U^{[N]}\rho_{a}^{[N]}  \\[1ex]
		U^{[N]}\rho_{s}^{[N]}\otimes\rho_{a}^{[N]}(U^{[N]})^{\dagger}  & U^{[N]}\rho_{s}^{[N]}(U^{[N]})^{\dagger}\otimes\rho_{a}^{[N]} 
	\end{pmatrix}.	
\end{equation}
After the controlled rotation ($c\mathcal{G}_{k}$) (the last gate of circuit in Fig.~\ref{fig:cU_unknownWR}), the resulting density matrix is
\begin{equation}
	\label{Eq.S6}
	\begin{split}
		\rho_{4}^{[2N+1]} & = 
		\begin{pmatrix} 
			\mathds{1}^{[2N]} & 0^{[2N]}  \\[1ex]
			0^{[2N]}  &\mathds{1}^{[N]}\otimes \mathcal G_k^{[N]}
		\end{pmatrix} \rho_{3}^{[2N+1]}
		\begin{pmatrix} 
			\mathds{1}^{[2N]} & 0^{[2N]}  \\[1ex]
			0^{[2N]}  & \mathds{1}^{[N]}\otimes \mathcal G_k^{[N]}
		\end{pmatrix}^{\dagger}\\
		& = \frac{1}{2}
		\begin{pmatrix} 
			\rho_{s}^{[N]}\otimes U^{[N]}\rho_{a}^{[N]}(U^{[N]})^{\dagger} 
		& 
		\rho_{s}^{[N]}(U^{[N]})^{\dagger}\otimes U^{[N]}\rho_{a}^{[N]}(\mathcal{G}_{k}^{[N]})^{\dagger}  \\[1ex]
		U^{[N]}\rho_{s}^{[N]}\otimes\mathcal{G}_{k}^{[N]}\rho_{a}^{[N]}(U^{[N]})^{\dagger}  & U^{[N]}\rho_{s}^{[N]}(U^{[N]})^{\dagger}\otimes\mathcal{G}_{k}^{[N]}\rho_{a}^{[N]}(\mathcal{G}_{k}^{[N]})^{\dagger} 
		\end{pmatrix}.
	\end{split} 
\end{equation}
After tracing out $N$ ancilla qubits $q_{1}^{a},\dots,q_{N}^{a}$ (indicated by the red barrier in the circuit depicted in Fig.~\ref{fig:cU_unknownWR}), the density matrix is
\begin{equation}
	\label{Eq.S7}
	\rho_{5}^{[N+1]} = 
	\frac{1}{2}
	\begin{pmatrix} 
		\rho_{s}^{[N]} & 
		\rho_{s}^{[N]}(U^{[N]})^{\dagger}\operatorname{tr}(U^{[N]}\rho_{a}^{[N]}(\mathcal{G}_{k}^{[N]})^{\dagger}) \\[1ex]
		U^{[N]}\rho_{s}^{[N]}\operatorname{tr}(\mathcal{G}_{k}^{[N]}\rho_{a}^{[N]}(U^{[N]})^{\dagger})  & U^{[N]}\rho_{s}^{[N]}(U^{[N]})^{\dagger} 
	\end{pmatrix}.	
\end{equation}
Here $\operatorname{tr}(\mathcal{G}_{k}^{[N]}\rho_{a}^{[N]}(U^{[N]})^{\dagger})$ is the scaling factor. This factor depends on the unknown process $U^{[N]}$ and the rotations $\mathcal{G}_{k}$ with different values of $k$. \\

If initially the system $q_1,\dots,q_{N}$ and the ancilla qubits $q_{1}^{a},\dots,q_{N}^{a}$ are both in a maximally mixed state, i.e., $\rho_{s}=\rho_{a}=\frac{1}{2^N}\mathds{1}^{[N]}$, Eq.~\eqref{Eq.S7} simplifies to:
\begin{equation}
	\label{Eq.S8}
	\tilde{\rho}_{5}^{[N+1]} = \frac{1}{2^{(N+1)}}
	\begin{pmatrix} 
		\mathds{1}^{[N]} 
		& 
		\epsilon^{*}_{k}(U^{[N]})^{\dagger}  \\[1ex]
		\epsilon_{k}U^{[N]}  & \mathds{1}^{[N]} 
	\end{pmatrix},
\end{equation}
which can be rewritten as:
\begin{equation}
	\label{Eq.S9}
	\rho_{U_{k}}^{[N+1]} = 
	\frac{1}{2^{N+1}}
	\begin{pmatrix} 
		\mathds{1}^{[N]}  &  (U_{k}^{[N]})^{\dagger}  \\[1ex]
		U_{k}^{[N]}  & \mathds{1}^{[N]} 
	\end{pmatrix}, 
\end{equation}
where
\begin{equation}
	\label{Eq.S10}
	U_{k}^{[N]} = \epsilon_{k}U^{[N]},
\end{equation}
and the scaling factor
\begin{equation}
	\label{Eq.S11}
	\epsilon_{k} = \frac{1}{2^{N}} \langle U^{[N]}|\mathcal{G}_{k}^{[N]}\rangle.
\end{equation}
Here $\epsilon^{*}_{k}$ is the complex conjugate of $\epsilon_{k}$. Eq.~\eqref{Eq.S8} shows the mapping of an $N$ qubit unknown process matrix $U^{[N]}$ onto an $N+1$ qubit density matrix up to a scaling factor $\epsilon_{k}$. Note that the density operator $\tilde{\rho}_{4}^{[N+1]}$ in Eq.~\eqref{Eq.11} is identical to $\tilde{\rho}_{5}^{[N+1]}$ when $\mathcal{G}_{k}^{[N]} = \mathds{1}^{[N]}$. 
\section{Details of the reconstruction algorithm}
\label{Supp:reconstruction_details}
\subsection{Correlation matrix}
\label{Supp:correlation_matrix}
Given the experimental scaled droplet functions $\hat{f}_{1}$, $\hat{f}_{2}$, $\hat{f}_{3}$, and $\hat{f}_{4}$, the correlation matrix $M$ can be computed as follows:
\begin{equation}
	\label{Eq.4.17}
	M = 
	\begin{bmatrix} 
		 \langle \hat{f}_{1}|\hat{f}_{1}\rangle & \langle \hat{f}_{1}|\hat{f}_{2}\rangle  & \langle \hat{f}_{1}|\hat{f}_{3}\rangle & \langle \hat{f}_{1}|\hat{f}_{4}\rangle\\[1ex]
		 \langle \hat{f}_{2}|\hat{f}_{1}\rangle & \langle \hat{f}_{2}|\hat{f}_{2}\rangle  & \langle \hat{f}_{2}|\hat{f}_{3}\rangle & \langle \hat{f}_{2}|\hat{f}_{4}\rangle\\[1ex]
		  \langle \hat{f}_{3}|\hat{f}_{1}\rangle & \langle \hat{f}_{3}|\hat{f}_{2}\rangle  & \langle \hat{f}_{3}|\hat{f}_{3}\rangle & \langle \hat{f}_{3}|\hat{f}_{4}\rangle\\[1ex]
		   \langle \hat{f}_{4}|\hat{f}_{1}\rangle & \langle \hat{f}_{4}|\hat{f}_{2}\rangle  & \langle \hat{f}_{4}|\hat{f}_{3}\rangle & \langle \hat{f}_{4}|\hat{f}_{4}\rangle\\[1ex]
	\end{bmatrix}.
\end{equation}
Here, each element of the matrix can be computed using the discretized scalar product between two spherical droplet functions as described in Ref.~\cite{Devra_WQST}. The matrix $M$ provides correlations between different droplet functions, allowing for the zero-order (i.e., iteration $i=0$) estimation of the quaternion components. The absolute values of zero-order ($i=0$) estimates of quaternion components $A$, $B$, $C$, and $D$ can be determined from the diagonal elements of $M$:
\begin{equation}
	\label{Eq.4.18}
	\begin{aligned}
		\abs{A_{0}} = {} & \sqrt{M(1,1)},\\
		\abs{B_{0}} = {} & \sqrt{M(2,2)},\\
		\abs{C_{0}} = {} & \sqrt{M(3,3)},\\
		\abs{D_{0}} = {} & \sqrt{M(4,4)}.
	\end{aligned}
\end{equation}
The relative signs of the zero-order ($i=0$) quaternion components $A_{0}, B_{0}, C_{0}$, and $D_{0}$ can be determined based on the signs of the off-diagonal elements of the matrix $M$. For example, one possible case of $M$ could be:
\begin{equation*}
	M = 
	\begin{bmatrix} 
		 \color{red} \langle \hat{f}_{1}|\hat{f}_{1}\rangle & \color{red}\langle \hat{f}_{1}|\hat{f}_{2}\rangle  & \color{green}\langle \hat{f}_{1}|\hat{f}_{3}\rangle & \color{red}\langle \hat{f}_{1}|\hat{f}_{4}\rangle\\[1ex]
		\color{red} \langle \hat{f}_{2}|\hat{f}_{1}\rangle & \color{red}\langle \hat{f}_{2}|\hat{f}_{2}\rangle  & \color{green}\langle \hat{f}_{2}|\hat{f}_{3}\rangle & \color{red}\langle \hat{f}_{2}|\hat{f}_{4}\rangle\\[1ex]
		  \color{green}\langle \hat{f}_{3}|\hat{f}_{1}\rangle & \color{green}\langle \hat{f}_{3}|\hat{f}_{2}\rangle  & \color{red}\langle \hat{f}_{3}|\hat{f}_{3}\rangle & \color{green}\langle \hat{f}_{3}|\hat{f}_{4}\rangle\\[1ex]
		   \color{red}\langle \hat{f}_{4}|\hat{f}_{1}\rangle & \color{red}\langle \hat{f}_{4}|\hat{f}_{2}\rangle  & \color{green}\langle \hat{f}_{4}|\hat{f}_{3}\rangle & \color{red}\langle \hat{f}_{4}|\hat{f}_{4}\rangle\\[1ex]
	\end{bmatrix}.
\end{equation*}
Here, the color of each matrix element corresponds to its respective sign. The red color indicates a positive sign, whereas the green color indicates a negative sign. In the presented case, assuming $D_{0}$ is positive, the quaternion components $A_{0}$ and $B_{0}$ are positive, while $C_{0}$ is negative. Using the approach developed based on these conditions, the signs of the quaternion components can be estimated. 
\subsection{Supporting plots for example in Fig.~\ref{fig:8}}
\label{Supp:fig:8_plots}
\begin{figure}[h]
	\centering
	\includegraphics[scale=0.5]{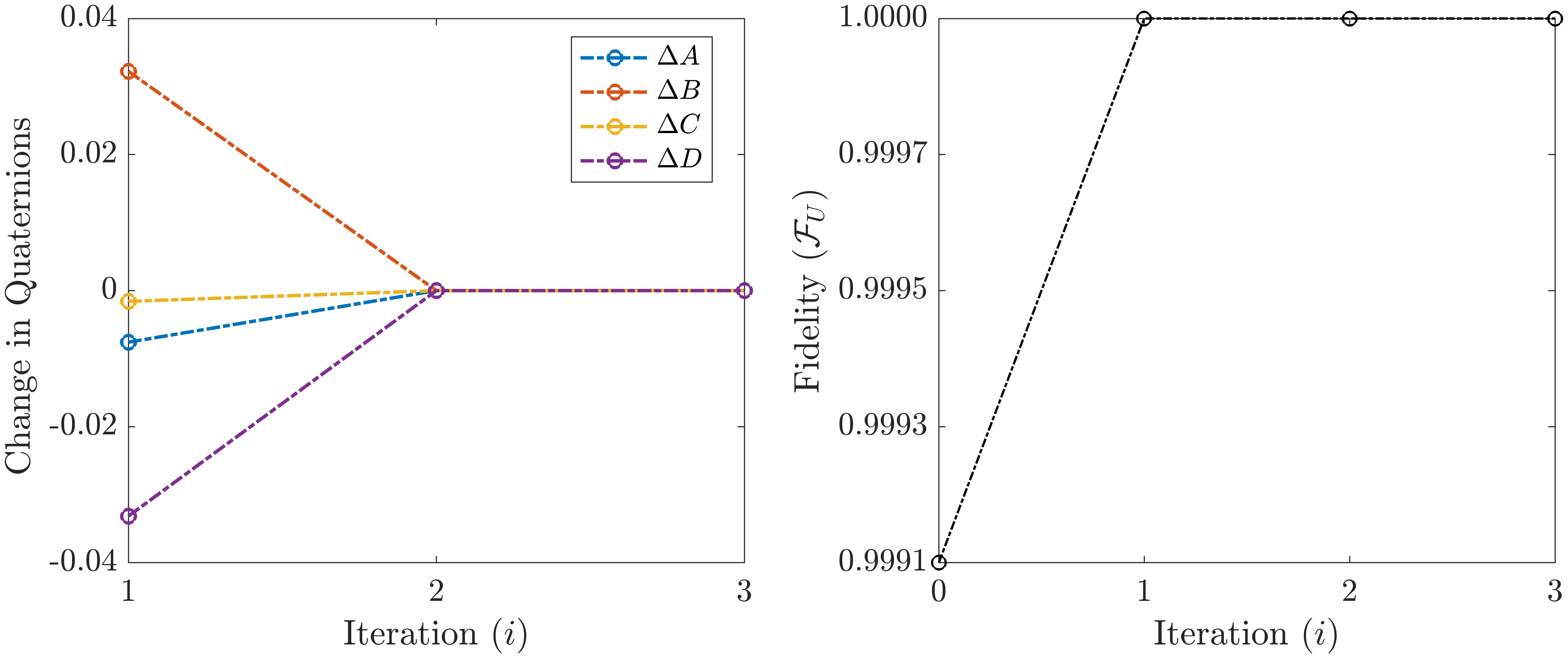}
	\caption{Plot of change in the values of quaternion components with iteration (on the left panel) and plot of fidelity ($\mathcal{F}_{U}$) with iteration (on the right panel) for the reconstruction example illustrated in Fig.~\ref{fig:8}. Here a change in quaternion value over iteration is computed using $\Delta{A} = A_{i+1}-A_{i}$, for example.}
	\label{fig:8_plots}
\end{figure}

The plots in Fig.~\ref{fig:8_plots} show the change in the values of reconstructed quaternion components with iteration (on the left panel) and fidelity of the reconstructed process with iteration (on the right panel) for the example presented in Fig.~\ref{fig:8}. The process fidelity ($\mathcal{F}_U$) is calculated using the following definition~\cite{glaser_unitaryControl,Devra_WQST}:  
\begin{equation}
    \label{Eq:D3}
    \mathcal{F}_U = \dfrac{|\text{tr}(U(U_{t})^{\dagger})|}{2},
\end{equation}
where $U$ is a reconstructed process matrix and $U_{t}$ is a target process matrix, which can be constructed based on Eq.~\eqref{Eq.19}.   

\subsection{Example of the reconstruction algorithm with optimization}
\label{Supp:recons_opt_example}
\begin{figure}[h]
	\centering
	\adjustbox{max width=\textwidth}{\includegraphics[scale=1.2]{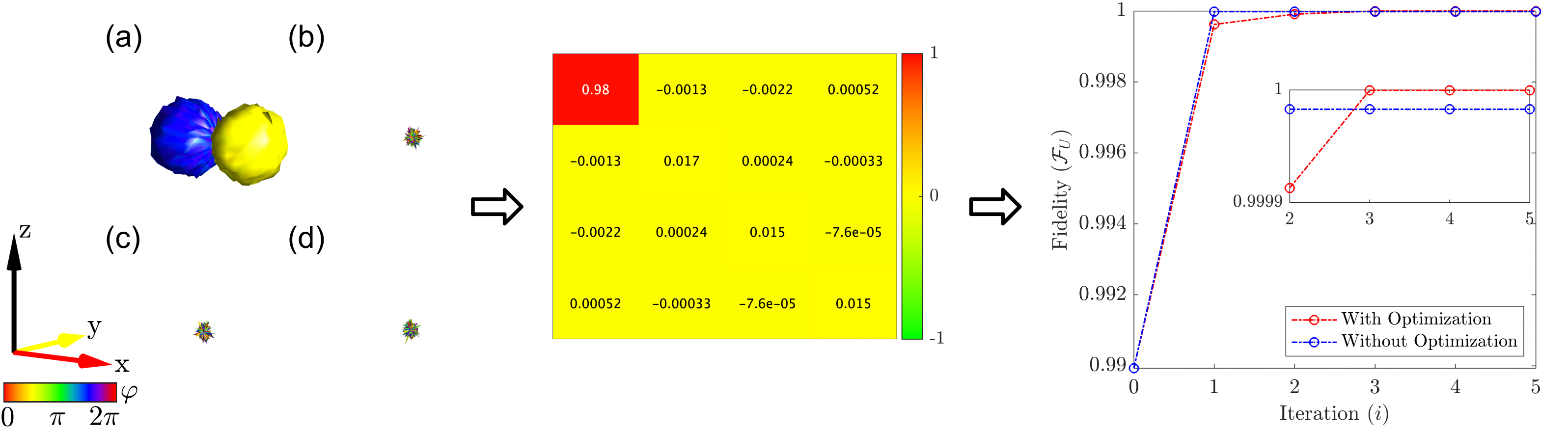}}
	\caption{Illustration of the reconstruction algorithm with and without optimization. The algorithms take as input tomographed scaled process droplets: (a) $\hat{f}_{1}$, (b) $\hat{f}_{2}$, (c) $\hat{f}_{3}$, and (d) $\hat{f}_{4}$ (see Eq.~\eqref{Eq.29}), displayed in the left panel. The zero-order estimates of the quaternion components are used as an initial guess to minimize the cost function $J$ given in Eq.~\eqref{Eq.33}. The fidelity with and without optimizations are shown in the right panel. The inset figure shows the variation of fidelity for iterations 2 to 5 between 0.9999 and 1. The plot of cost for this case is shown in Fig.~\ref{fig:14_supp}.}
	\label{fig:13_supp}
\end{figure}

\begin{figure}[h]
	\centering
	\includegraphics[scale=0.6]{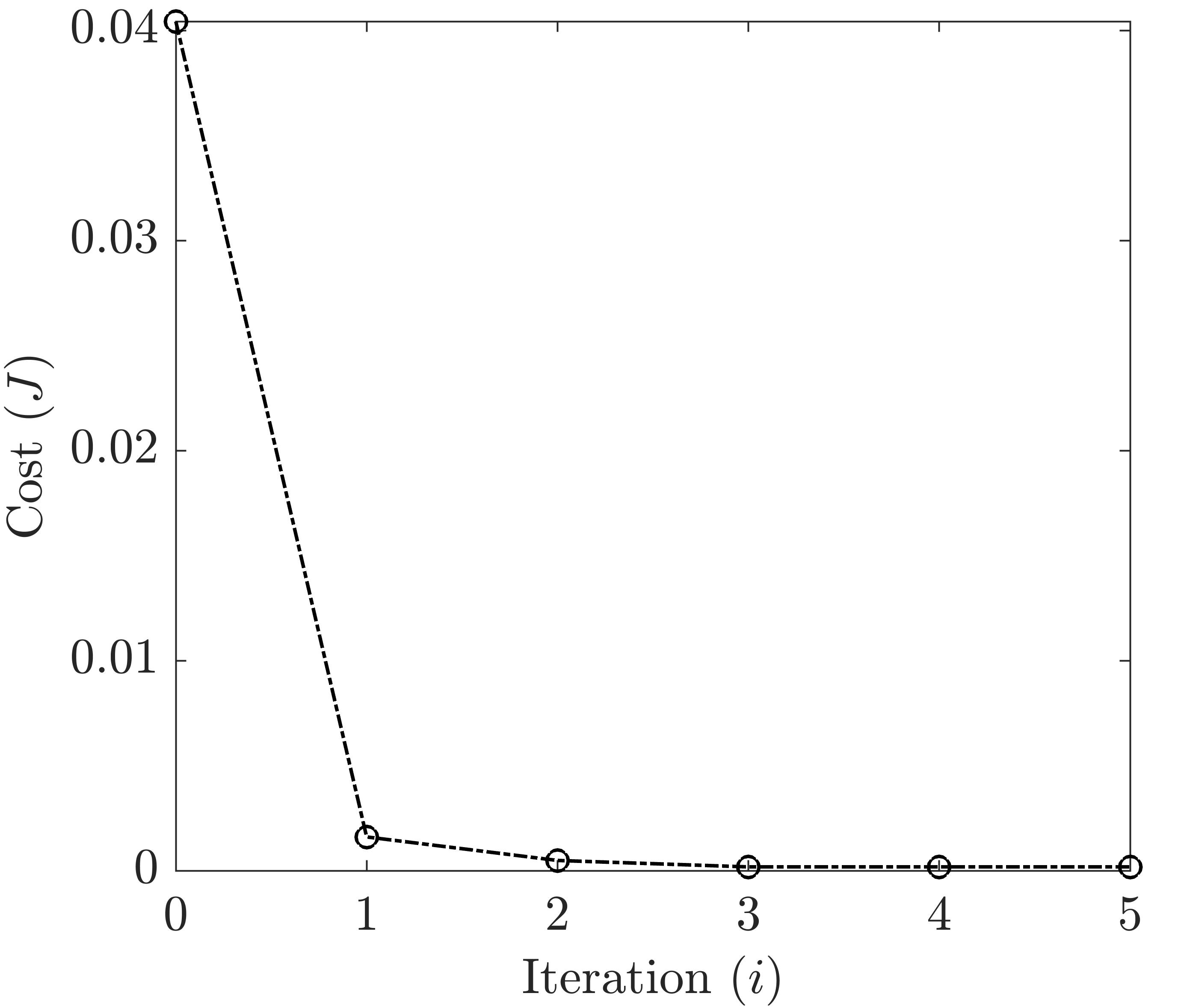}
	\caption{Illustration of cost function ($J$) used in the optimization (see Eq.~\eqref{Eq.33}) with iteration ($i$) for the case considered in Fig.~\ref{fig:13_supp}.}
	\label{fig:14_supp}
\end{figure}

We chose an example with quaternion components $A = 1$ and $B = C = D = 0$. The scaled tomographed droplets depicted in the left panel of Fig.~\ref{fig:13_supp} are obtained after $N_s = 500$ shots. The zero-order estimates computed from the correlation matrix, assuming $D_{0}$ is positive (note that $U$ is defined only upto the global phase), are $A_{0} = 0.9899$, $B_{0} = -0.1305$, $C_{0} = -0.1230$, and $D_{0} = 0.1211$. These zero-order estimates are used as an initial guess for the optimization. The fidelity ($\mathcal{F}_{U}$) per iteration is shown in the right panel of Fig.~\ref{fig:13_supp} for both approaches, i.e., with and without optimization. 
The quaternion components estimated after the final iteration ($i=5$) without optimization are $A_{5} = 0.99998$, $B_{5} = 0.0017$, $C_{5} = 0.0042$, and $D_{5} = 0.0038$, whereas, the quaternion components estimated after the final iteration ($i=21$) with optimization are: $A^{\text{opt}}_{21} = 0.99999$, $B^{\text{opt}}_{21} = 0.0005$, $C^{\text{opt}}_{21} = 0.0011$, and $D^{\text{opt}}_{21} = 0.0022$. The optimization terminated after 21 iterations following the defined tolerance limit. However, in Fig.~\ref{fig:13_supp}, the fidelity is displayed only up to 5 iterations because the change in fidelity with each iteration is not accurately visible on this scale. The process fidelity after the termination of the algorithm without optimization is 0.99998, whereas with optimization is 0.99999. The cost function with iteration for optimization is shown in Fig.~\ref{fig:14_supp}. Although the gain in fidelity in this case is very small, this shows that optimization does improve the estimation of the quaternion components.

\section{$\mathrm{U}_{3}$ gates}
\label{Supp:U3}
$\mathrm{U}_{3}(\theta,\phi,\lambda)$ gates act as a general~\cite{Devra_WQST} rotation matrix for a single qubit with three Euler angles $\theta,\phi$, and $\lambda$. The general expression is given by
\begin{equation}
	\label{eq.s16}
 \mathrm{U}_3\mathrm{(}\mathrm{\theta},\mathrm{\phi},\mathrm{\lambda})= 
	\begin{pmatrix} 
		\text{cos}(\theta/2) 
		& 
		-\text{e}^{i\lambda}\text{sin}(\theta/2) \cr
		\text{e}^{i\phi}\text{sin}(\theta/2) & \text{e}^{i\lambda+i\phi}\text{cos}(\theta/2)
	\end{pmatrix},
\end{equation}
which corresponds to:
\begin{equation}
	\label{eq.s17}
\mathrm{U}_3\mathrm{(}\mathrm{\theta},\mathrm{\phi},\mathrm{\lambda})
		=\textit{RZ}(\mathrm{\phi})\textit{RY}(\mathrm{\theta})\textit{RZ}(\mathrm{\lambda}).
\end{equation}
Hence, the rotation for scanning $\mathrm{U}_{3}(\beta,\alpha,0)$ corresponds to rotation of $\beta$ around $y$ axis followed a rotation of $\alpha$ around $z$ axis. Similarly, in detection-associated rotation ($\mathcal{D}$) step, $\mathrm{U}_{3}(-\frac{\pi}{2},0,0)$ corresponds to a rotation of angle $-\frac{\pi}{2}$ around $y$ axis which projects $x$ axis onto $z$ axis to measure $\langle{\sigma_{0x}}\rangle$ and $\langle{\sigma_{0x}\sigma_{1z}}\rangle$. Along the same lines, $\mathrm{U}_{3}(\frac{\pi}{2},0,\frac{\pi}{2})$ corresponds to a rotation of angle $\frac{\pi}{2}$ around $z$ axis followed by a $\frac{\pi}{2}$ around $y$ axis, which projects $y$ axis onto $z$ axis to measure $\langle{\sigma_{0y}}\rangle$ and $\langle{\sigma_{0y}\sigma_{1z}}\rangle$. 

\section{Supporting figures for the experimental results}
\label{Supp:experiments_add}
\subsection{Hadamard ($\mathrm{H}$) gate}
\label{Supp:Hadamard_gate}
\begin{figure}[h]
	\centering
	\adjustbox{max width=\textwidth}{\includegraphics[scale=1]{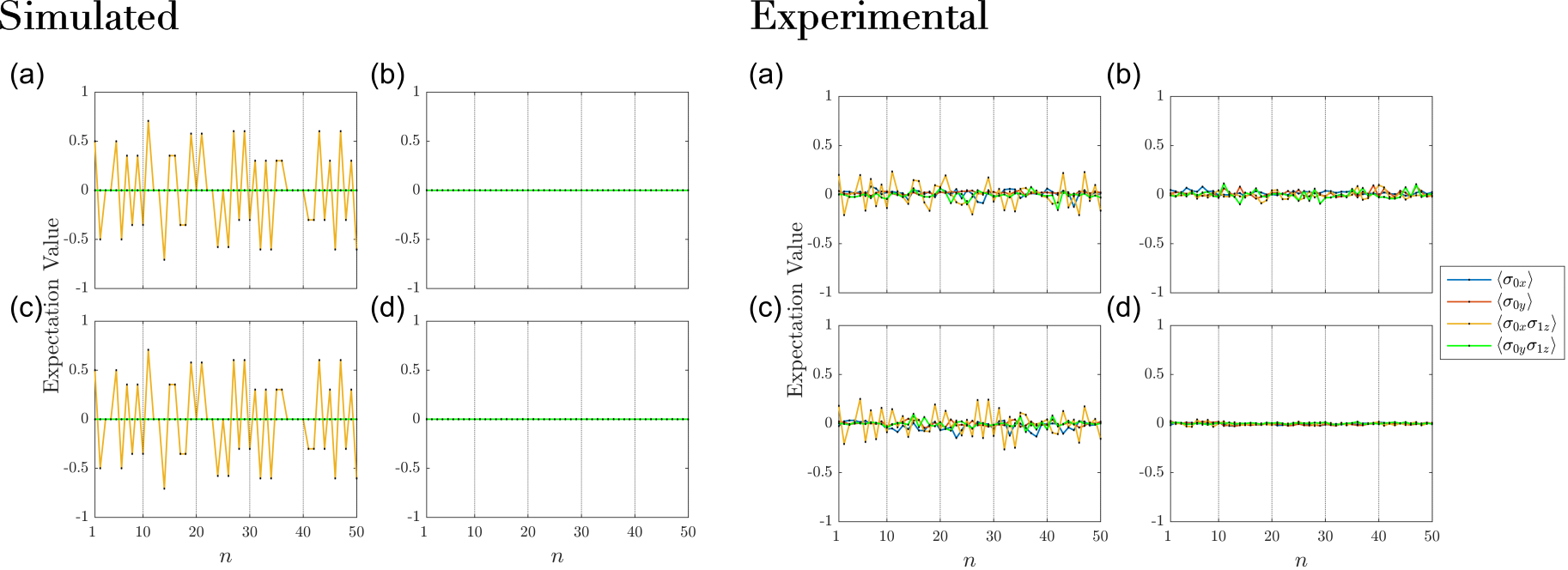}}
	\caption{Simulated and experimental expectation values corresponding to scaled process droplets (a) $\hat{f}_{1}$, (b) $\hat{f}_{2}$, (c) $\hat{f}_{3}$, and (d) $\hat{f}_{4}$ presented in Fig.~\ref{fig:13}. The expectation values were computed for the Lebedev 50 grid (depicted as black dots).}
	\label{fig:Had_expec}
\end{figure}

The section contains supporting information about the experimental results of the Hadamard gate presented in Fig.~\ref{fig:13}. The simulated and the experimental expectation values corresponding to different scaled process droplets are presented in Fig.~\ref{fig:Had_expec}. 
\\

\begin{figure}[h]
	\centering
	\adjustbox{max width=\textwidth}{\includegraphics[scale=1.1]{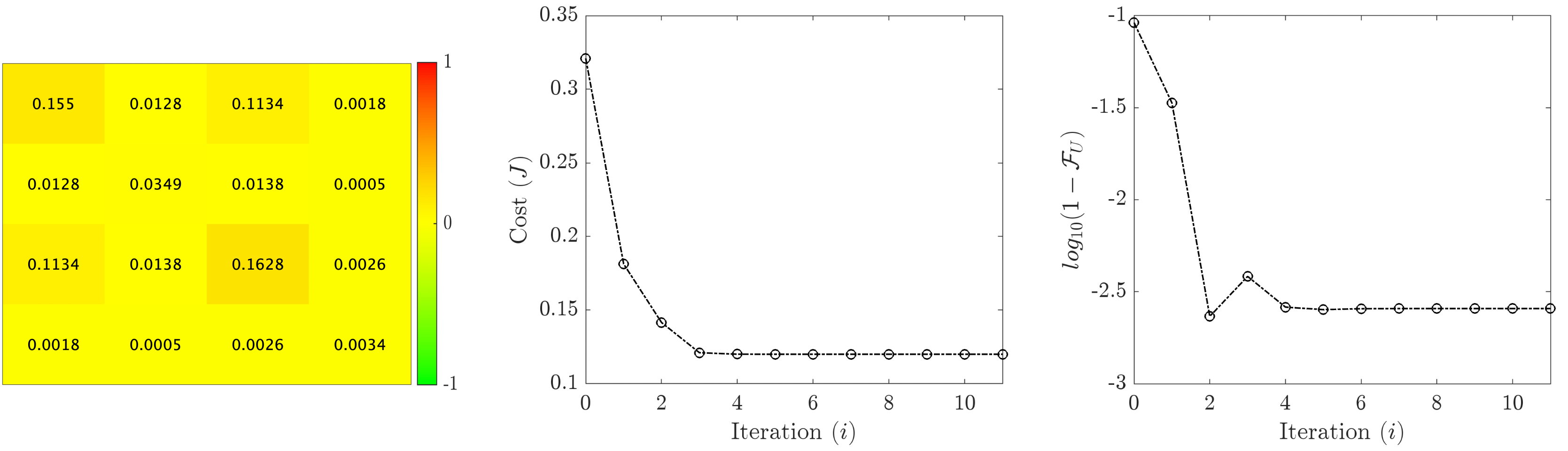}}
	\caption{Additional figures for the reconstruction using experimental tomography data of the Hadamard gate shown in Fig.~\ref{fig:13}. The figure shows the correlation matrix (left), plot of cost with iteration (middle), and plot of non-fidelity ($1-\mathcal{F}_{U}$) in logarithmic scale with iteration (right).}
	\label{fig: C.3}
\end{figure}

For the reconstruction algorithm, the initial guess computed from the correlation matrix is $A_{0} = 0.3937$, $B_{0} = 0.1868$, $C_{0} = 0.4035$, and $D_{0} = 0.0581$. The optimized quaternion components after reconstruction algorithm are $A^{\text{opt}}_{11} = 0.6566$, $B^{\text{opt}}_{11} = 0.0172$, $C^{\text{opt}}_{11} = 0.7540$, and $D^{\text{opt}}_{11} = 0.089$. The process fidelity achieved using the reconstruction algorithm with and without optimization is 0.9974 and 0.9934, respectively. See Fig.~\ref{fig: C.3} for the plot illustrating the cost and non-fidelity with optimization iterations, along with the correlation matrix.

\subsection{NOT ($\mathrm{X}$) gate}
\label{Supp:not_gate}
\begin{figure}[h]
	\centering
	\adjustbox{max width=\textwidth}{\includegraphics[scale=1]{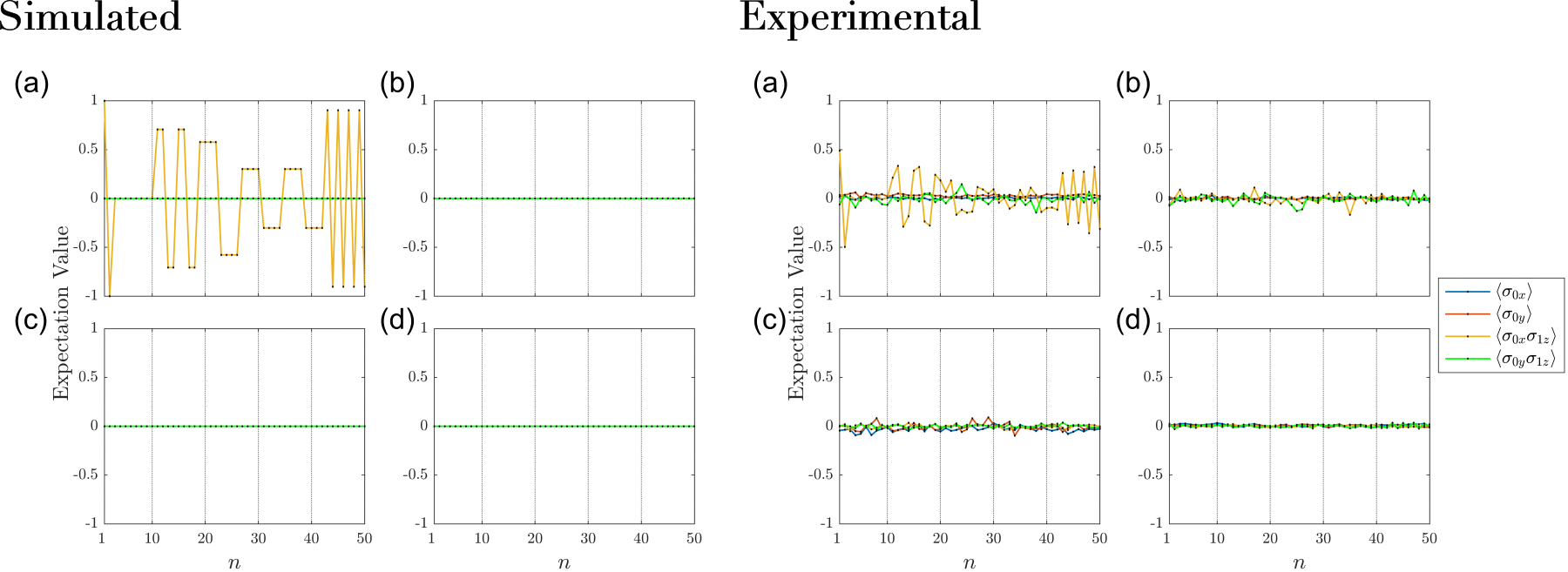}}
	\caption{Simulated and experimental expectation values corresponding to scaled process droplets (a) $\hat{f}_{1}$, (b) $\hat{f}_{2}$, (c) $\hat{f}_{3}$, and (d) $\hat{f}_{4}$ presented in Fig.~\ref{fig:14}. The expectation values were computed for the Lebedev 50 grid (depicted as black dots).}
	\label{fig:NOT_expec}
\end{figure}

The section contains supporting information about the experimental results of the NOT gate presented in Fig.~\ref{fig:14}. The simulated and the experimental expectation values corresponding to different scaled process droplets are presented in Fig.~\ref{fig:NOT_expec}. 
\\

\begin{figure}[h]
	\centering
	\adjustbox{max width=\textwidth}{\includegraphics[scale=1.1]{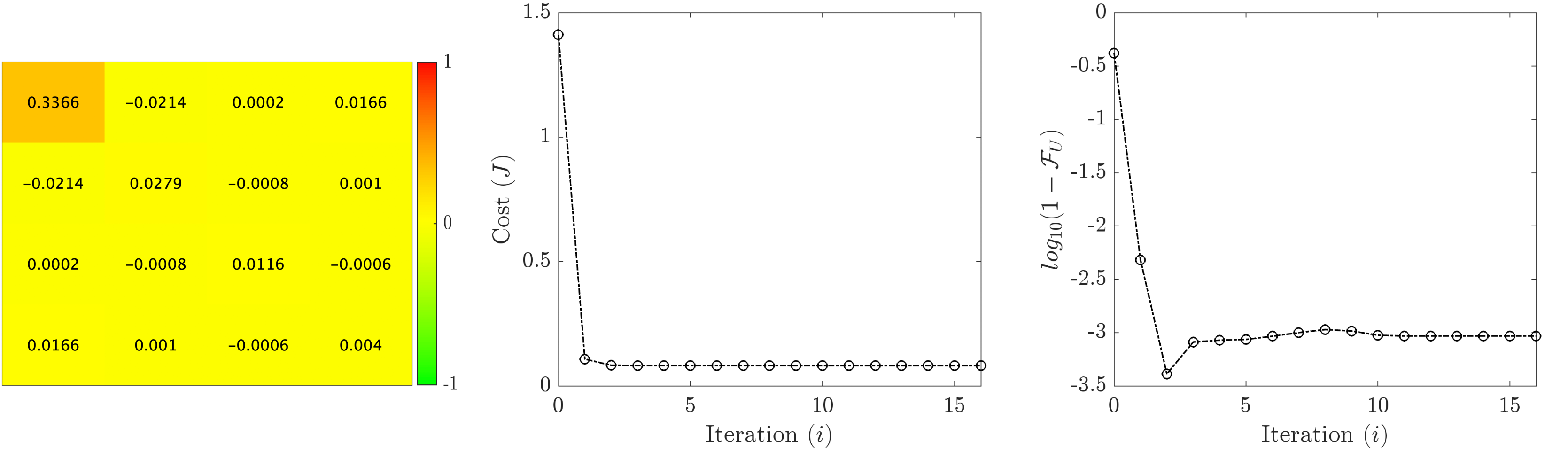}}
	\caption{Additional figures for the reconstruction using experimental tomography data of the NOT gate shown in Fig.~\ref{fig:14}. The figure shows the correlation matrix (left), plot of cost with iteration (middle), and plot of non-fidelity ($1-\mathcal{F}_{U}$) in logarithmic scale with iteration (right).}
	\label{fig: C.4}
\end{figure}

For the reconstruction algorithm, the initial estimations computed from the correlation matrix (see Fig.~\ref{fig: C.4}) are $A_{0} = 0.5802$, $B_{0} = -0.1672$, $C_{0} = 0.1075$, and $D_{0} = 0.0631$. After employing the reconstruction algorithm with optimization, the final quaternion values are $A^{\text{opt}}_{16} = 0.9991$, $B^{\text{opt}}_{16} = 0.0090$, $C^{\text{opt}}_{16} = -0.0407$, and $D^{\text{opt}}_{16} = -0.0110$. The process fidelity achieved using the reconstruction algorithm with and without optimization is 0.9991 and 0.9935, respectively. The plot illustrating the cost and non-fidelity with optimization iterations, along with the correlation matrix, is presented in Fig.~\ref{fig: C.4}.

\subsection{$\mathrm{Z}$ gate}
\label{Supp:Z_gate}
\begin{figure}[h]
	\centering
	\adjustbox{max width=\textwidth}{\includegraphics[scale=1]{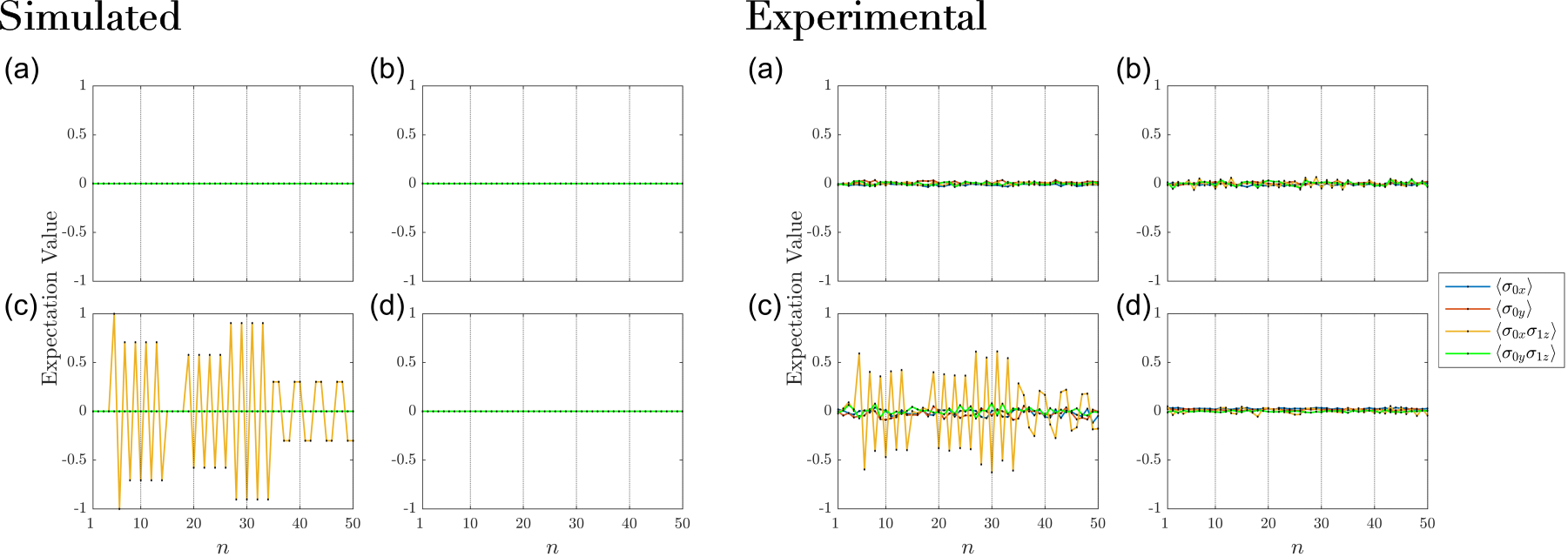}}
	\caption{Simulated and experimental expectation values corresponding to scaled process droplets (a) $\hat{f}_{1}$, (b) $\hat{f}_{2}$, (c) $\hat{f}_{3}$, and (d) $\hat{f}_{4}$ presented in Fig.~\ref{fig:15}. The expectation values were computed for the Lebedev 50 grid (depicted as black dots).}
	\label{fig:Z_expec}
\end{figure}

The section contains supporting information about the experimental results of the Z gate presented in Fig.~\ref{fig:15}. The simulated and the experimental expectation values corresponding to different scaled process droplets are presented in Fig.~\ref{fig:Z_expec}.
\\

\begin{figure}[h]
	\centering
	\adjustbox{max width=\textwidth}{\includegraphics[scale=1.1]{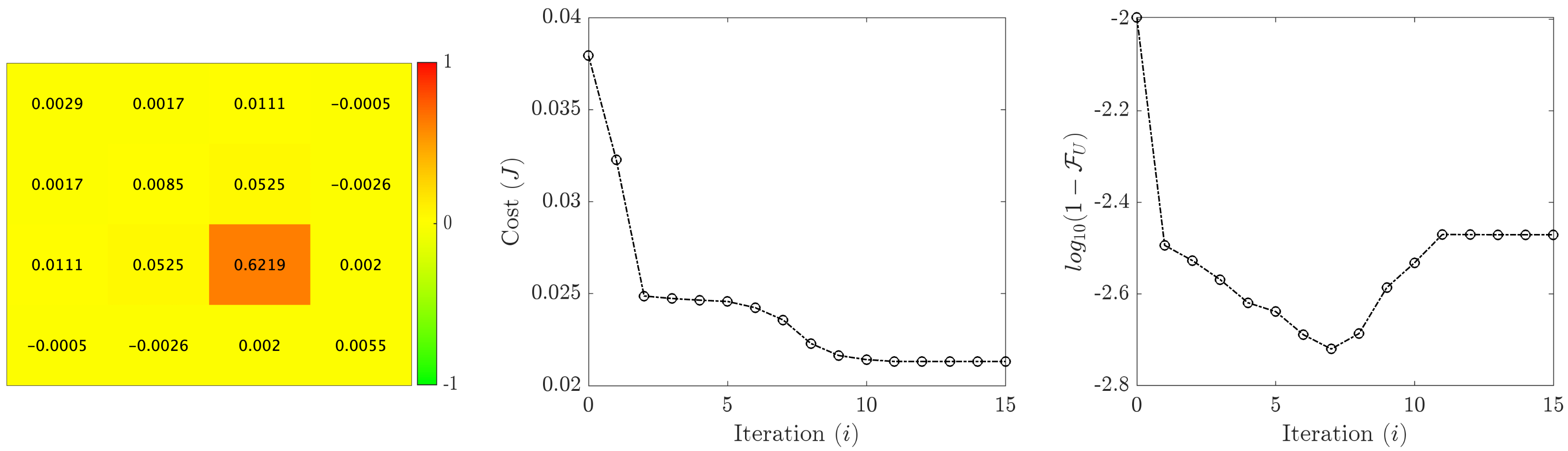}}
	\caption{Additional figures for the reconstruction using experimental tomography data of the NOT gate shown in Fig.~\ref{fig:15}. The figure shows the correlation matrix (left), plot of cost with iteration (middle), and plot of non-fidelity ($1-\mathcal{F}_{U}$) in logarithmic scale with iteration (right).}
	\label{fig: C.5}
\end{figure}

For reconstruction of the process from the experimental droplets, the initial estimations computed from the correlation matrix (see Fig.~\ref{fig: C.5}) are $A_{0} = 0.0543$, $B_{0} = 0.0920$, $C_{0} = 0.7886$, and $D_{0} = 0.0740$. The quaternion components obtained after employing reconstruction algorithm with optimization are $A^{\text{opt}}_{15} = -0.0023$, $B^{\text{opt}}_{15} = 0.0794$, $C^{\text{opt}}_{15} = 0.9966$, and $D^{\text{opt}}_{15} = 0.0210$. The process fidelity achieved using the reconstruction algorithm with and without optimization is 0.9966 and 0.9971, respectively. See Fig.~\ref{fig: C.5} for the plot illustrating the cost and non-fidelity with optimization iterations, along with the correlation matrix.
\clearpage
\section{Additional plots for the comparison of Wigner and standard process tomography approach}
\label{Supp:comparison}
\begin{figure}[h]
	\centering
	\adjustbox{max width=\textwidth}{\includegraphics[scale=0.5]{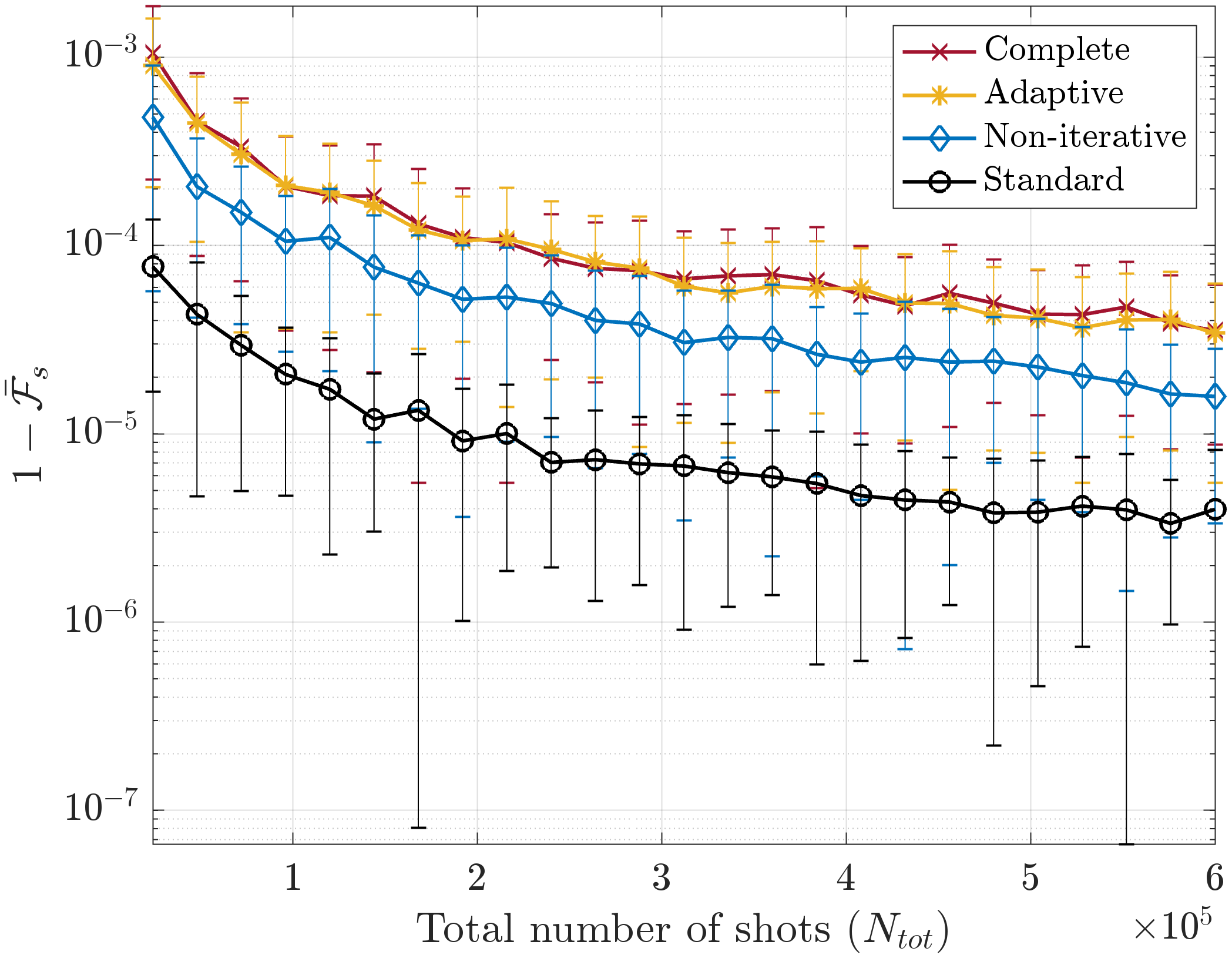}}
	\caption{Replot of Fig.~\ref{fig:18} including standard deviations, with vertical bars representing the standard deviation for each data point.}
	\label{fig: G.1}
\end{figure}

\begin{figure}[!]
	\centering
	\adjustbox{max width=\textwidth}{\includegraphics[scale=0.5]{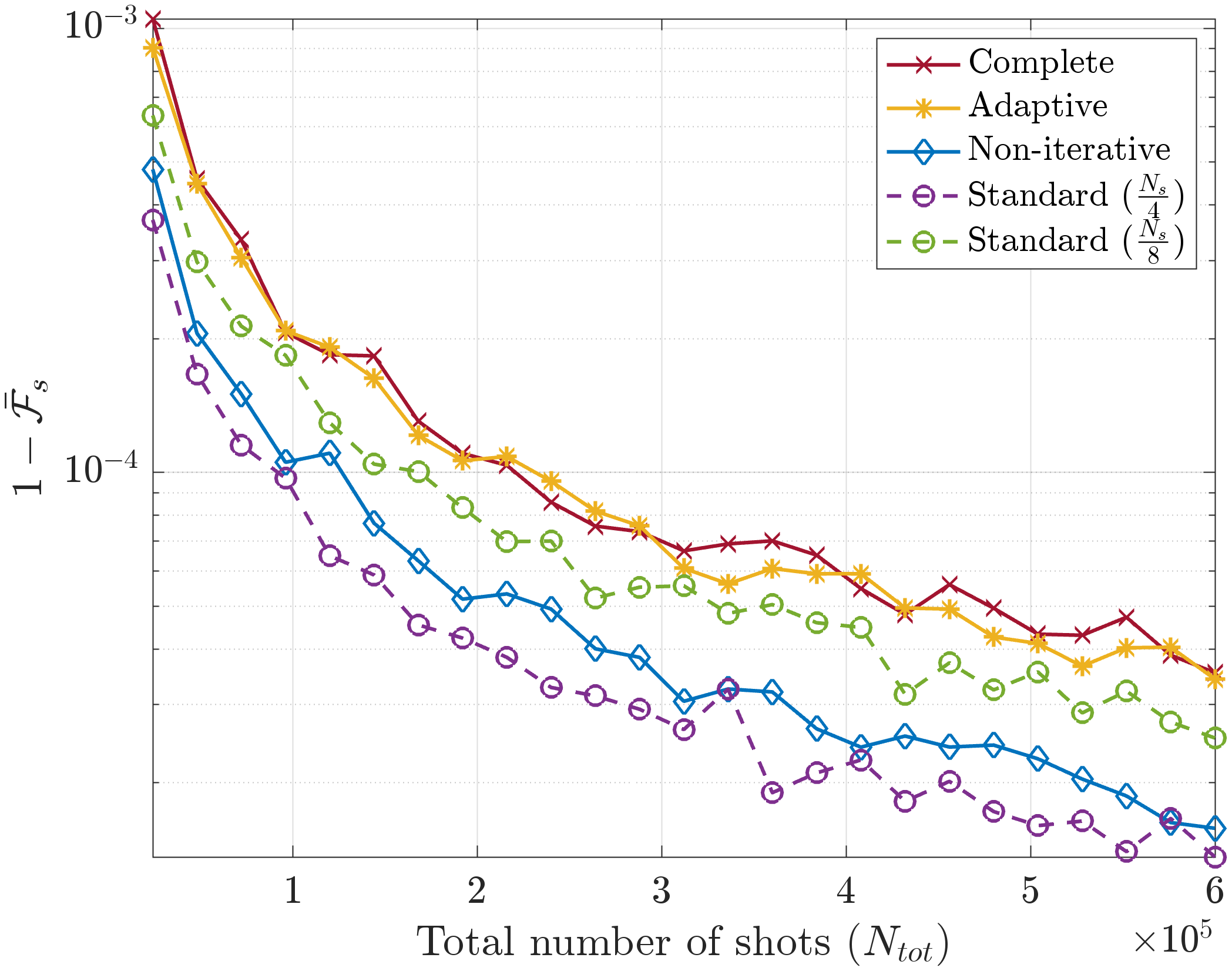}}
	\caption{Replot of Fig.~\ref{fig:18} with fewer shots $N_s$ for the standard approach (dashed lines), illustrating a performance comparison with the Wigner tomography approach (solid lines).}
	\label{fig: G.2}
\end{figure}

\twocolumngrid
\bibliography{WQPT_unknown.bib}
\end{document}